\pdfoutput=1
\documentclass[acmsmall,screen,natbib=false
,final,nonacm]{acmart}
\usepackage{etoolbox}
\newbool{arxiv}

\makeatletter
\@ifclasswith{acmart}{nonacm}{\booltrue{arxiv}}{\boolfalse{arxiv}}
\makeatother

\RequirePackage[
datamodel=acmdatamodel,
style=acmauthoryear, sortcites=false, maxcitenames=3, ]{biblatex}
\renewcommand\acksname{Acknowledgements}

\usepackage{mathtools}
\usepackage{algorithmic}
\usepackage{graphicx}
\usepackage{textcomp}
\usepackage{xcolor}
\usepackage{varwidth}
\usepackage{cprotect}
\usepackage[textsize=tiny,obeyFinal]{todonotes}
\usepackage{adjustbox}
\usepackage{mdframed}
\usepackage{xargs}

\usepackage{array}
\usepackage{booktabs}

\ifbool{arxiv}{
  \definecolor{dark-red}{rgb}{0.6,0,0}
  \definecolor{dark-blue}{rgb}{0,0.0,0.6}
  \definecolor{blue}{rgb}{0,0,0.8}
\def\pdfauthor{Éléonore Mangel, Paul-André Melliès and Guillaume Munch-Maccagnoni}

\hypersetup{
unicode=true,pdfusetitle,pdfauthor={\pdfauthor},
 bookmarks=true,bookmarksnumbered=false,bookmarksopen=false,
 breaklinks=true,pdfborder={0 0
 0},pdfborderstyle={},backref=page,colorlinks=true,
 ocgcolorlinks,linkcolor={dark-red},citecolor={dark-blue},urlcolor={blue}}

\definecolor{red}{rgb}{.8,0.2,0.2}
\definecolor{rouge}{rgb}{.8,0,0}
\definecolor{green}{rgb}{0,.5,0}
\definecolor{antigreen}{rgb}{.8,0,.8}
\definecolor{blue}{rgb}{0,0,1}
\definecolor{littleblue}{rgb}{.2,.2,.8}
\definecolor{LITTLEBLUE}{rgb}{.2,.2,.8}\definecolor{RED}{rgb}{.8,0.2,0.2}
\definecolor{graille}{rgb}{.01,.01,.00}
\definecolor{grispale}{rgb}{.6,.6,.6}
\definecolor{yellow}{rgb}{.5,.5,.2}
\definecolor{cyan}{rgb}{0.75,1,1}

}{}

\newcommand{\sepline}{\hspace*{\fill}\rule[-2ex]{0.618\textwidth}{0.4pt}\hspace*{\fill}\vspace*{2ex}\linebreak{}}

\ifbool{arxiv}{
  \newmdenv[shadow=true,shadowsize=2pt,shadowcolor=black!8]{framed}
}{
\newenvironment{framed}{}{}
}

\usepackage{comment}

\ifbool{arxiv}{
\includecomment{arxiv}
  \excludecomment{popl}
\newcommand{\pagebreakopportunitypopl}{}
\newcommand{\capitalize}[1]{#1}
}{
\excludecomment{arxiv}
  \includecomment{popl}
\newcommand{\pagebreakopportunitypopl}{\pagebreak[3]}
\usepackage{titlecaps}
  \Addlcwords{as a an of and to the in}\newcommand{\capitalize}[1]{\titlecap{#1}}
}

\newcommand{\sectioncaps}[1]{\section{\capitalize{#1}}}
\newcommand{\subsectioncaps}[1]{\subsection{\capitalize{#1}}}
\newcommand{\paragraphcaps}[1]{\paragraph{\capitalize{#1}}}

\usepackage[utf8]{inputenc}
\usepackage[T1]{fontenc}
\usepackage{amsmath}
\usepackage{amsthm}
\usepackage{stmaryrd}

\usepackage[scr=boondoxo]{mathalpha}

\usepackage{nicefrac}
\let\frac\nicefrac

\usepackage{perfectcut}
\usepackage{ebproof}
\usepackage{tikz-cd}
\usepackage[
]{cleveref}
\crefname{theorem}{thm.}{thms.}\crefname{figure}{fig.}{figs.}\crefname{example}{ex.}{exs.}\crefname{equation}{eq.}{eqs.}\crefname{lemma}{lem.}{lems.}\crefname{definition}{def.}{defs.}\crefname{proposition}{prop.}{props.}\Crefname{theorem}{Theorem}{Theorems}\Crefname{figure}{Figure}{Figures}\Crefname{example}{Example}{Examples}\Crefname{equation}{Equation}{Equations}\Crefname{lemma}{Lemma}{Lemmas}\Crefname{definition}{Definition}{Definitions}\Crefname{proposition}{Proposition}{Propositions}

\makeatletter
\@ifundefined{showcaptionsetup}{}{\PassOptionsToPackage{caption=false}{subfig}}
\usepackage{subfig}
\makeatother

\usepackage{microtype}

\makeatletter
\g@addto@macro\bfseries{\boldmath}
\makeatother

\makeatletter
\newcommand{\raisemathsymbol}[4]{\mathchoice
  {\raisebox{#2}{$\displaystyle#1$}}
  {\raisebox{#2}{$#1$}}
  {\raisebox{#3}{$\scriptstyle#1$}}
  {\raisebox{#4}{$\scriptscriptstyle#1$}}
}
\newcommand{\starup}{\raisemathsymbol\star{0.1pt}{0.066pt}{0.05pt}}

\newcommand{\ostar}{\mathbin{\mathpalette\make@circled\starup}}

\newcommand{\make@circled}[2]{\ooalign{$\m@th#1\smallbigcirc{#1}$\cr\hidewidth$\m@th#1#2$\hidewidth\cr}}
\newcommand{\smallbigcirc}[1]{\vcenter{\hbox{\scalebox{0.77778}{$\m@th#1\bigcirc$}}}}
\makeatother

\usepackage{rotating}
\global\long\def\rotxc#1{\begin{sideways}#1\end{sideways}}
\global\long\def\invert#1#2{\hbox{\rotxc{\rotxc{\ensuremath{#1#2}}}}}
\global\long\def\parr{\mathbin{\mathpalette\invert{\&}}}

\ifbool{arxiv}{
  \newcommand{\define}[1]{{\bf #1}}
}{
\newcommand{\define}[1]{\emph{#1}}
}

\newcommand{\dom}{\mathsf{dom}\hspace{0.5mm}}
\newcommand{\fv}{\upharpoonright\mathbf{fv}\hspace{0.4mm}}
\newcommand{\fvwo}{\mathbf{fv}\hspace{0.4mm}}
\newcommand{\fcv}{\upharpoonright\mathbf{fcv}\hspace{0.4mm}}
\newcommand{\fcvwo}{\mathbf{fcv}\hspace{0.4mm}}
\newcommand{\tsigma}{\tilde\sigma}
\newcommand{\tmu}{\tilde\mu}
\newcommand{\negN}{-^*}

\newcommand{\negP}{+^*}

\newcommand{\veps}{\varepsilon}
\newcommand{\1}{\mathsf{1}}
\newcommand{\nUparrow}{\mathord{\Uparrow}}
\newcommand{\nDownarrow}{\mathord{\Downarrow}}
\newcommand{\op}{{\mathsf{op}}}
\newcommand{\dupl}{\mathcal Dupl}
\newcommand{\Dduploid}{\mathcal D}
\newcommand{\Eduploid}{\mathcal E}
\newcommand{\inclusion}{\hookrightarrow}
\newcommand{\Id}{\mathrm{Id}}
\newcommand{\push}{\mathord{\cdot}}
\newcommand{\vdashf}{\vdash_{\scriptscriptstyle{f}}}
\let\iso\cong

\newcommand{\chirdual}[1]{#1^{*}}
\newcommand{\chirdoubledual}[1]{#1^{**}}

\newcommand{\dupldual}[1]{#1^{*}}
\newcommand{\dupldoubledual}[1]{#1^{**}}
\newcommand{\inv}[1]{{#1}^{-1}}

\let\cut\perfectcut

\global\long\def\mkernsp#1{\csname ltx@ifnextchar\endcsname.{\csname mkern\endcsname#1}{}}
\global\long\def\plus{^{{\mathord{+}\mkern-2mu}}\mkernsp{-2mu}}
\global\long\def\moins{^{{\mathord{-}}\mkern-1mu}\mkernsp{-3mu}}
\global\long\def\eps{^{{\varepsilon}\mkern-1mu}\mkernsp{-3mu}}

\global\long\def\defeq{\stackrel{\vphantom{.}\smash{\mathrm{\scriptscriptstyle{def}}}}{=}}
\global\long\def\orth#1{\overline{#1}}
\global\long\def\fold#1{\underline{#1}}

\global\long\def\graphto{\rightharpoonup}
\global\long\def\componentwise#1{\dot{#1}}
\global\long\def\linarrow{\mathbin{\rightarrowtriangle}}
\global\long\def\ev{\mathrm{ev}}
\global\long\def\Psh#1{\widehat{#1}}

\newcommand{\graphadjoint}[4]{{#1:#2\rightleftharpoons#3:#4}}

\newcommand{\command}[1]{\mathsf{#1}}

\newcommand{\pcomp}{\mathbin{\bullet}}
\newcommand{\ncomp}{\mathbin{\circ}}
\newcommand{\ccomp}{\mathbin{\mathrlap{\mathchoice{\mspace{2mu}}{\mspace{2.1mu}}{\mspace{0.95mu}}{\mspace{0.9mu}}\mathord{\cdot}}\mathord{\circ}}}

\newcommand{\keyword}[1]{\operatorname{\mathbf{#1}}}

\def\letinspace{3mu}
\newcommand{\letin}[3]{\keyword{let}\mkern\letinspace#1\mkern\letinspace\mathbin{=}\mkern\letinspace#2\mkern\letinspace\keyword{in}\mkern\letinspace#3}
\newcommand{\letinx}[4]{\keyword{let}\mkern\letinspace#1\mkern\letinspace\mathbin{\stackrel{#4}{=}}\mkern\letinspace#2\mkern\letinspace\keyword{in}\mkern\letinspace#3}
\newcommand{\letinplus}[3]{\letinx{#1}{#2}{#3}{\oplus}}
\newcommand{\letinminus}[3]{\letinx{#1}{#2}{#3}{\ominus}}
\newcommand{\letinepsilon}[3]{\letinx{#1}{#2}{#3}{\varepsilon}}
\newcommand{\letinepsilonprime}[3]{\letinx{#1}{#2}{#3}{\varepsilon'}}
\newcommand{\nifthenelse}[3]{\textbf{if} \hspace{.4em} {#1} \hspace{.4em} \textbf{then} \hspace{.4em} {#2} \hspace{.4em} \textbf{else} \hspace{.4em} {#3}}
\newcommand{\true}{\textbf{true}}
\newcommand{\false}{\textbf{false}}

\newcommandx\psum[4][usedefault, 1=]{\sum_{#2}\,#3\,#1|\, #4#1\rangle}
\newcommandx\dirac[2][usedefault, 1=]{1\,#1|\, #2#1\rangle}

\newcommand{\Acategory}{\mathscr{A}}
\newcommand{\Bcategory}{\mathscr{B}}
\newcommand{\Ccategory}{\mathscr{C}}

\newcommand{\Ecategory}{\mathscr{E}}
\newcommand{\Mcategory}{\mathscr{M}}
\newcommand{\Ncategory}{\mathscr{N}}
\newcommand{\Pcategory}{\mathscr{P}}
\newcommand{\Ncategoryl}{\mathscr{N}_l}
\newcommand{\Pcategoryt}{\mathscr{P}_t}
\newcommand{\twocategory}{\mathbf{2}}
\newcommand{\kleisli}[2]{\mathbf{Kl}{[#1,#2]}}
\newcommand{\cokleisli}[2]{\mathbf{coKl}{[#1,#2]}}

\newcommand{\collage}[2]{\mathbf{coll}_{#1,#2}}

\newcommand{\LC}{\textbf{LC}}
\newcommand{\LKT}{\textbf{LKT}}
\newcommand{\LKQ}{\textbf{LKQ}}
\newcommand{\polsystem}[1]{\textbf{#1}^\eta_p}

\newcommand{\tensorialand}{\varowedge}
\newcommand{\tensorialor}{\varovee}
\newcommand{\tensorialtrue}{\mathbf{\mathit{true}}}
\newcommand{\tensorialfalse}{\mathbf{\mathit{false}}}

\newcommand{\duploid}[2]{\mathbf{dupl}_{#1,#2}}

\newcommand{\median}{\mathbf{trans}}
\newcommand{\id}[1]{\mathrm{id}_{#1}}
\newcommand{\tensor}{\otimes}

\newcommand{\Set}{\mathit{Set}}

\newcommand{\bmid}{\mathrel{\boldsymbol{\nthleft{1}|}}}
\newcommand{\stretcharray}{\arraycolsep=0.5ex\def\arraystretch{1.5}}
\newcommand{\stretcharraymore}{\arraycolsep=2em\def\arraystretch{3}}

\newcommand{\raisesymbol}[2]{
  \setbox0=\hbox{\mbox{\ensuremath{#2 x}}}
  \setbox1=\hbox{\mbox{\ensuremath{#1}}}
  \dimen0=\dimexpr(\dp1 + \ht0) - (\dp0 + \ht1)\relax
  \dimen0=0.5\dimen0
  \raisebox{\dimen0}{\usebox1}
}
\newcommand{\smallsymbol}[1]{\mkern1mu\mathord{\mathchoice
  {\raisesymbol{\scriptstyle{#1}}{\displaystyle}}
  {\raisesymbol{\scriptstyle{#1}}{}}
  {\raisesymbol{\scriptscriptstyle{#1}}{\scriptstyle}}
  {\raisesymbol{\scriptscriptstyle{#1}}{\scriptscriptstyle}}}\mkern1mu}
\newcommand{\smallotimes}{\smallsymbol{\otimes}}
\newcommand{\smallwith}{\smallsymbol{\&}}
\newcommand{\smallparr}{\mathord{\mathpalette\invert{\smallwith}}}

\newcommand{\ltensortimes}{\ltensor}
\newcommand{\rtensortimes}{\rtensor}
\newcommand{\lparrtimes}{\ltimes}
\newcommand{\rparrtimes}{\rtimes}

\RequirePackage{graphicx}
\RequirePackage{calc}

\makeatletter
\newsavebox\gadmm@boxsrc
\newsavebox\gadmm@boxtrg
\newcommand{\gadmm@resize}[4]{\sbox{\gadmm@boxtrg}{$\m@th#4$}\raisebox{-#3}{\resizebox{#1}{#2+#3}{\raisebox{\dp\gadmm@boxtrg}{\usebox{\gadmm@boxtrg}}}}}
\newcommand{\gadmm@resizeto}[2]{\sbox{\gadmm@boxsrc}{$\m@th#1$}\gadmm@resize{\wd\gadmm@boxsrc}{\ht\gadmm@boxsrc}{\dp\gadmm@boxsrc}{#2}
}
\newcommand{\gadmm@resizeadjust}[3]{\sbox{\gadmm@boxsrc}{$\m@th#1$}\gadmm@resize{\wd\gadmm@boxsrc-#3-#3-#3}{\ht\gadmm@boxsrc-#3-#3-#3}{\dp\gadmm@boxsrc+#3}{#2}
}

\def\blacktriangleleftadjust#1{\gadmm@resizeadjust{#1{\rtimes}}{#1{\mathbin{\mkern6mu{\blacktriangleleft}}}}{0.45pt}}
\newcommand{\rtensornormal}{\mkern2mu\blacktriangleleftadjust{}}
\newcommand{\rtensorscript}{\mkern2.5mu\blacktriangleleftadjust\scriptstyle}
\newcommand{\rtensorsscript}{\mkern3mu\blacktriangleleftadjust\scriptscriptstyle}

\newcommand{\rtensor}{\mathbin{\mathrlap{\mathchoice{\rtensornormal}{\rtensornormal}{\rtensorscript}{\rtensorsscript}}\rtimes}}

\newcommand{\ltensornormal}{\mkern2.3mu\gadmm@resizeadjust{\ltimes}{\mathbin{\mkern-2.5mu{\blacktriangleright}\mkern5mu}}{0.95pt}}
\newcommand{\ltensorscript}{\mkern2.5mu\gadmm@resizeadjust{\scriptstyle{\ltimes}}{\mathbin{\scriptstyle{\mkern-3mu\blacktriangleright\mkern5mu}}}{0.65pt}}
\newcommand{\ltensorsscript}{\mkern2.8mu\gadmm@resizeadjust{\scriptscriptstyle{\ltimes}}{\mathbin{\scriptscriptstyle{\mkern-4.5mu\blacktriangleright\mkern8mu}}}{0.4pt}}

\newcommand{\ltensor}{\mathbin{\mathrlap{\mathchoice{\ltensornormal}{\ltensornormal}{\ltensorscript}{\ltensorsscript}}\ltimes}}

\makeatother

\newcommand{\joyallemma}{Joyal's obstruction theorem}

\usepackage{tikz}
\usepackage{tikz-cd}
\usetikzlibrary{matrix}
\usetikzlibrary{arrows}
\usetikzlibrary{arrows.meta}
\usetikzlibrary{decorations.pathmorphing,shapes}
\usetikzlibrary{decorations.markings}

\tikzset{spanmap/.style={
            decoration={markings,
            mark= at position 0.5 with {
                  \node[transform shape] (tempnode) {$|$};
}
              },
              postaction={decorate}
}
}
\def\diagrowsep{1em}

\ifbool{arxiv}{}{\setcopyright{cc}
\setcctype{by-nc-nd}
\acmDOI{10.1145/3776715}
\acmYear{2026}
\acmJournal{PACMPL}
\acmVolume{10}
\acmNumber{POPL}
\acmArticle{73}
\acmMonth{1}
\received{2025-07-10}
\received[accepted]{2025-11-06}
}

\begin{document}

\def\FH{Hasegawa-Thielecke}

\newcommand{\breakiftex}{\texorpdfstring{\\}{}}

\ifbool{arxiv}{\def\extver{(extended version)}}{\def\extver{}}
\title[\capitalize{Classical notions of computation and the \FH{} theorem \extver}
]{\ifbool{arxiv}{Classical notions of computation and the\breakiftex{}
  \FH{} theorem \extver }{Classical Notions of Computation\breakiftex{}
  and the \FH{} Theorem}}

\thanks{\ifbool{arxiv}{\today. This is a slightly extended version with more illustrations
and proofs of a paper published in PACMPL
(\url{https://doi.org/10.1145/3776715}).
}{A slightly extended version of this paper with more illustrations
and proofs is available on arXiv~\citep{MMMM2025}.
}}

\author{\'El{\'e}onore Mangel}
\orcid{0009-0001-2530-2318}
\affiliation{\institution{Univ. Paris Cit{\'e}, CNRS, INRIA}\city{Paris}\country{France}}
\email{eleonore.mangel@irif.fr}

\author{Paul-André Melli{\`e}s}
\orcid{0000-0001-6180-2275}
\affiliation{\institution{CNRS, Univ. Paris Cit{\'e}, INRIA}\city{Paris}\country{France}}
\email{mellies@irif.fr}

\author{Guillaume Munch-Maccagnoni}
\orcid{0009-0001-7914-7165}
\affiliation{\institution{INRIA, LS2N CNRS}\city{Nantes}\country{France}}
\email{guillaume.munch-maccagnoni@inria.fr}

\begin{abstract}
In the spirit of the Curry-Howard correspondence between proofs 
and programs, we define and study a syntax and semantics
for classical logic equipped with a computationally involutive negation,
using a polarised effect calculus, the linear classical $L$-calculus.
A main challenge in designing a denotational semantics for the calculus
is to accommodate both call-by-value and call-by-name evaluation strategies, 
which leads to a failure of associativity of composition.
In order to tackle this issue, we define a notion of adjunction
between graph morphisms on non-associative categories, which we use to
formulate polarized and non-associative notions of \emph{symmetric
monoidal closed duploid} and of \emph{dialogue duploid}. We show that
they provide a direct style counterpart to adjunction models:
\emph{linear effect adjunctions} for the (linear) call-by-push-value calculus
and \emph{dialogue chiralities} for linear continuations,
respectively.
In particular, we show that the syntax of the linear classical
$L$-calculus can be interpreted in any dialogue duploid, and that it
defines in fact a syntactic dialogue duploid.
As an application, we establish, by semantic as well as syntactic
means, the \FH{} theorem, which states that the notions
of central map and of thunkable map coincide in any dialogue duploid (in
particular, for any double negation monad on a symmetric monoidal category).
\end{abstract}




\maketitle

\sectioncaps{Introduction}

\subsectioncaps{Emergence of non-associativity between call-by-value and call-by-name}

In this paper, we combine methods from
proof theory and programming language semantics to investigate
the meaning of expressions
for proofs or programs,
starting with those of the form\begin{equation}\label[equation]{equation/letin}
\letin{a}{u}{t}
\end{equation}
where $t$ and $u$ are effectful
expressions\ifbool{arxiv}{, and where~$t$ possibly contains
free instances of the variable~$a$}.
One main difficulty we face is that there are
two canonical ways of assigning meaning to the $\keyword{let}$ construct~\eqref{equation/letin}
depending on the evaluation paradigm at work:

\medbreak
\noindent
\emph{In the call-by-value (CBV) paradigm,} the expression~$u$
performs a number of actions and returns a value $v$~; the value~$v$ 
is then substituted for every free instance of the variable~$a$ in the expression~$t$~;
it is then the turn of the expression $t[a:=v]$ to perform its actions and to return a value.

\medbreak
\noindent
\emph{In the call-by-name (CBN) paradigm,}
the expression~$t$ performs its actions and is evaluated while
the expression~$u$ is ``frozen'' and substituted for each free instance /
of the variable~$a$ in~$t$~;~a
new copy of the expression~$u$ performs its actions and is evaluated
each time a free instance of the variable~$a$
appears as head variable during the evaluation of the expression $t[a:=u]$.
\medbreak

By way of illustration, consider the probabilistic program
$$
\letin{x}{(\,\frac{1}{3}\,\true+\frac{2}{3}\,\false\,)}{t}
$$
which returns a probabilistic distribution of values of a given type
computed by the pure program~$t$ defined as\vspace{-1.3ex}
\begin{center}
\begin{tabular}{ccc}	
$\textbf{if}$ $x=\true$ & \hspace{-.5em} $\textbf{then}$ \hspace{-.5em} & 
${(\nifthenelse{x=\true}{a}{b})}$
\\
& \hspace{-.5em} $\textbf{else}$ \hspace{-.5em} & ${(\nifthenelse{x=\true}{c}{d})}$
\end{tabular}\ifbool{arxiv}{\vspace{1ex}}{\vspace{0.3ex}}
\end{center}
where~$a,b,c,d$ are values of the given type.
The meaning of the probabilistic program is 
equal to the distribution\vspace{-1.5ex}
\begin{align*}
u \quad &= \quad \frac{1}{3}\,a+\frac{2}{3}\,d && \mspace{-100mu}\text{in call by value, and}\\
u \quad &= \quad \frac{1}{9}\,a+\frac{2}{9}\,b+\frac{2}{9}\,c+\frac{4}{9}\,d && \mspace{-100mu}\text{in call by name.}
\end{align*}
Note that the value of~$a$ is sampled once and for all during the call-by-value evaluation,
while it is sampled twice during the call-by-name evaluation.
This explains that the summands $b$ and $\command c$ do not appear in the former case,
and that fractions over $9=3\times 3$ appear in the latter case.

\paragraphcaps{Kleisli categories}
The seminal work on computational effects
by \citet{Moggi89computationallambda-calculus, Moggi1991} initiated a
well-established tradition of interpreting CBV expressions of type
$a:A\vdash t:B$ as maps $t:A\to B$ in the Kleisli
category~$\kleisli{\Ccategory}{T}$ associated to a
monad~$(T,\mu,\eta)$ on a category~$\Ccategory$
(\cite{Filinski94representingmonads,PowerRobinson97,fuhrmanndirectmodels,Levy99CBPV,Plotkin_2002,power2002premonoidal,ThieleckeFuhrmann04,Lindley_2005,Katsumata2005}
among others).
Recall that a map $f:A\to B$ in the Kleisli category is a map $f:A\to TB$ in the original category~$\Ccategory$
and that two maps~$f:A\to TB$ and~$g:B\to TC$ are composed 
using the multiplication~$\mu$ of the monad:
\[
g\pcomp f \quad = \quad A\xlongrightarrow{\smash{f}} TB \xlongrightarrow{\smash{tg}} TTC \xlongrightarrow{\mu_C} TC
\]
Symmetrically, there is a well-established tradition after \citet{Gir87}
of interpreting CBN expressions of type $a:A\vdash t:B$ as maps $t:A\to B$ 
in the co-Kleisli category~$\cokleisli{\Ccategory}{K}$ associated to a computational comonad~$(K,\delta,\varepsilon)$
on a given category~$\Ccategory$ of types and pure programs.
Recall that a map $f:A\to B$ in the co-Kleisli category is a map $f:KA\to B$ in the original category $\Ccategory$
and that two maps $f:A\to B$ and $g:B\to C$ are composed in the co-Kleisli category
using the comultiplication~$\delta$ of the comonad:\vspace{-0.5ex}
\[
g\ncomp f \quad = \quad KA\xlongrightarrow{\smash{\delta_A}}  KKA \xlongrightarrow{\smash{Kf}} KB \xlongrightarrow{\smash{g}} C
\]
\noindent
The mathematical property
that composition is \emph{associative} in $\kleisli{\Ccategory}{T}$ and $\cokleisli{\Ccategory}{K}$,
that is:
\[
\begin{array}{ccc}
h\pcomp (g\pcomp f) \, = \, (h\pcomp g)\pcomp f
& & 
h\ncomp (g\ncomp f) \, = \, (h\ncomp g)\ncomp f
\end{array}
\]
reflects the {computational property} that for all effectful expressions
$\vdash f:A$ and $a:A\vdash g:B$ and $b:B\vdash h:C,$
the two effectful expressions~$(i)$ and~$(ii)$ defined below
\begin{center}
\fbox{
\begin{tabular}{ccc}
\vspace{-1.4em}
\\
$(i)$ & $\letinepsilon{a}{f}{(\letinepsilonprime{b}{g}{h})}$ &
\\
\vspace{-1.2em}
\\
$(ii)$ & $\letinepsilonprime{b}{(\letinepsilon{a}{f}{g})}{h}$ &
\\
\vspace{-1.2em}
\\
\end{tabular}
}
\end{center}
are equal whenever the \emph{polarities}
$\varepsilon,\varepsilon'\in\{\oplus,\ominus\}$
of the $\keyword{let}$ constructs are the same.
Here, we use the polarity $\varepsilon\in\{\oplus,\ominus\}$ to
indicate in which style $\letinepsilon{a}{u}{t}$ should be evaluated:
CBV ($\varepsilon=\oplus$) or CBN ($\varepsilon=\ominus$).
The fact that the expressions~$(i)$ and~$(ii)$ behave in the same way
implies in particular that they evaluate $f$, $g$ and~$h$ in the same order
in CBV as well as in CBN, as shown below.
\pagebreakopportunitypopl
\vspace{1ex}
\[
\begin{array}{ccc}
\toprule
\textrm{Composition style} & & \textrm{Order of evaluation} \\
\midrule
(\varepsilon,\varepsilon') = (\oplus,\oplus) & &
\begin{array}{cc}
\vspace{-1em}
\\
(i) = (ii) &
\mbox{$f$ then $g$ then $h$} 
\\
\vspace{-1.2em}
\\
\end{array}
\\
(\varepsilon,\varepsilon') = (\ominus,\ominus) & &
\begin{array}{cc}
\vspace{-1em}
\\
(i) = (ii) & 
\mbox{$h$ then $g$ then $f$}
\\
\vspace{-1.2em}
\\
\end{array}
\\
\bottomrule
\end{array}
\]

\paragraphcaps{Mixing call-by-name and call-by-value}
In many concrete
situations, the programmer would like to control and reason about the
order of evaluation.
This can be modelled by letting
both styles of $\keyword{let}$ constructs appear inside expressions.
Inspecting the two effectful expressions~$(i)$ and~$(ii)$ again in that hybrid scenario,
we see that the two expressions~$(i)$ and~$(ii)$ behave in the same way when $(\varepsilon,\varepsilon') = (\ominus,\oplus)$
but behave differently when $(\varepsilon,\varepsilon') = (\oplus,\ominus)$.
In particular, in that latter case, the expression~$f$ is evaluated before $h$ and then $g$ in~$(i)$
whereas the expression~$h$ is evaluated before $f$ and then $g$ in~$(ii)$.\vspace*{0.5ex}
\[
\begin{array}{ccc}
\toprule
\textrm{Composition style} & & \textrm{Order of evaluation} \\
\midrule
(\varepsilon,\varepsilon') = (\ominus,\oplus) & &
\begin{array}{cc}
\vspace{-1em}
\\
(i) = (ii) & \mbox{$g$ then $f$ then $h$} \\
\vspace{-1.2em}
\\
\end{array}
\\
(\varepsilon,\varepsilon') = (\oplus,\ominus) & &
\begin{array}{cc}
\vspace{-1em}
\\
\hspace*{1.1em} (i) \hspace*{1.3em}
& 
\mbox{$f$ then $h$ then $g$}
\\
\hspace*{1.1em} (ii) \hspace*{1.3em}
&
\mbox{$h$ then $f$ then $g$}
\\
\vspace{-1.2em}
\\
\end{array}
\\
\bottomrule
\end{array}\vspace*{1ex}
\]
A natural question is how we could develop a mathematical framework
that considers seriously the combination of evaluation paradigms,
without a priori biases towards monads nor comonads.
In order to reflect these equations,
such a framework needs to integrate both Kleisli and co-Kleisli categories,
where the former associativity equation holds
$$
(h\pcomp g)\ncomp f \hspace{.5em} = \hspace{.5em} h\pcomp (g\ncomp f)
$$
but where the latter associativity equation
$$
(h\ncomp g)\pcomp f \hspace{.5em} = \hspace{.5em} h\ncomp (g\pcomp f)
$$
does not necessarily hold in general.
There is no hope of defining categories and we thus need to move 
to ``non-associative'' forms of categories.
This is the direction taken by the third author~\cite*{munchduploids}
based on a non-associative and polarized notion of \emph{duploid}.

The idea of non-associativity is far from new:
it appeared for the first time in Girard's \emph{``constructive''}
classical logic $\LC$, which introduced a formal distinction between
``positive'' and ``negative'' formulae~\cite{girardnew}.
The idea then resurfaced with the ``Blass problem'' in game semantics~\cite{Blass1992a,Abramsky2003}, 
whose origin was traced back to the existence of an adjunction between 
categories of ``positive'' and ``negative'' games~\cite{Mellies2005ag3}. 
However, non-associativity was mainly perceived as an anomaly
until the introduction of duploids and their \emph{computational account of adjunctions}
where it was shown that having ``three fourths'' of the associativity
equations captures directly effectful computation integrating both
monadic and comonadic effects.

\subsectioncaps{The non-associative category associated to an adjunction}

\paragraphcaps{Adjunctions}
In order to intertwine the interpretations of CBV 
and CBN evaluation in a single mathematical structure including the Kleisli and co-Kleisli categories,
a good starting point is indeed to consider a pair of adjoint functors\begin{equation}\label[equation]{equation/adjunction-LR}
\begin{tikzcd}[column sep = 1em]
\Acategory \arrow[rr,"L", bend left] & \bot & \Bcategory\arrow[ll,"R",bend left]
\end{tikzcd}
\end{equation}
Incidentally, shifting attention from Moggi's monads to adjunctions became
\pagebreakopportunitypopl
standard after the pioneering works of Power and
Robinson~\cite*{PowerRobinson97}, Fiore~\cite*{Fiore94PhD},
Thielecke~\cite*{Thielecke97Thesis}, and Benton~\cite*{benton1996linear}
in the 1990's, and after Levy's Call by Push
Value~\cite*{Levy99CBPV,Levy2004,Levy05adjunctionmodels} which built
upon these works.

\medbreak

The adjunction induces a monad $T=R\circ L$ on the category~$\Acategory$ 
and a comonad~$K=L\circ R$ on the category~$\Bcategory$.
In order to mix the CBV style
and the CBN style
we need to combine the Kleisli category~$\kleisli{\Acategory}{T}$ and
the co-Kleisli category~$\cokleisli{\Bcategory}{K}$ in a single algebraic structure.

\paragraphcaps{The collage category of an adjunction}
It is well-known that an adjunction $L\dashv R$
can equivalently be seen as 
a bifibration 
$p:\Ecategory\to\twocategory$ over the order category $\twocategory={0\to 1}$
with two objects~$0$ and~$1$ and a unique map $\median:0\to 1$.
Here, the category $\Ecategory=\collage{L}{R}$
is defined as the \emph{collage} of the adjunction $L\dashv R$:
its objects are the pairs $(0,A)$ where $A$ is an object of $\Acategory$
and the pairs $(1,B)$ where $B$ is an object of $\Bcategory$, and
\begin{itemize}
\item its maps $(0,A)\to(0,A')$ are the maps $A\to A'$ in~$\Acategory$,
\item its maps $(1,B)\to(1,B')$ are the maps $B\to B'$ in~$\Bcategory$,
\item its maps $(0,A)\to(1,B)$ are the maps $A\to RB$ in $\Acategory$
or equivalently $LA\to B$ in $\Bcategory$,
\item there are no maps of the form $(1,B)\to(0,A)$.
\end{itemize}
The bifibration $p:\Ecategory\to\twocategory$ 
transports every object of the form~$(0,A)$ to $0$ 
and of the form $(1,B)$ to~$1$.
Note that the category~$\Ecategory$ comes equipped with
two injective on objects and fully faithful functors
\begin{center}
\begin{tikzcd}[column sep = 2em]
\Acategory \arrow[rr,"{inj_{\Acategory}}"] && \Ecategory 
&&
\Bcategory \arrow[ll,"{inj_{\Bcategory}}"{swap}]
\end{tikzcd}
\end{center}
identifying~$\Acategory$ and~$\Bcategory$ as the fibers over~$0$ and~$1$ respectively.
We find convenient to write $A$ for $(0,A)$ and $B$ for $(1,B)$ when there are no ambiguities.
We also call \emph{transverse} a map of the form $f:A\to B$ with image $p(f)=\median$.
Note that every object~$A\in\Acategory$ induces a transverse map 
\begin{equation}\label[equation]{equation/positive-negative-transverse-maps-0}
\begin{tikzcd}[column sep = 1.5em]
A_{01} \hspace{.5em} : \hspace{.5em} A \arrow[rr,"\Ecategory"] && LA
\end{tikzcd}
\end{equation}
obtained by ``pushing'' the object~$(0,A)\in\Ecategory$
along the map $\median:0\to 1$.
Symmetrically, every object~$B\in\Bcategory$ induces a transverse map
\begin{equation}\label[equation]{equation/positive-negative-transverse-maps-1}
\begin{tikzcd}[column sep = 1.5em]
B_{01} \hspace{.5em} : \hspace{.5em} RB \arrow[rr,"\Ecategory"] && B
\end{tikzcd}
\end{equation}
obtained by ``pulling'' the object~$(1,B)\in\Ecategory$
along the map $\median:0\to 1$.
Note also that every transverse map $f:(0,A)\to (1,B)$ in the collage category
can be equivalently be seen as a map $f^{\lhd}:A\to RB$ in~$\Acategory$
or as a map $f^{\rhd}:LA\to B$ in~$\Bcategory$.
The two maps~$f^{\lhd}$ and $f^{\rhd}$ can be characterized
as the unique maps of~$\Ecategory$ making the diagram below
\begin{center}
\begin{small}
\begin{tikzcd}[column sep = 3em, row sep = \diagrowsep]
{RB} \arrow[rr,"B_{01}"]
&& {B}
\\
\\
{A} \arrow[rr,"A_{01}"{swap}] \arrow[rruu,"f"] \arrow[uu,dashed,"f^{\,\lhd}"] 
&& {LA} \arrow[uu,dashed,"f^{\,\rhd}"{swap}]
\end{tikzcd}
\end{small}
\end{center}
commutative in the collage category.

\paragraphcaps{The duploid associated to an adjunction}
Every adjunction~$L\dashv R$ induces a non-associative category~$\duploid{L}{R}$
which contains the Kleisli category $\kleisli{\Acategory}{RL}$ of the monad~$RL$ and 
the Kleisli category $\cokleisli{\Bcategory}{LR}$ of the comonad~$LR$ as full subcategories.
The non-associative category~$\duploid{L}{R}$ is constructed in~\citet{munchduploids}
where it is called a \emph{duploid} because it comes equipped with a polarity structure recalled in~\S\ref{section/duploids}.
The objects of~$\duploid{L}{R}$ are the objects $(0,A)$ and $(1,B)$ of the collage category,
for~$A\in\Acategory$ and~$B\in\Bcategory$.
Its maps are defined as follows.
Every object $X=(0,A)$ or $X=(1,B)$ of the duploid induces a transverse map
\begin{equation}\label[equation]{equation/positive-negative-transverse-maps-X}
\begin{tikzcd}[column sep = 1.5em]
X_{01} \hspace{.5em} : \hspace{.5em} X_0 \arrow[rr,"\Ecategory"] && X_1
\end{tikzcd}
\end{equation}
defined as $X_{01}=A_{01}$ in \eqref{equation/positive-negative-transverse-maps-0} when $X=(0,A)$
and as $X_{01}=B_{01}$ in \eqref{equation/positive-negative-transverse-maps-0} when $X=(1,B)$.
A morphism $f:X\to Y$ between two objects $X$ and $Y$ of the duploid $\duploid{L}{R}$
with associated transverse maps
\begin{center}
\begin{tikzcd}[column sep = 1.2em]
X_{01} \hspace{.5em} : \hspace{.5em} X_0 \arrow[rr,"\smash{\Ecategory}"] && X_1
&&&&
Y_{01} \hspace{.5em} : \hspace{.5em} Y_0 \arrow[rr,"\smash{\Ecategory}"] && Y_1
\end{tikzcd}
\end{center}
is simply defined as a transverse map
\begin{center}
\begin{tikzcd}[column sep = 1.2em]
f  \hspace{.5em} : \hspace{.5em} X_0 \arrow[rr,"\Ecategory"] && Y_1
\end{tikzcd}
\end{center}
in the collage category~$\Ecategory=\collage{L}{R}$.
The situation may be depicted as follows:
\begin{center}
\begin{small}
\begin{tikzcd}[column sep = 2.5em, row sep = \diagrowsep]
{Y_0} \arrow[rr,"Y_{01}"] && {Y_1}
\\
\\
{X_0} \arrow[rr,"X_{01}"{swap}] \arrow[rruu,"f"] 
&& {X_1}
\end{tikzcd}
\end{small}
\end{center}
The exercise of defining the composite
\begin{center}
\begin{tikzcd}[column sep = 1.2em]
g\ccomp f \hspace{.5em} : \hspace{.5em} X \arrow[rr] && Z
\end{tikzcd}
\end{center}
of two morphisms~$f:X\to Y$ and~$g:Y\to Z$ of the duploid $\duploid{L}{R}$
looks somewhat challenging when one considers the diagram 
of the adjunction:
\begin{center}
\begin{small}
\begin{tikzcd}[column sep = 2.5em, row sep = \diagrowsep]
{Z_0} \arrow[rr,"Z_{01}"] && {Z_1}
\\
\\
{Y_0} \arrow[rr,"{Y_{01}}"] \arrow[rruu,"g"] 
&& {Y_1}
\\
\\
{X_0} \arrow[rr,"X_{01}"{swap}] \arrow[rruu,"f"] 
&& {X_1}
\end{tikzcd}
\end{small}
\end{center}
in the collage category~$\Ecategory=\collage{L}{R}$.
The composite~$g\ccomp f:X\to Z$ is defined in two different ways,
depending on the polarity of the intermediate object~$Y$:
\begin{itemize}
\item when $Y=(0,A)$ is positive,
the transverse map $Y_{01}=A_{01}$
is of the form~\eqref{equation/positive-negative-transverse-maps-0}
and there exists for that reason a unique map $g^{\,\rhd}:LA\to Z_1$ 
such that $g=g^{\,\rhd}\circ_{\Ecategory} A_{01}$ ;
the composite $g\ccomp f$ is defined in that case 
as $g\ccomp f=g^{\,\rhd}\circ_{\Ecategory} f:X_0\to Z_1$ in the collage category.
\item when $Y=(1,B)$ is negative, the transverse map $Y_{01}=B_{01}$
is of the form~\eqref{equation/positive-negative-transverse-maps-1}
and there exists for that reason a unique map 
$f^{\,\lhd}:X_0\to RB$ such that $f=B_{01}\circ_{\Ecategory} f^{\,\rhd}$ ;
the composite $g\ccomp f$ is defined in that case as 
$g\ccomp f=g\circ_{\Ecategory} f^{\,\lhd}:X_0\to Z_1$ in the collage category.
\end{itemize}
The recipe used to define the composite~$g\ccomp f:X\to Z$ in the collage category
is depicted below, with the case when $Y=(0,A)$ is positive on the lefthand side,
and the case when $Y=(0,B)$ is negative on the righthand side:
\begin{center}
\begin{small}
\begin{tabular}{ccc}
\begin{tikzcd}[column sep = 2.5em, row sep = \diagrowsep]
{Z_0} \arrow[rr,"Z_{01}"] && {Z_1}
\\
\\
{A} \arrow[rr,"A_{01}"] \arrow[rruu,"g"] 
&& {LA} \arrow[uu,dashed,"g^{\,\rhd}"{swap}]
\\
\\
{X_0} \arrow[rr,"X_{01}"{swap}] \arrow[rruu,"f"] 
&& {X_1}
\end{tikzcd}
& \hspace{12em}&
\begin{tikzcd}[column sep = 2.5em, row sep = \diagrowsep]
{Z_0} \arrow[rr,"Z_{01}"] && {Z_1}
\\
\\
{RB} \arrow[rr,"B_{01}"] \arrow[rruu,"g"] 
&& {B}
\\
\\
{X_0} \arrow[rr,"X_{01}"{swap}] \arrow[uu,dashed,"f^{\,\lhd}"] \arrow[rruu,"f"] 
&& {X_1}
\end{tikzcd}
\end{tabular}
\end{small}
\end{center}
\medbreak
\noindent
The construction
establishes~$\duploid{L}{R}$ 
as a simple and canonical way to integrate the Kleisli and co-Kleisli 
categories $\kleisli{\Acategory}{RL}$ and $\cokleisli{\Bcategory}{LR}$
in a larger overarching mathematical structure.
Indeed, an easy computation shows that
\begin{itemize}
\item $\kleisli{\Acategory}{RL}$ coincides with the full subcategory
of positive objects (= objects of~$\Acategory$)
\item $\cokleisli{\Bcategory}{LR}$ coincides with the full subcategory
of negative objects (= objects of~$\Bcategory$)
\end{itemize}
in the non-associative category~$\duploid{L}{R}$.
For that reason, it makes sense to write the composite $g\ccomp f$ 
as $g\pcomp f$ when $Y=(0,A)$ is positive, and as $g\ncomp f$ when $Y=(1,B)$ is negative.
The fact that composition
is not associative in~$\duploid{L}{R}$ comes from the fact that three maps
\begin{center}
\begin{tikzcd}[column sep = 1em]
A'\arrow[rr,"f"] && A\arrow[rr,"g"] && B \arrow[rr,"h"] && B'
\end{tikzcd}
\end{center}
defining a path of length 3
in $\duploid{L}{R}$ may induce different composite maps
\begin{center}
$(i): \,\,\, (h\ncomp g)\pcomp f$ \hspace{2em} and \hspace{2em} $(ii):\,\,\, h\ncomp(g\pcomp f)$
\end{center}
The explicit computation of the two maps in the collage category~$\Ecategory=\collage{L}{R}$
is depicted below:
\begin{center}
\begin{small}
\begin{tabular}{ccccc}
\begin{tikzcd}[column sep = 2.5em, row sep = \diagrowsep]
{RB'} \arrow[rr,"B'_{01}"] && {B'}
\\
\\
{RB} \arrow[rr,"B_{01}"] \arrow[rruu,"h"] 
&& {B}
\\
\\
&& {LA} \arrow[uu,dashed,"g^{\,\rhd}"]
\\
\\
{A'} \arrow[rr,"A'_{01}"{swap}] \arrow[rruu,"f"] 
\arrow[uuuu,dashed,"(g^{\,\rhd}\circ_{\Ecategory} f)^{\,\lhd}"{swap}]
&& {LA'}
\end{tikzcd}
&
\hspace{-1em}
\begin{tikzcd}[column sep=2em]
\,
&&
\,\arrow[ll,|->,"h\ncomp(g\pcomp f)"{swap}]
\end{tikzcd}
\hspace{-1em}
&
\begin{tikzcd}[column sep = 2.5em, row sep = \diagrowsep]
{RB'} \arrow[rr,"B'_{01}"] && {B'}
\\
\\
{RB} \arrow[rr,"B_{01}"] \arrow[rruu,"h"] 
&& {B}
\\
\\
{A} \arrow[rr,"A_{01}"] \arrow[rruu,"g"] 
&& {LA}
\\
\\
{A'} \arrow[rr,"A'_{01}"{swap}] \arrow[rruu,"f"] 
&& {LA'}
\end{tikzcd}
&
\hspace{-1em}
\begin{tikzcd}[column sep=2em]
\,\arrow[rr,|->,"(h\ncomp g)\pcomp f"]
&&
\,
\end{tikzcd}
\hspace{-1em}
&
\begin{tikzcd}[column sep = 2.5em, row sep = \diagrowsep]
{RB'} \arrow[rr,"B'_{01}"] && {B}
\\
\\
{RB} \arrow[rruu,"h"] 
&& 
\\
\\
{A} \arrow[rr,"A_{01}"]
\arrow[uu,dashed,"g^{\,\rhd}"{swap}] 
&& {LA} \arrow[uuuu,dashed,"(h\circ_{\Ecategory} g^{\,\rhd})^{\,\lhd}"]
\\
\\
{A'} \arrow[rr,"A'_{01}"{swap}] \arrow[rruu,"f"] 
&& {LA'}
\end{tikzcd}
\end{tabular}
\end{small}
\end{center}
The two maps 
give rise to maps of the collage category
$$(h\ncomp g)\pcomp f = (h\circ_{\Ecategory} g^{\,\rhd})^{\,\lhd}\circ_{\Ecategory} f
 \quad\quad\quad 
 h\ncomp (g\pcomp f) = h\circ_{\Ecategory}(g^{\,\rhd}\circ_{\Ecategory} f)^{\,\lhd}
$$
which are in general different for good reasons, as we illustrate below.

\subsectioncaps{Non-associativity seen as a blessing: thunkable and linear maps}
\label{subsection/intro-thunkable-linear}
The fact that duploids are non-associative forms of categories
should not be seen as a defect but rather as an opportunity
to characterize two specific classes of maps: thunkable and linear maps,
conveying important computational intuitions.
It is natural to ask when an expression in an effectful language is pure.
One possible definition is that it can be substituted like a value,
a notion called \emph{algebraic value} or
\emph{thunkable}~\cite{Thielecke97Thesis} expression.
One benefit of working in a duploid or more generally in a non-associative category,
is that thunkability for a map $f:X\to Y$ can be formulated as an associativity property
of composition~\cite{munchduploids} capturing the following syntactic equation:
\begin{equation}\label[equation]{equation/thunkable}
\def\letinspace{1mu}
\letinplus{y}{f}{\letinminus{z}{g_y}{h_z}}\hspace{0.6em}
= \hspace{0.6em}\letinminus{z}{(\letinplus{y}{f}{g_y})}{h_z}
\end{equation}
for all expressions~$g_y,h_z$.
This leads us to the following definition of thunkable maps
as well as the symmetric notion of linear maps in any 
non-associative category.

\paragraphcaps{Thunkable and linear maps}
A map $f$ is called \emph{thunkable}
when every path of length 3 that starts with $f$:
\[
    X \xlongrightarrow{\smash{f}} Y \xlongrightarrow{\smash{g}} Z \xlongrightarrow{\smash{h}} W
\]
associates, that is, the following holds:
\[
  h \ccomp (g \ccomp f) = (h\ccomp g) \ccomp f
\]
Dually, a morphism $h$ is called \emph{linear}
if every path of length 3 that ends with $h$ associates.

\paragraphcaps{Thunkable and linear maps: illustration}\label{paragraph/proba-example-i}
We illustrate the benefits of introducing these concepts
with probabilistic computations and the finite distribution monad.
Recall that the finite distribution monad $T:\Set\to\Set$
\pagebreakopportunitypopl
associates to every set~$A$ the set $TA$
of finite probability distributions $\psum{i}{p_i}{a_i}$ on~$A$,
that is finite families $(a_i)_{i\in I}$ of elements~$a_i\in A$
equipped with a function~$p:I\to [0,1]$ to the interval $[0,1]$ of reals
between $0$ and~$1$, such that $\sum_i p_i = 1$.
The Kleisli category~$\Set_T$ associated to the monad~$T$
is the category of \emph{stochastic maps} whose morphisms
$f:B\to B'$ between sets~$B,B'$ are defined as functions $f:B\to TB'$
which transport every element $b\in B$ to a finite probability
distribution $f(b)=\psum{i}{p_i(b)}{f_i(b)}$ of~$B'$.

The duploid construction applied to the adjunction
between $\Set$ and its Kleisli category $\Set_T$
\[
\begin{tikzcd}[column sep = 1em]
\Set \arrow[rr,"\smash{L}\vphantom{x}", bend left] & \bot & \Set_T\arrow[ll,"R",bend left]
\end{tikzcd}
\]
induces a duploid whose objects are sets $(0,A)$ and $(1,B)$
annotated with a polarity $0$ for positive and $1$ for negative.
\begin{itemize}
\item its maps $(0,A)\to(0,A')$ are the stochastic maps $A\to A'$,
\item its maps $(1,B)\to(1,B')$ are the stochastic maps $TB\to B'$,
\item its maps $(0,A)\to(1,B)$ are the stochastic maps $A\to B$,
\item its maps $(1,B)\to(0,A)$ are the stochastic maps $TB\to A$.
\end{itemize}
The composite $g\ccomp f:X\to Z$ of two morphisms $f:X\to Y$ and $g:Y\to Z$
in the duploid depends on the polarity of the intermediate object~$Y$.

\smallbreak
\noindent
\emph{$\rhd$ when $Y=(0,A)$ is positive}, we compose 
the stochastic maps~$f$ and~$g$ in the usual way,
by distributing the probabilities.
For instance, 
\[
  \letinplus{y}{\psum{i}{p_i}{f_i(x)}}{\psum{j}{q_j}{g_j(y)}}
  \quad = \quad \psum{i,j}{p_i q_j}{g_j(f_i(x))}
\]
when the coefficients~$q_j$ do not depend on~$y$.

\smallbreak
\noindent
$\rhd$ 
\emph{when $Y=(1,B)$ is negative}, we substitute the expression
$\Sigma_i \, p_i \,|\, u_i(x)\rangle$ in the expression $\Sigma_j q_j \,| \, v_j(y)\rangle$
without distributing the probabilities.
For instance,
\[
  \letinminus{y}{\psum{i}{p_i}{f_i(x)}}{\psum{j}{q_j}{g_j(y)}} 
  \quad = \quad \psum[\Big]{j}{q_j}{g_j\big(\psum{i}{p_i}{f_i(x)}\big)}
\]
when the coefficients~$q_j$ do not depend on~$y$.

Now that the duploid associated to the finite distribution monad $T$ has been defined,
we characterize the thunkable maps $f : X \to Y$ in the duploid. 
It is not difficult to see that $f:X\to Y$ is always thunkable when $Y = (1, B)$ is negative.
We thus focus on the interesting case where $Y = (0,A)$ is positive.
In that case, the map $f:X\to (0,A)$ is a stochastic map to the set~$A$ of the general form 
$$f(x) := \psum{i}{p_i(x)}{f_i(x)}$$
with $\sum_i p_i(x) = 1$ and each $f_i(x)\in A$, for all $x\in X$.
At this stage, we introduce the two maps
which will play the role of ``effectful context'' testing the behavior of the map~$f$ in the duploid:
\[
  \begin{array}{c} g : (0,A) \to (1,A) \\ g : a \mapsto \dirac{a}\end{array} \quad\quad \mbox{and} \quad\quad \begin{array}{c} h : (1,A) \to (0, TA) \\ h : d \mapsto \dirac d \end{array},
\]
defined as (1) the stochastic map~$g:A\to A$ which associates to any $a\in A$ the Dirac distribution $\dirac a\in TA$ and as (2) the stochastic map $h:TA\to TA$ which associates to any probability distribution $d\in TA$ the Dirac distribution of distributions $\dirac{d}\in TTA$.
This defines a path
\[
    X \xlongrightarrow{\smash{f}} (0,A) \xlongrightarrow{g} (1,A) \xlongrightarrow{\smash{h}} (0,TA)
\]
of length~$3$ which can be composed in two different ways, left-to-right or right-to-left as follows\ifbool{arxiv}{:}{.}\pagebreakopportunitypopl
\begin{gather*}
  \letinminus{d}{\big(\letinplus{a}{f(x)}{\dirac a}\big)}{\dirac d} \quad\quad \mbox{reducing to} \quad\quad \dirac[\Big]{\psum{i}{p_i(a)}{f_i(a)}}\\
  \letinplus{a}{f(x)}{\big(\letinminus{d}{\dirac a}{\dirac d}\big)} \quad\quad \mbox{reducing to} \quad\quad \psum[\big]{i}{p_i(x)}{\dirac{f_i(x)}}
\end{gather*}
When the map~$f$ is thunkable, these two expressions must be equal by definition.
From this follows that the stochastic map~$f$ must transport every element~$x\in X$
into a Dirac distribution of the form $\dirac{b(x)}\in TA$.
Conversely, any such map~$f$ is thunkable.
This shows that in this duploid associated to the probability
adjunction, thunkable maps coincide with \emph{values} in the sense
of \citet{Moggi1991}.

\subsectioncaps{Continuations, dialogue duploids, and classical notions of computation}
One important aspect of duploids is that they exhibit
and preserve the perfect symmetry between the monadic and comonadic effects
of an adjunction, by treating on an equal footing the CBV and CBN evaluation policies.
Our main goal in the present paper is to explore how this symmetric 
and non-associative account of effects can benefit the long quest 
for a perfectly symmetric computational account of classical logic,
in the spirit and philosophy of the Curry-Howard correspondence.
\paragraphcaps{The self-adjunction of negation}
For our purposes,
we find convenient and evocative to work with the notion
of \define{dialogue category} introduced by the third
author~\citep{melliesdialogue, MelliesTabareau10} as a categorical
semantics of linear continuations.
Recall that a dialogue category is defined 
as a symmetric monoidal category~$(\Ccategory,\otimes,1)$
equipped with a return object~$\bot$ in the following sense:
\begin{definition}\label[definition]{definition/return-object}
An object $\bot$ is called a \emph{return object}
in a symmetric monoidal category~$(\Ccategory,\otimes,1)$
when it comes equipped with an object~$\bot^{A}$ and a family of bijections 
$$
\varphi_{A,B} \quad : \quad \Ccategory(A\tensor B,\bot) \quad \xlongrightarrow{\iso} \quad \Ccategory(B,\bot^A)
$$
natural in~$B$, for every object~$A$ of the category~$\Ccategory$.
\end{definition}
An easy categorical argument shows that every dialogue category 
comes equipped with a functor
$$
\begin{tikzcd}[column sep = 1em]
\lnot \quad : \quad \Ccategory\arrow[rr] && \Ccategory^{\op}
\end{tikzcd}
$$
which transports every object~$A$ to the object~$\bot^A$ which
can be seen as a negation of~$A$ and written accordingly as $\lnot\, A$.
A well-known fact is that the negation functor defines an adjunction with itself:
\begin{equation}\label[equation]{equation/adjunction-neg-neg}
\begin{tikzcd}[column sep = 1em]
\Ccategory \arrow[rr,"{L\,=\,\neg}", bend left] & \bot & \Ccategory^{\mathrlap{\op}}\arrow[ll,"{R\,=\,\neg}",bend left]
\end{tikzcd}
\end{equation}
This observation, dating back to \citet{Kock1970}, was given emphasis
in Thielecke's Ph.D. thesis~\cite*{Thielecke97Thesis} on the structure
of CPS translations.
This self-adjunction plays also a central role in the the foundations of functorial game semantics~\cite{Mellies12}.

\paragraphcaps{Dialogue duploids}
We have seen that
the construction of the duploid~$\duploid{L}{R}$ associated to an adjunction~$L\dashv R$ 
amounts to building a direct computational interpretation combining the CBV and the CBN models and preserving the symmetry between them.
Our purpose in the present paper is to uncover the structural properties
of the duploids~$\duploid{L}{R}$ associated to a dialogue category.
In order to better understand these structures, we start from the symmetric reformulation
(up to equivalence) of dialogue categories as \emph{dialogue chiralities}
defined below:
\begin{definition}[\citet{Mellies12,melliesdialogue}]\label[definition]{definition/chirality}
A \define{dialogue chirality} is a pair of symmetric monoidal categories
$(\Acategory,\tensorialand,\tensorialtrue)$ and $(\Bcategory,\tensorialor,\tensorialfalse)$
equipped with an
adjunction~${L:\Acategory\rightleftarrows\Bcategory:R}$ as depicted
\pagebreakopportunitypopl
in~\eqref{equation/adjunction-LR}
together with a symmetric monoidal equivalence:
\begin{equation}\label[equation]{equation/chiral-duality}
\begin{tikzcd}[column sep = 1em]
  (\Acategory,\tensorialand,\tensorialtrue) \arrow[rr,"\smash{\chirdual{(-)}}\vphantom{j}", bend left] & \simeq &
  (\Bcategory,\tensorialor,\tensorialfalse)^{\mathrlap{\op}}\arrow[ll,"\chirdual{(-)}",bend left]
\end{tikzcd}
\end{equation}
and a family of bijections (called currifications)
\ifbool{arxiv}{
\[
  \chi_{A_1,A_2,B} : \Acategory(A_1 \tensorialand A_2, RB)
  \xlongrightarrow{\smash{\iso}} \Acategory(A_1, R(\chirdual{A_2}\tensorialor B)) 
\]
}{
$
  \chi_{A_1,A_2,B} : \Acategory(A_1 \tensorialand A_2, RB)
  \xlongrightarrow{\smash{\iso}} \Acategory(A_1, R(\chirdual{A_2}\tensorialor B)) 
$
}
natural in $A_1$, $A_2$ and $B$ and satisfying a coherence diagram.
\end{definition}
\noindent
In order to understand the specific nature of duploids associated
to dialogue chiralities, we will develop a general theory 
of duploids equipped with different forms of monoidal structures,
in link with classical logic and linear as well as non linear continuations.
In particular, we will define the notion of \emph{dialogue duploid}
which describes the structure of a duploid associated to a dialogue chirality.

\paragraphcaps{A categorical account of Girard's \LC}
Our work is guided by the observation that dialogue duploids
provide a compelling categorical and semantic foundation to the proof-theoretic
account of classical logic based on $\LC$ developed in~\citet{girardnew}.
Indeed, it appears that the notion of dialogue duploid provides
the categorical semantics of a linear and two-sided multiplicative
variant of $\LC$, what can be summarized as follows:\vspace*{0.5ex}
\begin{center}
\fbox{\begin{tabular}{c}
Linear two-sided multiplicative $\LC$ \quad = \quad duploids + dialogue chiralities
\end{tabular}}\vspace*{0.5ex}
\end{center}
In that sense, the duploid construction provides
in the case of dialogue categories a precise mathematical
and denotational counterpart to the multiplicative fragment
of the new form of double-negation translation 
implemented by $\LC$
which contains the traditional CBV and the
CBN computational models as its \emph{positive} and \emph{negative subcategories}
respectively.

\subsectioncaps{The \FH{} theorem}\label{section/introFH} 
We have seen in~\eqref{equation/thunkable}
that thunkability of a map~$f:X\to Y$
captures a concept of purity which can be expressed 
by the syntactic equation
\[
\def\letinspace{1mu}
\letinplus{y}{f}{\letinminus{z}{g_y}{h_z}}\hspace{0.6em}
= \hspace{0.6em}\letinminus{z}{(\letinplus{y}{f}{g_y})}{h_z}
\]
for all expressions~$g_y,h_z$. As we will see, this corresponds to an
equality between derivations in sequent calculus of the following
form:
\newcommand{\stackhypo}[1]{\hypo{&\mathclap{#1}}\infer[no rule]1}
\begin{equation}
\label[equation]{eq:FH-thunk}
\begin{prooftree}
\stackhypo{\scriptstyle{f}}{\Gamma'' & \vdash P, \Delta''}
\stackhypo{\scriptstyle{h_z}}{\Gamma,P & \vdash N, \Delta }
\stackhypo{\scriptstyle{g_y}}{\Gamma', N & \vdash \Delta' }
\infer2{\Gamma,\Gamma'\!,P & \vdash \Delta,\Delta' }
\infer2[$\hspace{2.6em}{=}$]{\Gamma,\Gamma'\!,\Gamma'' & \vdash \Delta,\Delta'\!,\Delta''}
\end{prooftree}
\quad
\begin{prooftree}
\stackhypo{\scriptstyle{f}}{\Gamma'' & \vdash P, \Delta''}
\stackhypo{\scriptstyle{h_z}}{\Gamma,P & \vdash N, \Delta}
\infer2{\Gamma,\Gamma'' & \vdash N, \Delta,\Delta''}
\stackhypo{\scriptstyle{g_y}}{\Gamma'\!,N & \vdash \Delta'}
\infer2{\Gamma,\Gamma'\!,\Gamma'' & \vdash \Delta,\Delta'\!,\Delta''}
\end{prooftree}
\end{equation}

\noindent A weaker concept of purity,
\emph{centrality}~\cite{PowerRobinson97}, captures the idea of
irrelevance of order of evaluation with another property of commutation
(for all $g,h_{x,y}$):\[
\def\letinspace{2mu}
\letinplus{x}{f}{\letinplus{y}{g}{h_{x,y}}}\hspace{0.6em}
= \hspace{0.6em}\letinplus{y}{g}{\letinplus{x}{f}{h_{x,y}}}
\]
or in sequent calculus:
\begin{equation}
\label[equation]{eq:FH-central}
\begin{prooftree}
\stackhypo{\scriptstyle{f}}{\Gamma'' & \vdash P, \Delta''}
\stackhypo{\scriptstyle{g}}{\Gamma' & \vdash Q, \Delta'}
\stackhypo{\scriptstyle{h_{x,y}}}{\Gamma,P,Q & \vdash \Delta}
\infer2{\Gamma,\Gamma'\!,P & \vdash \Delta,\Delta'}
\infer2[$\hspace{2.6em}{=}$]{\Gamma,\Gamma'\!,\Gamma'' & \vdash \Delta,\Delta'\!,\Delta''}
\end{prooftree}
\quad
\begin{prooftree}
\stackhypo{\scriptstyle{g}}{\Gamma' & \vdash Q, \Delta'}
\stackhypo{\scriptstyle{f}}{\Gamma'' & \vdash P, \Delta''}
\stackhypo{\scriptstyle{h_{x,y}}}{\Gamma,P,Q & \vdash \Delta}
\infer2{\Gamma,\Gamma''\!,Q & \vdash \Delta,\Delta''}
\infer2{\Gamma,\Gamma'\!,\Gamma'' & \vdash \Delta,\Delta'\!,\Delta''}
\end{prooftree}
\end{equation}

\noindent Note that, strikingly, these two instances of commutations
between $f$ and $g$ are the same up to duality in the sequent calculus.
For the classical notions of computation we are considering, another
ingredient makes them actually coincide: the presence of a negation
connective which is involutive at the level of proof denotation, whose
rules in sequent calculus provide a way to exchange between the
left-hand and right-hand sides without loss of information.
We formalize this idea with our proof of the Hasegawa-Thielecke
theorem (\cref{theorem/fh-syntactic}) both semantically and
syntactically, using the linear classical $L$-calculus
(\S\ref{section/classical-L-calculus}).

The conceptually simple proof of the Hasegawa-Thielecke
theorem illustrates the benefits of using duploids to reason
about effectful programs.
This result was noticed as a characterisation of centrality
by \citet{Thielecke97Thesis} in the context of categorical semantics
for continuations, in which it plays an important role
(see \citet{Thielecke97Thesis,Sel01Control,Hasegawa_2002} among
others). The essential status of thunkability as a concept distinct
from centrality became apparent in the works on the direct axiomatic
theory of monadic effects
by \citet{Fuhrmann2000PhD,fuhrmanndirectmodels}.
Our conceptual reformulation of the proof suggests that expressing
thunkability as an associativity property is a key part of the result,
yet one which remains true and useful beyond continuation models.

It follows from this theorem that in any dialogue category,
\emph{a map is thunkable if
and only if it is central}. In particular, \emph{the double-negation
monad is commutative if and only if it is idempotent}. The latter
condition corresponds to the case where the duploid is a category
(hence, in the case of cartesian dialogue categories, to a boolean
algebra by \joyallemma{} (\S\ref{section/classical-history})).
The refinement of this property from the cartesian to the symmetric
monoidal setting was suggested by Hasegawa and played a key role
in the second author's analysis of the
Blass problem in game semantics as a non-commutativity of the
double-negation monad~\cite{Mellies2005ag3}. We are not aware of a
previously-published proof of this theorem in the symmetric monoidal
case.

\subsectioncaps{Summary and main contributions}
After this long but necessary introduction,
we explain in~\S\ref{section/non-associative-categories}
how to reason diagrammatically in a non-associative category,
and then recall in~\S\ref{section/duploids} the notion of duploid.
We then start our journey towards the linear classical $L$-calculus by
introducing in~\S\ref{section/sm-duploids}
and~\S\ref{section/sm-duploids-ii} the notion of symmetric monoidal
duploid, alongside the crucial notion of \emph{adjunction between
graph morphisms} of non-associative categories. We use this notion
in~\S\ref{sec:SMCD} to define symmetric monoidal closed duploid and
show a correspondence with linear call-by-push-value adjunction
models. We also recall in~\S\ref{section/CBPV-L-calculus}
the \emph{linear call-by-push-value $L$-calculus} of \citet{CFM2015}
and establish a soundness theorem of the interpretation in symmetric
monoidal closed duploid.

From this intuitionistic basis, we consider models with an
involutive negation, leading to the notion of dialogue duploid
in~\S\ref{section/dialogue-duploids} and to the \emph{linear classical
$L$-calculus} in~\S\ref{section/classical-L-calculus}. We show that
dialogue duploids are in correspondence with dialogue chiralities and
soundly interpret the linear classical $L$-calculus.
Building on this result, we illustrate the relevance and robustness of
our approach by defining the syntactic dialogue duploid
in \S\ref{section/syntactic-duploid} and by proving
in \S\ref{section/fh-theorem} the \FH{} theorem using both semantic
and syntactic methods. \ifbool{arxiv}{In this extended version of the
paper, we also explain in~\S\ref{section/onesided} the relationship
between the two-sided and the \emph{one-sided} variant of the linear
classical $L$-calculus as a coherence result.}{}
We then give a historical perspective in~\S\ref{section/classical-history}
on the different Curry-Howard approaches to classical logic,
and finally conclude and give directions for future work in \S\ref{section/conclusion}.

\ifbool{arxiv}{This extended version contains an appendix with additional
illustrations, and some details of proofs. A table of contents and a
list of figures are provided on the final page.}{}

\sectioncaps{Non-associative categories}\label{section/non-associative-categories}
We have observed in the introduction
that the composition of effectful programs ${g,f}\mapsto {g\ccomp f}$
is not associative in general when one wants to make positive and
negative types coexist in the same overarching mathematical structure.
This justifies to develop a good theory and practice of non-associative categories.
As we will see, working with non-associative categories is not only possible,
it is also a highly compelling exercise,
which sheds light on the fundamental nature of effects, commutative or not commutative.

\begin{definition}
A \define{unital magmoid} or \define{non-associative category} $\mathcal M$
is defined as a reflexive graph equipped with a composition law
  $$
\begin{tikzcd}[row sep=-.3em,column sep=1em]
\ccomp_{X,Y,Z} \quad : \quad \mathcal M(Y, Z) \times \mathcal M(X, Y) \arrow[rr] && \mathcal M(X, Y)
\end{tikzcd}
$$
which associates to every pair of maps $f:X\to Y$ and $g:Y\to Z$ a composite map $g\ccomp f:X\to Z$,
such that the neutrality equations below are satisfied
  \[
    f \ccomp \mathsf{id}_X = f = \mathsf{id}_Y \ccomp f
  \]
for every map~$f:X\to Y$, where $\mathsf{id}_X\in\mathcal M(X, X)$
denotes the chosen map at object~$X$ of the reflexive graph.
We use the notation $|\mathcal M|$ for the set (or more generally the class) of objects of $\mathcal M$.
\end{definition}

\noindent
Given a non-associative category $\mathcal M$, we define $\mathcal M^{\op}$
as the non-associative category with the same objects as $\mathcal M$ 
and with the orientation of maps reversed, in the sense that $\mathcal M^{\op}(X,Y):=\mathcal M(Y,X).$

\medbreak
\noindent
In a non-associative category $\mathcal M$, we declare that a path $(f,g,h)$ of length~$3$
\[
    X \xlongrightarrow{f} Y \xlongrightarrow{g} Z \xlongrightarrow{h} W
\]
associates when the associativity equation below is satisfied:
$$(h \ccomp g) \ccomp f = h \ccomp (g \ccomp f)$$
We recall the definitions of linear and thunkable maps given in the introduction:
\begin{itemize}
\item A map $h$ is called \define{linear} when every path of the form $(f,g,h)$ associates,
\item A map $f$ is called \define{thunkable} when every path of the form $(f,g,h)$ associates.
\end{itemize}

\medbreak
\noindent
Note also that the usual definitions of epis and monos
as right and left cancellable maps
in an associative category 
immediately extend to non-associative categories:
\begin{itemize}
\item a map $e:X\to Y$ is called \define{epi} when $f\ccomp e = g\ccomp e$ implies $f=g$
for all maps $f,g:Y\to Z$,
\item a map $m:Y\to Z$ is called \define{mono} when $m\ccomp f = m\ccomp g$ implies $f=g$
for all maps $f,g:X\to Y$.
\end{itemize}

\noindent
In the same way as reasoning within associative categories can be performed
by chasing commutative diagrams, reasoning within non-associative categories
generally reduces to chasing "triangulated" commutative diagrams,
and rewriting them using the usual and well-studied notion of flip of triangulation~\cite{Pournin201413}.
Typically, the fact that the path $(f,g,h)$ associates
means that one can ``flip'' the triangular commutative decomposition of the square
on the left into the triangular commutative decomposition of the same square on the right:
\begin{center}
\begin{small}
\begin{tabular}{ccc}
\begin{tikzcd}[column sep = 1.8em, row sep = 1.4em]
Y\arrow[rr,"g"]\arrow[rrdd] && Z\arrow[dd,"h"]
\\
\\
X\arrow[uu,"f"]\arrow[rr,"k"{swap}] &&  W
\end{tikzcd}
& \hspace{2em} &
\begin{tikzcd}[column sep = 1.8em, row sep = 1.4em]
Y\arrow[rr,"g"] && Z\arrow[dd,"h"]
\\
\\
X\arrow[uu,"f"]\arrow[rruu]\arrow[rr,"k"{swap}] && W
\end{tikzcd}
\end{tabular}
\end{small}
\end{center}
and conversely, from the commutative diagram on the left
to the commutative diagram on the right.
Similarly, one can flip the commutative diagram on the left 
into the commutative diagram on the right whenever the path $(f,h,g')$ associates:
\begin{center}
\begin{small}
\begin{tabular}{ccc}
\begin{tikzcd}[column sep = 1.8em, row sep = 1.4em]
Y\arrow[rr,"g"]\arrow[rrdd,"h"] && Z
\\
\\
X\arrow[uu,"f"]\arrow[rr,"f'"{swap}] &&  W\arrow[uu,"{g'}"{swap}]
\end{tikzcd}
& \hspace{2em} &
\begin{tikzcd}[column sep = 1.8em, row sep = 1.4em]
Y\arrow[rr,"g"] && Z
\\
\\
X\arrow[uu,"f"]\arrow[rruu]\arrow[rr,"f'"{swap}] && W\arrow[uu,"{g'}"{swap}]
\end{tikzcd}
\end{tabular}
\end{small}
\end{center}
The notion of flip between triangular decompositions
plays a central role in the combinatorial study of associativity.
It is thus striking to see the same combinatorial idea turned here
into an equational reasoning method for effectful programs.

At the same time, it is remarkable that very basic principles
of usual associative categories are not necessarily true anymore in non-associative categories.
Typically, 
the fact that the triangulated diagram below on the left commutes,
in the sense that $h_Y\ccomp f = f' \ccomp h_X$ and $h_Z\ccomp g = g' \ccomp h_Y$,\begin{equation}\label[equation]{equation/two-squares}
\begin{tabular}{ccccc}
\begin{tikzcd}[row sep = 1.05em, column sep = 1.7em]
X\arrow[rr,"f"] \arrow[dd,"h_X"{swap}]
\arrow[rrdd]
&&
Y \arrow[dd,"h_Y"]\arrow[rr,"g"]
\arrow[rrdd]
&& 
Z \arrow[dd,"h_Z"]
\\
\\
X'\arrow[rr,"f'"{swap}] 
&&
Y'\arrow[rr,"g'"{swap}] && Z'
\end{tikzcd}
& \quad\quad &
\begin{tikzcd}[row sep = 1.05em, column sep = 3.0em]
X\arrow[rr,"{g\circ f}"] \arrow[dd,"h_X"{swap}]
\arrow[rrdd]
&&
Z \arrow[dd,"h_Z"]
\\
\\
X'\arrow[rr,"{g'\circ f'}"{swap}] && Z'
\end{tikzcd}
\end{tabular}
\end{equation}
\emph{does not} imply that the triangulated diagram on the right commutes,
in the sense that $h_Z\ccomp (g\ccomp f) = (g'\ccomp f') \ccomp h_X$.
However, the triangulated diagram commutes when the three paths below associate:
\begin{align*}
& (h_X,f',g') && (f,h_Y,g') && (f,g,h_Z)
\end{align*}
The property can be established either equationally or
diagrammatically, by chasing and flipping commutative triangulations.
More details can be found
in \ifbool{arxiv}{Appendix \ref{section/chasing}}{the extended version~\citep{MMMM2025}}.

\sectioncaps{Duploids}\label{section/duploids}
We have seen in the introduction that every adjunction~$L\dashv R$
induces a non-associative category~$\duploid{L}{R}$ where every object
is polarized either positive or negative.
In this section, we recall the notion of \emph{duploid} introduced in \citet{munchduploids}
as non-associative categories equipped with a polarity structure defined as
a pair of shift operators $X\mapsto\nDownarrow X$ and $X\mapsto\nUparrow X$.
The notion of duploid is justified by the fact, established in~\citet{munchduploids},
that it characterizes the class of non-associative categories~$\duploid{L}{R}$
associated to an adjunction~$L\dashv R$, up to an equivalence, 
as per Thm~\ref{theorem/adjunctions-duploids} below.

We first recall the observation made in \citet{newduploids} that every non-associative category~$\mathcal M$
comes with an intrinsic notion of polarity on objects.

\begin{definition}[Polarity]\label[definition]{definition/polarity}
An object~$X \in |\mathcal M|$ is called \define{$\mathcal M$-positive} when, for all $Y \in |\mathcal M|$,
all maps of $\mathcal M(X,Y)$ are linear.
Symmetrically, an object $Y$ of $\mathcal M$ is \define{$\mathcal M$-negative} when, 
for all $X \in |\mathcal M|$, all maps of $\mathcal M(X,Y)$ are thunkable.
\end{definition}

\noindent
Note that an object~$X$ may be both $\mathcal M$-positive and $\mathcal M$-negative:
this is the case in particular for every object~$X$ of an associative category.
Note also that, if a map $f$ is linear in the non-associative category $\mathcal M$, then $f$ is thunkable
in the opposite non-associative category~$\mathcal M^\op$, and conversely. 
From this follows that $(-)^\op$ reverses the polarities.

\begin{definition}
A \define{positive shift} on a non-associative category $\mathcal M$ consists
of an object~${\nDownarrow X}$ equipped with a thunkable epi $\omega_X : X \to \nDownarrow X$ 
for every object~$X$, satisfying the following lifting property:
for every map~$f:X\to Y$, there exists a unique linear map $f^{\dagger}:\nDownarrow X\to Y$ 
making the diagram (\ref{eq:pos-shift}) on the left commute, that is, $f=f^{\dagger}\ccomp \omega_X$. A \define{negative shift} $(\nUparrow,\delta)$ is a positive shift on $\mathcal M^{\op}$.\begin{align}\label{eq:pos-shift}
&
\begin{tikzcd}[column sep = 2.2em, row sep = 1.6em, ampersand replacement=\&]
X\arrow[d,"{\omega_X}"{swap}]
\arrow[rd,"f"]
\\
\nDownarrow X \arrow[r,dashed,"\mkern-9muf^{\dagger}\text{ lin.}"{swap}] \& Y
\end{tikzcd}
&&
\begin{tikzcd}[column sep = 2.2em, row sep = 1.6em, ampersand replacement=\&]
X\arrow[dr,"f"]\arrow[d,dashed,"f^{\dagger}\text{ thunk.}"{swap}]
\\
\nUparrow Y \arrow[r,"{\delta_Y}"{swap}] \& Y
\end{tikzcd}
\end{align}
\end{definition}

A nice and instructive exercise in non-associative categories
is to show that the object~${\nDownarrow X}$ is $\mathcal M$-positive,
that the lifting $(-)^{\dagger}$ transports thunkable maps
into thunkable maps,
and that the map ${\overline\omega}_X:=\id{X}^{\dagger}:\nDownarrow X\to X$ 
defined as the lift of the identity map~$\id{X}$ is both linear and thunkable,
\pagebreakopportunitypopl
and satisfies the two equations:
\begin{align*}
{\overline\omega}_X \ccomp \omega_X &= \id{X} &
&\mathclap{\text{and}} &
\omega_X\ccomp {\overline\omega}_X &= \id{\nDownarrow X}
\end{align*}
defining a left and right inverse to the map~$\omega_X$.

Despite respecting all the usual conditions to be an isomorphism in a
usual associative category, the map~$\omega_X$ should not be
considered as an isomorphism in non-associative categories. Indeed,
the correct notion for an isomorphism in a non-associative category is
an inversible morphism such that the morphism and its inverse are
thunkable as well as linear---this reflects and formalises an
important observation made on the notion of contextual isomorphisms
in \citet{levy2017contextual}. The objects $X$ and $\nDownarrow X$ are
isomorphic in this sense exactly when $X$ is positive.

\medbreak
A positive shift defines a function $X\mapsto \nDownarrow X$ on objects which can be extended to a function which transports
every map $f \in \mathcal M(X,Y)$ to the unique map $\nDownarrow f \in \mathcal M(\nDownarrow X, \nDownarrow Y)$
making the triangulated diagram below commute:\vspace*{-1.5ex}
\begin{center}
\begin{small}
\begin{tikzcd}[column sep = 1.6em, row sep = 1.0em]
X\arrow[rr,"f"]\arrow[dd,"\omega_X"{swap}]\arrow[rrdd]  && Y\arrow[dd,"\omega_Y"] 
\\
\\
{\nDownarrow X} \arrow[rr,dashed,"{\nDownarrow f}"{swap}] && {\nDownarrow Y}
\end{tikzcd}
\end{small}
\end{center}
Note that this unique map $\nDownarrow f$
can be defined as the map $\nDownarrow f:=(\omega_Y \ccomp f)^{\dagger}$ obtained by lifting.

The positive shift just defined transports thunkable maps into thunkable maps and preserves identities.
On the other hand, it is important to stress that it \emph{does not} preserve composition,
in the sense that the functoriality diagram below does not commute in general:\vspace{-0.5ex}
\begin{equation}\label[equation]{equation/non-functoriality-of-shift}
\begin{tikzcd}[column sep = 1.0em, row sep = 0.9em]
&&
\nDownarrow X'
\arrow[rrdd,"\nDownarrow f'"]
\\
\\
\nDownarrow X\arrow[rruu,"\nDownarrow f"] \arrow[rrrr,"\nDownarrow(f'\ccomp f)"] &&&& \nDownarrow X''
\end{tikzcd}
\end{equation}
This lack of functoriality is a direct consequence of the phenomenon observed in~\eqref{equation/two-squares}
that glueing two commutative triangulated squares do not necessarily produce a commutative triangulated square.
On the other hand, and this is the whole beauty and simplicity of non-associative categories,
one recovers functoriality precisely when a specific path of length 3 associates:  
\begin{proposition}
The diagram~\eqref{equation/non-functoriality-of-shift}
commutes precisely when the path below associates
\[
X\xlongrightarrow{\smash{f}} X' \xlongrightarrow{\smash{f'}} X'' \xlongrightarrow{\omega_{X''}} \nDownarrow X''
\]
in the sense that $\omega_{X''}\ccomp(f'\ccomp f)=(\omega_{X''}\ccomp f')\ccomp f$.
\end{proposition}
\noindent
This non-functoriality of the shift operator should not be seen as a defect.
On the contrary, it provides a simple combinatorial explanation for 
an important and subtle phenomenon in effectful programming.
\ifbool{arxiv}{
This is illustrated in the
Appendix~\ref{section/non-functoriality-illustration} with the example
of the duploid associated to the finite distribution monad
$T:\Set\to\Set$, already discussed in the introduction.}{}

At this stage, we are ready to recall (a slight variant of) the definition of duploid from~\citet{munchduploids}.

\begin{definition}
  A \define{duploid} is a non-associative category equipped with a positive and a negative shift,
  and where every object is either positive or negative (or both).
\end{definition}

Note that $\Dduploid^\op$ is a duploid whenever $\Dduploid$ is a duploid.

\noindent
Given a duploid~$\Dduploid$, we find convenient to introduce below
notations for usual (associative) subcategories of $\Dduploid$:
\begin{itemize}
    \item $\Dduploid_l$ is the subcategory of linear maps,
    \item $\Dduploid_t$ is the subcategory of thunkable maps,
    \item $\Pcategory$ is the full subcategory of $\Dduploid$-positive objects,
    \item $\Ncategory$ is the full subcategory of $\Dduploid$-negative objects,
    \item $\Pcategoryt$ is the subcategory of thunkable maps of $\mathcal P$,
    \item $\Ncategoryl$ is the subcategory of linear maps of $\mathcal N$.
\end{itemize}
Given $f \in \mathcal M(X, Y)$ and $g \in \mathcal M(Y,Z)$, we find convenient to write~$g\ccomp f$
as $g \pcomp f$ when $Y$ is $\mathcal{M}$-positive and as $g \ncomp f$ when $Y$ is  $\mathcal{M}$-negative.

\begin{definition}
  A duploid functor $F : \Dduploid \to \Eduploid$ between duploids
consists of a function $F:{|\Dduploid|}\longrightarrow{|\Eduploid|}$
which preserves polarities of objects, together with a family of functions
\[
{F_{X,Y}} \;\; : \;\; {\Dduploid(X,Y)} \;\longrightarrow\; {\Eduploid(FX,FY)}
\]
which preserves compositions and identities as well as linearity and thunkability.
\end{definition}
\begin{proposition}
  Duploids, duploid functors and thunkable linear natural transformations form a 2-category $\mathcal Dupl$.
\end{proposition}
\medbreak
\noindent
The notion of duploid is justified in \citet{munchduploids}
by the following characterization result:
\begin{theorem}[\citet{munchduploids}]\label{theorem/adjunctions-duploids}
Every non-associative category~$\duploid{L}{R}$ associated to an
adjunction $L\dashv R$
comes equipped with a duploid structure, where $\Pcategory$ is
equivalent to the Kleisli category on the monad $T=R\circ L$, and
$\Ncategory$ is equivalent to the co-Kleisli category on the comonad
$K=L\circ R$. Moreover, $\duploid{L}{R}$ is associative if and only if
the monad, or equivalently the comonad, is idempotent.
Conversely, every duploid~$\Dduploid$ induces an
adjunction
\begin{equation}\label[equation]{equation/adjunction-PN}
\begin{tikzcd}[column sep = 1em]
{\Pcategoryt} \arrow[rr,"L", bend left] & \bot & {\Ncategoryl}\arrow[ll,"R",bend left]
\end{tikzcd}
\end{equation}
where $L=\nUparrow$ and $R=\nDownarrow$ are defined
by the negative and positive shift operators,
whose associated duploid $\duploid{L}{R}$ is equivalent to $\Dduploid$.
\end{theorem}

\sectioncaps{Symmetric monoidal duploids}\label{section/sm-duploids}
We have just seen how the notion
of duploid enables one to characterize the non-associative categories
associated to an adjunction $L\dashv R$ (\cref{theorem/adjunctions-duploids}).
We want to extend this characterisation to relate adjunctions giving
rise to continuation models to a duploidal axiomatisation of classical (linear)
logic reflecting the dualities of sequent calculus. In this section we
start with the general case of
the structure inherited by a
duploid~$\duploid{L}{R}$ associated to an adjunction~$L\dashv R$ of
the form~\eqref{equation/adjunction-LR} where the
category~$\Acategory$ is equipped with a symmetric monoidal structure
$(\Acategory,\tensorialand,\tensorialtrue)$ and where the monad
$T=R\circ L$ is strong.

A monad on such a category is
\define{strong} if it is equipped with a pair of left and right strengths
related by symmetry:\begin{align*}
&
\mathrm{rstr}_{A_1,A_2} \, : \, TA_1\tensorialand A_2 \longrightarrow T(A_1\tensorialand A_2)
&&
\mathrm{lstr}_{A_1,A_2} \, : \, A_1\tensorialand TA_2 \longrightarrow T(A_1\tensorialand A_2)
\end{align*}
In that case, the Kleisli category~$\kleisli{\Acategory}{T}$ comes
equipped with a premonoidal structure compatible 
with the original tensor product.
The tensor product $f\ltimes A_2$ of a Kleisli map $f:A_1\to TA_1'$ and an object~$A_2$ is defined as
\[
f\ltimes A_2 \;\; : \;\; A_1\tensorialand A_2 \xlongrightarrow{\;\smash{f\tensorialand A_2}\;} TA_1'\tensorialand A_2 \xlongrightarrow{\;\smash{\mathrm{rstr}}\;} T (A_1'\tensorialand A_2)
\]
and symmetrically, the tensor product of an object~$A_1$ and a Kleisli map $g:A_2\to TA_2'$ is defined as
\[
A_1\rtimes g \;\; : \;\; A_1\tensorialand A_2 \xlongrightarrow{\;\smash{A_1\tensorialand}g\;} A\tensorialand TA_2' \xlongrightarrow{\;\smash{\mathrm{lstr}}\;} T (A_1\tensorialand A_2')
\]
The compatibility between the monoidal structure on~$\Acategory$
and the premonoidal structure on~$\kleisli{\Acategory}{T}$
is witnessed by the fact that the identity-on-object functor
$\iota : \Acategory\to \kleisli{\Acategory}{T}$
transports (strictly) the symmetric monoidal 
structure of~$\Acategory$ to the symmetric
premonoidal structure of~$\kleisli{\Acategory}{T}$.
Recall that given two maps $f:A_1\to A_1'$ and $g:A_2\to A_2'$
in a premonoidal category~$\Pcategory$,
the diagram
\begin{equation*}\label[equation]{equation/premonoidal-square}
\ifbool{arxiv}{
\begin{tikzcd}[column sep = 1.8em,row sep = 1.2em]
}{
\begin{tikzcd}[column sep = 1.6em,row sep = 0.9em]
}
A_1\tensor A_2
\arrow[rr,"f\ltimes A_2"]\arrow[dd,"A_1\rtimes g"{swap}]
&&
A_1'\tensor A_2
\arrow[dd,"A_1'\rtimes g"]
\\
\\
A_1\tensor A_2'\arrow[rr,"f\ltimes A_2'"]
&&
A_1'\tensor A_2'
\end{tikzcd}
\end{equation*}
does not necessarily commute.
A map~$f$ is called \emph{central} when this
diagram
commutes for all maps~$g$.
One shows that the functor $\iota$ transports every morphism in~$\Acategory$
into a central morphism in~$\kleisli{\Acategory}{T}$.
The importance of this structure in the semantics on effects has been
recognised and intensively
studied~\cite{PowerRobinson97,power2002premonoidal,Staton14}.
\begin{definition}
A \define{symmetric monoidal Freyd structure} (also called symmetric
premonoidal $[\rightarrow,\Set]$-category
in \citet{power2002premonoidal})
\[
\iota\;\;:\;\;(\Mcategory,\tensor,1)\;\to\;(\Pcategory,\tensor,1)
\]
is given by a symmetric premonoidal category~$(\Pcategory,\tensor,1)$,
a symmetric monoidal
category~$(\Mcategory,\tensor,1)$, and an identity-on-object functor
$\iota : \Mcategory\rightarrow\Pcategory$
which transports (strictly) the symmetric monoidal structure of~$\Mcategory$
to the symmetric premonoidal structure of~$\Pcategory$,
and such that every morphism~$\iota(f):A\to A'$ in $\Pcategory$
coming from a morphism~$f:A\to A'$ in $\Mcategory$
is central in $\Pcategory$.
\end{definition}

We have seen
in the reconstruction with \cref{theorem/adjunctions-duploids} of a
given duploid~$\mathcal D$, the positive subcategory~$\Pcategory$ of $\Dduploid$
plays the role of the Kleisli category~$\kleisli{\Acategory}{T}$,
whereas the subcategory~$\Pcategoryt$ of thunkable morphisms of
$\Pcategory$ plays the role of the category~$\Acategory$.
This leads us to the following definition:
\begin{definition}\label[definition]{definition/sm-duploid}
A (positive) \define{symmetric monoidal structure}~$(\tensor,1)$ on a
duploid $\Dduploid$ is given by a symmetric monoidal Freyd structure
$(\Pcategoryt,\tensor,1)\to(\Pcategory,\tensor,1)$
for the inclusion functor $\Pcategoryt\inclusion\Pcategory$.
\end{definition}
The De Morgan duality of classical logic will also imply to consider later on the
dual notion of \emph{negative} symmetric monoidal duploid structure
$(\Dduploid,\parr,\bot)$, that is, a symmetric monoidal Freyd category
structure $(\Ncategoryl,\parr,\top)\to(\Ncategory,\parr,\top)$ for the
inclusion functor $\Ncategoryl\inclusion\Ncategory$.

The notion of symmetric monoidal duploid is justified by the following theorem:
\begin{theorem}\label{theorem/adjunctions-monoidal-duploids}
Every non-associative category~$\duploid{L}{R}$ 
associated to an adjunction ${L:\Acategory\rightleftarrows\Bcategory:R}$
where $\Acategory$ is symmetric monoidal
and the monad~$T=R\circ L$ is strong
comes equipped with a (positive) symmetric monoidal duploid structure.
Conversely, every (positive) symmetric monoidal duploid~$\Dduploid$ 
induces an adjunction~\eqref{equation/adjunction-PN}
where $\Pcategoryt$ is equipped with a symmetric monoidal structure
for which the associated monad on $\Pcategoryt$ is strong.
\end{theorem}
\sectioncaps{Graph morphisms and adjunctions between them}\label{section/sm-duploids-ii}
In this more technical section, we will describe the structure induced
in a symmetric monoidal duploid by its symmetric monoidal
Freyd category
$(\Pcategoryt,\tensor,1)\to(\Pcategory,\tensor,1)$ on the rest of the
duploid. The intuition is that a sequent $A_1,\dots,A_n\vdash B$ will
be interpreted by a duploid morphism $A_1\otimes\cdots\otimes
A_n\rightarrow B$.

By using the positive shift $\nDownarrow$, we can generalize the
premonoidal structure on $\Pcategory$ to all objects and morphisms of
$\Dduploid$.
However, this extended tensorial structure is not simply premonoidal,
as it inherits the non-functoriality of the shift. We are therefore
going to consider (reflexive) \emph{graph morphisms} (notation
$F:{\mathscr G}\graphto{\mathscr H}$), which map objects to objects
and morphisms to morphisms while preserving source, target and
identities, but not necessarily composition. We now introduce a notion
of binoidal graph to describe the tensorial structure on $\Dduploid$.

The asynchronous product ${\mathscr G}\boxtimes {\mathscr H}$ of two
reflexive graphs~${\mathscr G}$ and ${\mathscr H}$ is defined as the
reflexive graph whose objects are pairs $(X,Y)$ of objects~$X$ of
${\mathscr G}$ and $Y$ of ${\mathscr H}$ and whose maps are of the
form
$$
(f,Y) : (X,Y)\to (X',Y)
\quad 
(X,g) : (X,Y)\to (X,Y')
$$
with the maps $(\id{X},Y)$ and $(X,\id{Y})$ identified and defining
the identity map $\id{(X,Y)}$ of the object~$(X,Y)$. 
A binoidal graph~${\mathscr G}$ is defined as a reflexive graph
equipped with a graph morphism
\[
\begin{tikzcd}
\otimes \quad : \quad {\mathscr G}\boxtimes{\mathscr G} \arrow[rr,rightharpoonup] &&  {\mathscr G}
\end{tikzcd}
\]
We write $f\ltensortimes Y:X\tensor Y\to X'\tensor Y$ and
$X\rtensortimes g:X\tensor Y\to X\tensor Y'$ the image of $(f,Y)$ and
$(X,g)$, respectively.

One
important observation is that every symmetric monoidal duploid in
the sense of \cref{definition/sm-duploid} comes equipped with a
binoidal structure on objects of any polarity.
The tensor product is extended to every pair of objects $X$ and $Y$ as 
the tensor product of their positive shifts:
\begin{equation}\label[equation]{equation/definition-of-tensor}
X\tensor Y \quad := \quad \nDownarrow X \otimes^+ \nDownarrow Y
\end{equation}
Accordingly, writing $\rtensortimes\plus$ and
$\ltensortimes\plus$ for the premonoidal structure between positive
objects in~$\Pcategory$, given a map $f:X\to X'$ and an object~$Y$, we
define $ f\ltensortimes Y \, = \, \nDownarrow
f \ltensortimes\plus \nDownarrow Y $,
and symmetrically for $\rtensortimes$.

One side-consequence of the definition is that shifting positively
coincides in the monoidal duploid $(\mathcal D,\tensor,1)$ 
with the operation of tensoring with the unit~$1$,
up to a thunkable and linear isomorphism, what can be written:
$$
\nDownarrow X \quad \iso \quad X\tensor 1.
$$
\noindent
In preparation for the \FH{} theorem in~\S\ref{section/fh-theorem}, we
establish that every symmetric monoidal
duploid~$(\Dduploid,\otimes,1)$ still satisfies in this new sense the
following property:
\begin{proposition}\label[proposition]{lemma/thunkableimpliescentral}
Every thunkable map is central.
\end{proposition}
\noindent
Indeed, the positive shift $\nDownarrow$ preserves thunkability, so
$\nDownarrow f$ is also thunkable and thus, central for $\tensor^+$.

The converse
property is not
true in general: consider for instance the symmetric monoidal
duploid~$(\Dduploid,\otimes,1)$ associated to the finite distribution
monad~$T:\Set\to\Set$
already introduced in~\S\ref{subsection/intro-thunkable-linear},
which maps every set~$A$ to the set~$TA$ of its finite probability distributions.
The monad~$T$ is commutative, and every map in~$\Dduploid$ is thus central.
On the other hand, as proved in~\S\ref{subsection/intro-thunkable-linear}, maps into positive objects are thunkable if and only
they are of the form $x \mapsto \dirac{b(x)}$.

Working with graph morphisms is particularly fruitful, as we will see
using the next notion of adjunction between graph morphisms. As a
preliminary observation, note that the composition in any
non-associative category $\Dduploid$ defines a
\emph{graph hom-morphism}
$\Dduploid^\op\boxtimes\Dduploid\graphto\Set$.

\begin{definition}
Let $F:\Dduploid\graphto\Eduploid$ and $G:\Eduploid\graphto\Dduploid$
two graph morphisms between non-associative categories.
An \define{adjunction between graph morphisms} (notation
$\graphadjoint F \Dduploid \Eduploid G$) is given by an isomorphism of
graph morphisms $\Dduploid^{\op}\boxtimes\Eduploid\graphto\Set$:
\[
\varphi\quad:\quad{\Eduploid}(F-,=)\quad\xlongrightarrow{\iso}\quad{\Dduploid}(-,G{=})
\]
natural component-wise, that is to say\ifbool{arxiv}{
\begin{align*}
Gg\ccomp\varphi(f) &=\varphi(g\ccomp f) \\
\varphi(f)\ccomp h &=\varphi(f\ccomp Fh)
\end{align*}
}{
$
Gg\ccomp\varphi(f)=\varphi(g\ccomp f)
$
and
$
\varphi(f)\ccomp h=\varphi(f\ccomp Fh)
$
}
for every $f\in\Eduploid(FX,Y)$ and every morphisms $g$ of $\Eduploid$
and $h$ of $\Dduploid$.
\end{definition}
It is an instructive exercise to check the following:
\begin{proposition}
\label[proposition]{prop/graph-adjunctions-props-1}Let such an adjunction between graph morphisms
${\Eduploid}(F-,=)\iso{\Dduploid}(-,G{=})$.
\begin{enumerate}
\item \label{enu:adj-pres-thunk-lin-1}$F$ preserves thunkability, and
$G$ preserves linearity.
\item \label{enu:G-functorial-1}For $f,g$ morphisms in $\Eduploid$, one
has $G(f\ccomp g)=Gf\ccomp Gg$ if and only if $f\ccomp
g\ccomp\varepsilon_{X}$ associates where
$\varepsilon_X\defeq\inv{\varphi}(\id{GX})$ (for instance, whenever
$f$ is linear or the domain of $g$ is negative). (And dually for $F$
and $\eta_X=\varphi(\id{FX})$.)
\end{enumerate}
\end{proposition}
The first property can be thought of as a property of focusing as seen
in proof search: left-invertible connectives are focused on the
right-hand side of $\vdash$, whereas right-invertible connectives are
focused on the left-hand side of $\vdash$. We have previously
described an example of the second property and given computational
intuitions in the case of $F=\nDownarrow$ and $\eta=\omega$. This
leads us to our first examples of adjunctions:
\begin{proposition}
A duploid $\Dduploid$ is the same thing as a non-associative category
$\Dduploid$ where all objects are either positive or negative (or
both), together with a left adjoint $\nDownarrow$ and a right adjoint
$\nUparrow$ to the identity graph morphism $\Id_{\Dduploid}$, such
that for every object $A\in\Dduploid$, $\nDownarrow A$ is positive and
$\nUparrow A$ is negative.
\end{proposition}

\sectioncaps{Symmetric monoidal closed duploids}\label{sec:SMCD}

Important logical connectives (such as the implication)
turn out to give rise to adjunctions, and can moreover be
characterised in this way, as we show by investigating the closed
structure on symmetric monoidal duploids.
\begin{definition}
A (positive) symmetric monoidal duploid is \define{closed} when
each graph morphism $\mathord{-}\otimes A:\Dduploid\graphto\Dduploid$ for $A\in \Dduploid$ has a right
adjoint $A\linarrow\mathord{-}$, such that every object $A\linarrow B$
is negative.\end{definition}
From this definition we recover a standard model of linear call by
value: every symmetric monoidal closed structure on a duploid
$\Dduploid$ gives rise to a closed structure on the symmetric monoidal
Freyd structure of its positive category, in the sense
of \citet{power2002premonoidal}, namely a right adjoint to the functor
$\iota\mathord{-}\otimes P:\Pcategory_{t}\rightarrow\Pcategory$ for
every object $P$ of $P_t$. The closed Freyd structure is obtained with
the definition
$P\linarrow^{+}\mathord{-}\defeq\nDownarrow(P\linarrow\nUparrow-)$,
recalling the one for call-by-value arrows in polarised logics and in
call-by-push-value.

However, not every closed duploid structure can be recovered from its
symmetric monoidal closed Freyd structure with the converse definition
$\nUparrow(P\linarrow^{+}\nDownarrow N)$. Symmetric monoidal closed
duploids correspond in fact to the more general notion of \emph{linear
effect calculus} model from \citet{Mellies2012parametric},
or \emph{linear call-by-push-value} (without sums) from
\citet{CFM2015}---a linear version of the \emph{effect calculus} model
of \citet{Egger2012}.
The next definition uses the notions of presheaf-enriched category and
adjunction.
\begin{definition}[\citep{Mellies2012parametric,CFM2015}]
A \emph{linear effect adjunction} is given by a symmetric monoidal
category $\Acategory$, an $\Psh{\Acategory}$-category $\underline{\Bcategory}$, a
$\Psh{\Acategory}$-adjunction
${\underline{L}:\underline{\Acategory}\rightleftarrows\underline{\Bcategory}:\underline{R}}$, and powers of
representable presheaves on $\underline{\Bcategory}$.
\end{definition}
We also recall the following characterisation of linear effect
adjunctions given in \citet{Mellies2012parametric} which less
common, but more convenient in this context as we will see.
\begin{proposition}[\citep{Mellies2012parametric}]
A linear effect adjunction is the same thing as a symmetric
monoidal category $\Acategory$
together with
an adjunction ${L:\Acategory\rightleftarrows\Bcategory:R}$, a
pseudo-action
$\mathord{\linarrow}:\Acategory^{\op}\times\Bcategory\rightarrow\Bcategory$
of $\Acategory^{\op}$ on $\Bcategory$, and a family of
adjunctions
\begin{equation}
L(-\otimes A):\Acategory\rightleftarrows\Bcategory:R(A\linarrow-)\;.\label[equation]{eq:param-adj-main}
\end{equation}
\end{proposition}

Suppose given such a linear effect adjunction, and write
$\Dduploid=\duploid{L}{R}$ for the associated symmetric monoidal duploid,
with $\otimes_\Dduploid$ its premonoidal tensor product extended to a
graph morphism ${\Dduploid\boxtimes\Dduploid\graphto\Dduploid}$ as
previously. 
Define for any $A,B\in\Dduploid$ the negative object
${A\linarrow_\Dduploid B\defeq \nDownarrow A\linarrow\nUparrow B}$. We
observe that the family of bijections between hom-sets underlying
(\ref{eq:param-adj-main}) is exactly a family
\pagebreakopportunitypopl
of bijections (with
${A,B,C\in\Dduploid}$):
\[
  \Dduploid(A\otimes_{\Dduploid}B,C)\iso\Dduploid(A,B\linarrow_{\Dduploid}C)
\]
This family of bijections extends, for each $B\in\Dduploid$, into an
adjunction $\graphadjoint {-\otimes_{\Dduploid} B} \Dduploid \Dduploid
{B\linarrow_\Dduploid -}$. This follows from a general principle:

\begin{proposition}\label[proposition]{prop:adjunction-right-adjoint}
Let $\Dduploid,\Eduploid$
two non-associative categories and $F$ a graph morphism
$\Eduploid\graphto\Dduploid$ that \emph{preserves thunkability}.
Assume that for each $A\in\Eduploid$ there exists a \emph{negative}
object $N\in\Dduploid$ and a family of bijections
$\Dduploid(FA,B)\iso\Eduploid(A,N)$ natural in
$A$. Then these families of bijections assemble into a right adjoint
$\graphadjoint F \Eduploid \Dduploid G$.
\end{proposition}
\begin{theorem}\label{theorem/adjunction-closed-duploid}
For every linear effect
adjunction~${L:\Acategory\rightleftarrows\Bcategory:R}$, its
associated non-associative category~$\duploid{L}{R}$ comes equipped
with a symmetric monoidal closed duploid structure.
Conversely, every symmetric monoidal closed duploid~$\Dduploid$ 
induces an adjunction $L\dashv R$ equipped with the structure of
a linear effect adjunction.

\end{theorem}

\begin{example}
$\Set$ with the finite distribution monad, or any symmetric monoidal
closed category $\Acategory$ with a strong monad $T$, gives rise
to linear effect adjunctions $F_{T}\dashv G_{T}:\Acategory_{T}\rightarrow\Acategory$
and $F^{T}\dashv U^{T}:\Acategory^{T}\rightarrow\Acategory$ \citep{melliesparametric,CFM2015},
and thus to symmetric monoidal closed duploids.
\end{example}

\sectioncaps{The linear call-by-push-value $L$-calculus}\label{section/CBPV-L-calculus}

\begin{figure*}[!t]
\begin{framed}
\begin{small}
\centering
\hspace*{\fill}\subfloat[Grammar]{
$
 \stretcharray
 \begin{array}{lrcccccccccccccccc}
   \mbox{Values:} & V, W & ::= & a &\bmid& \mu\alpha\moins.\command c &\bmid& () &\bmid& V \smallotimes W &\bmid& \mu(a \push \beta).\command c\\
   \mbox{Expressions:} & t,u & ::= & V &\bmid& \mu\alpha\plus.\command c\\
   \mbox{Stacks:} & S & ::= & \alpha &\bmid& \tmu a\plus.\command c &\bmid& \tmu().\command c &\bmid& \tmu(a\smallotimes b).\command c &\bmid& V \push S\\
   \mbox{Contexts:} & e & ::= & S &\bmid& \tmu a\moins.\command c\\
   \mbox{Commands:} & \command c & ::= & \perfectcut{V}{e}\moins &\bmid& \perfectcut{t}{S}\plus
 \end{array}
$
}\hspace*{\fill}

\hspace*{\fill}\subfloat[Reduction and expansion rewriting rules]{
 $
 \stretcharray
 \begin{array}{lccl}
   (R\tmu\eps) & \perfectcut{V}{\tmu a\eps.\command c}\eps & \triangleright_R & c[V/a]\\
   (R\mu\eps) & \perfectcut{\mu \alpha\eps.\command c}{S}\eps & \triangleright_R & c[S/\alpha]\\
   (R1) & \perfectcut{()}{\tmu().\command c}\plus & \triangleright_R & c\\
   (R\otimes) & \perfectcut{V\smallotimes W}{\tmu(a\smallotimes b).\command c}\plus & \triangleright_R & c[V/a,W/b]\\
   (R\linarrow) & \perfectcut{\mu(a \push \beta).\command c}{V\push S}\moins & \triangleright_R & c[V/a, S/\alpha]
 \end{array}
 \quad\quad\quad\quad
 \begin{array}{lrcl}
   (E\tmu\eps) & e & \triangleright_E & \tmu a\eps.\perfectcut{a}{e}\eps\\
   (E\mu\eps) & t & \triangleright_E & \mu\alpha\eps.\perfectcut{t}{\alpha}\eps \\
   (E1) & S & \triangleright_E & \tmu().\perfectcut{()}{S}\plus \\
   (E\otimes) & S & \triangleright_E & \tmu(a\smallotimes b).\perfectcut{a\smallotimes b}{S}\plus \\
   (E\linarrow) & V & \triangleright_E & \mu(a \push \beta).\perfectcut{V}{a \push \beta}\moins
 \end{array}
 $
}\hspace*{\fill}

\hspace*{\fill}\subfloat[Judgements]{
$\begin{gathered}
 \command c : (\Gamma \vdash \Delta) \quad\quad\quad \Gamma \vdash t : A\ |\ \Delta\quad\quad\quad \Gamma\ |\ e : A \vdash \Delta\\
\mbox{where }\quad\quad\Gamma=x_1:A_1,\dots,x_n:A_n\quad\quad\Delta=\alpha_1:B_1,\dots,\alpha_m:B_m
\end{gathered}
$
}\hspace*{\fill}

\hspace*{\fill}\subfloat[Typing rules]{
\begin{varwidth}{\linewidth}
\centering
 $
   \begin{prooftree}
     \infer0[$(\vdash\mathbf{ax})$]{a:A\vdash a : A\ |}
   \end{prooftree}
   \quad\quad
   \begin{prooftree}
     \infer0[$(\mathbf{ax}\vdash)$]{|\ \alpha:A\vdash \alpha : A}
   \end{prooftree}
 $
 \bigbreak
 $
   \begin{prooftree}
     \hypo{\command c:(\Gamma, a:A_\veps\vdash \Delta)}
     \infer1[$(\boldsymbol{\tmu}\eps\vdash)$]{\Gamma\ |\ \tmu a\eps.\command c:A_\veps\vdash\Delta}
   \end{prooftree}
   \quad\quad
   \begin{prooftree}
     \hypo{\command c:(\Gamma \vdash\alpha : A_\veps, \Delta)}
     \infer1[$(\vdash\boldsymbol{\mu}\eps)$]{\Gamma\vdash \mu \alpha\eps.\command c:A_\veps\ |\ \Delta}
   \end{prooftree}
   \quad\quad
   \begin{prooftree}
     \hypo{\Gamma\ |\ e : A_\veps\vdash \Delta}
     \hypo{\Gamma' \vdash t : A_\veps\ |\ \Delta'}
     \infer2[$(\mathbf{cut}\eps)$]{\perfectcut{t}{e}\eps : (\Gamma, \Gamma' \vdash \Delta, \Delta')}
   \end{prooftree}
 $
\sepline
 $
 \text{For any }\sigma\in\Sigma(\Gamma,\Gamma')\text{ and }\forall\tsigma\in\Sigma(\Delta,\Delta'):
 $
  \bigbreak
$
 \begin{prooftree}
   \hypo{\Gamma \vdash t : A\ |\ \Delta}
   \infer1[$\mkern-4mu(\vdash\sigma, \tsigma)$]{\Gamma'\vdash t[\sigma,\tsigma] : A\ |\ \Delta'}
 \end{prooftree}
 \quad \quad
 \begin{prooftree}
   \hypo{\Gamma\ |\ e : A\vdash\Delta}
   \infer1[$\mkern-4mu(\sigma, \tsigma\vdash)$]{\Gamma'\ |\ e[\sigma,\tsigma] : A\vdash\Delta'}
 \end{prooftree}
 \quad \quad
 \begin{prooftree}
   \hypo{\command c:(\Gamma \vdash \Delta)}
   \infer1[$\mkern-4mu(\sigma, \tsigma)$]{\command{c}{[\sigma,\tsigma]}:(\Gamma'\vdash \Delta')}
 \end{prooftree}
 $
 \sepline
 \vspace*{-2ex}\linebreak{}\vspace*{-1ex}$
 \stretcharraymore
 \begin{array}{cc}
 \begin{prooftree}
   \hypo{}
   \infer1[$(\vdash 1)$]{\vdash() : 1\ |}
 \end{prooftree}
 &
 \begin{prooftree}
   \hypo{\command c : (\Gamma\vdash\Delta)}
   \infer1[$(1 \vdash)$]{\Gamma\ |\ \tmu().\command c : 1 \vdash \Delta}
 \end{prooftree}
 \\
 \begin{prooftree}
   \hypo{\Gamma\vdash V : A\ |\ \Delta}
   \hypo{\Gamma'\vdash W : B\ |\ \Delta'}
   \infer2[$(\vdashf\otimes)$]{\Gamma,\Gamma'\vdash V \smallotimes W:A\otimes B\ |\ \Delta, \Delta}
 \end{prooftree}
 &
 \begin{prooftree}
   \hypo{\command c : (\Gamma, a : A, b : B\vdash\Delta)}
   \infer1[$(\otimes\vdash)$]{\Gamma\ |\ \tmu(a \smallotimes b).\command c : A\otimes B\vdash\Delta}
 \end{prooftree}
 \\
 \begin{prooftree}
  \hypo{\command c : (\Gamma, a : A \vdash \beta : B, \Delta)}
  \infer1[$(\vdash\linarrow)$]{\Gamma \vdash \mu(a \push \beta).\command c : A \linarrow B\ |\ \Delta}
 \end{prooftree}
 &
 \begin{prooftree}
   \hypo{\Gamma \vdash V : A\ |\ \Delta}
   \hypo{\Gamma' \ |\ S : B \vdash \Delta'}
   \infer2[$(\linarrow\vdashf)$]{\Gamma, \Gamma'\ |\ V \push S : A \linarrow B \vdash \Delta, \Delta'}
 \end{prooftree}
 \end{array}
 $

\end{varwidth}
}\hspace*{\fill}
\end{small}
\end{framed}
\caption{Syntax of the linear call-by-push-value $L$-calculus \citep{CFM2015}}
\label{figure/syntaxCBPV}
\end{figure*}

$L$-calculi are $\lambda$-calculi (higher-order rewriting systems)
that present themselves in the form of abstract machines (with an
explicit context), and whose typing rules closely match those of
sequent calculus, subsuming the rich relationship between CPS, abstract
machines, focusing, etc. They are the final ingredient in our correspondence.

In this section, we recall the \emph{linear call-by-push-value}
$L$-calculus introduced in \citep{CFM2015} under the name
$\polsystem{IMLL}$.
It is presented in
\cref{figure/syntaxCBPV}. Later on (\S\ref{section/classical-L-calculus}),
we will extend this calculus with a involutive negation to obtain a
linear classical sequent calculus.

\paragraphcaps{Syntax}

The terms of the $L$-calculus come in five syntactic categories:
expressions, values, contexts, stacks and commands.
Values $V$ and stacks $S$ are particular expressions and contexts,
respectively, which can be understood as pure or effect-free.
Variables (noted $a$, $b$, $c$, ...) stand for values $V$, and dually,
co-variables (noted $\alpha$, $\beta$, $\gamma$, ...) stand for stacks
$S$.
\ifbool{arxiv}{Types are as follows, each one equipped with a polarity, including atoms $X\plus,X\moins$}{Each type is equipped with a polarity}:
\begin{align*}
\text{Types } A, B, A_\veps &::=
\underbrace{X\plus\;\bmid\; 1\;\bmid\; A\otimes B}_{\text{positive } (P, Q, A_+)} \;\bmid\;
\underbrace{X\moins\;\bmid\; A \linarrow B}_{\text{negative } (N, M, A_-)}
\end{align*}

Variables are bound by $\tmu$ to form a stack~$\tmu a\plus.\command c$
when the variable~$a$ has a positive type, and to form a context~$\tmu
a\moins.\command c$ when the variable~$a$ has a negative type.
Dually, co-variables are bound by $\mu$ to form a
value~$\mu \alpha\moins.\command c$ when the co-variable~$\alpha$ has
a negative type, or an expression~$\mu \alpha\plus.\command c$ when
the co-variable~$\alpha$ has a positive type.
The value $()$ and the nullary binder $\tmu().\command c$
are associated with the unit of the conjunction $\1$. We can construct
conjunctive terms with either the binary binder $\tmu(a \smallotimes
b).\command c$ or the constructor $V \smallotimes W$.
For the closure, we have the binary binder $\mu(a \push \beta).\command c$
and the constructor $V \push S$.

\paragraphcaps{Reductions and expansions}

\noindent \Cref{figure/syntaxCBPV} defines a reduction relation $\triangleright_R$
($\beta$-like) and an expansion relation $\triangleright_E$
($\eta$-like) between terms. First, the toplevel reduction
$\triangleright_R$ defines an operational semantics because its
(well-typed) normal forms for terms without free variables are of the
form $\perfectcut{V}{\alpha}$, denoting a computation ending
with a value $V$.

We note $\to_{RE}$ the contextual closure of ($\beta\eta$)
\emph{reduction} $\triangleright_R \cup \triangleleft_E$, and
$\simeq_{RE}$ the symmetric, transitive and reflexive closure of
$\to_{RE}$. This rewriting theory is well-behaved and
meaningful, as follows from standard techniques adapted to the
$L$-calculus~\citep{CurienMunch09DualFocus,Munch-Maccagnoni2017curry}:
for instance, $\to_{R}$ is confluent and
strongly normalising on typed terms, and the $\to_{R}$-normal,
$\to_{E}$-long forms correspond to focused proofs in the
proof-theoretic sense. No additional commuting conversion is
necessary.

\paragraphcaps{Example with the composition}

In this calculus, the $\keyword{let}$ construct is defined as:
\[
  \letinepsilon a t u \;\;{}\defeq{}\;\; \mu\alpha^{\veps'}.\perfectcut{t}{\tmu a\eps.{\perfectcut u \alpha}^{\veps'}}^{\veps}\;.
\]
The behavior of this $\keyword{let}$ matches the intuitive
account given in the introduction. The positive $\keyword{let}$
($\veps=+$) follows the call-by-value paradigm (first reducing $t$
until obtaining a value, then making the substitution in $u$) whereas
the negative $\keyword{let}$ ($\veps=-$) follows the
call-by-name paradigm (substituting $t$ in $u$). This follows
from the fact that $\veps$ determines whether the inner command is of
the form $\cut{V}{e}$ or $\cut{t}{S}$. Whether the whole expression
$\letinepsilon a t u$ is evaluated strictly or lazily depends on the
polarity $\veps'$, which determines whether the whole expression is a
value or not.

\paragraphcaps{Typing derivations}

The judgements are of the form $\Gamma\vdash \Delta$. As is usual for
$L$-calculi, expressions and values have a type taken from the
right-hand side of $\vdash$ while contexts and stacks have a type
taken from the left-hand side of $\vdash$. Commands do not have any
type.
A context on the left-hand side $\Gamma$ (resp. context on the
right-hand side $\Delta$) is a map from an ordered finite set of
variables (resp. co-variables) to types of any polarity. The notations
$\Gamma,\Gamma'$ and $\Delta,\Delta'$ imply that the contexts have
disjoint domains. In the case of the linear
call-by-push-value $L$-calculus, we can in fact see (by induction)
that \emph{there is always exactly one formula on the right-hand-side
of $\vdash$}. This will no longer be true in the linear classical
$L$-calculus from \S\ref{section/classical-L-calculus}.

\paragraphcaps{Structural rules and linearity}

A single structural rule is given for each kind of judgement, as
in \citep{CFM2015} (see also \citet{Atkey2006}). Structural rules let
us \emph{rename} the (co-)variables of the contexts and \emph{reorder}
them. To this effect, we define
\define{$\Sigma(\Gamma,\Gamma')$} the set of \emph{bijective} maps $\sigma :
\dom \Gamma \to \dom \Gamma'$ that preserve types (\emph{i.e.} $\Gamma'(\sigma(a)) =
\Gamma(a)$ for all $a\in\dom \Gamma$).
The \emph{(non-linear) call-by-push-value $L$-calculus} (without sums)
is defined by simply relaxing the bijection requirement for structural
maps $\sigma\in\Sigma(\Gamma,\Gamma')$---thus
allowing \emph{weakening} and
\emph{contraction}.

\begin{theorem}[Subject reduction \citep{CFM2015,Munch-Maccagnoni2017curry}]
  If $\command c \to_{RE} \command c'$ and $\command c :
  (\Gamma \vdash \Delta)$, then $\command c' :
  (\Gamma \vdash \Delta)$.
\end{theorem}

\paragraphcaps{Recovering non-focused rules}
Non-restricted constructors $t\smallotimes u$ and $t\push e$, and
corresponding non-focused rules $(\vdash\otimes)$ and
$(\linarrow\vdash)$, can be defined from $V\smallotimes W$, $V\push
S$, $(\vdashf\otimes)$ and $(\linarrow\vdashf)$ by introducing cuts.
The constructions are detailed in \citep{CFM2015}. These definitions
imply the existence of generalized substitutions $\command c[t/a]$ and
$\command c[e/\alpha]$ which appears in the characterisation of
thunkability.

\paragraphcaps{Soundness of the calculus}
\begin{theorem}[Soundness for symmetric monoidal closed duploids]\label{thm/soundness-CBPV}
\ifbool{arxiv}{
Given some assigment $\rho$ of atoms to objects respecting the
polarity, the interpretation of typed terms of the linear
call-by-push-value $L$-calculus in any symmetric monoidal closed
duploid, sending sequents $A_1,\dots,A_n\vdash B$ to hom-sets
$\Dduploid(A^\rho_1\otimes\cdots\otimes A^\rho_n,B^\rho)$, is
invariant modulo typed reductions and expansions.
}{
The interpretation of typed terms of the linear call-by-push-value
$L$-calculus in any symmetric monoidal closed duploid is invariant
modulo typed reductions and expansions.
}
\end{theorem}
The soundness of an interpretation into linear effect adjunctions was
previously established in \citep{CFM2015}.
In fact:
\begin{theorem}
The interpretation of the linear call-by-push-value $L$-calculus
(without sums) in \citep{CFM2015,Munch-Maccagnoni2017curry} factors
into the previous construction of a symmetric monoidal closed duploid
from a linear effect adjunction
(\cref{theorem/adjunction-closed-duploid}), and the direct
interpretation of the calculus into the symmetric monoidal closed
duploid.
\end{theorem}
\ifbool{arxiv}{The result can be extended to the definition of non-focused constructors, in which case the definition of
$V\push e$ matches the construction of $\linarrow$ as a right adjoint
using \cref{prop:adjunction-right-adjoint}.}{}

\sectioncaps{Dialogue duploids}\label{section/dialogue-duploids}
In this section, we will describe the structure inherited by a duploid
associated to a dialogue chirality~\cite{melliesdialogue} described in
the introduction (\cref{definition/chirality}) as a rephrasing of
continuation models. A dialogue duploid is going to be a duploid given with
positive and negative symmetric monoidal structures $(\otimes,1)$ and
$(\parr,\bot)$ related by a \emph{duality functor} $\dupldual{(-)}$.
The intuition is that a classical sequent $A_1,\dots,A_n\vdash
B_1,\dots,B_m$ will be interpreted by a duploid morphism
$A_1\otimes\cdots\otimes A_n\rightarrow B_1\parr\cdots\parr B_m$ and
$\dupldual{(-)}$ will interpret negation.

Before formally defining the notion of dialogue duploid, we find convenient
to define a notion of strong monoidal functor between symmetric monoidal
duploids:
\begin{definition}
A strong monoidal functor
\begin{center}
\begin{tikzcd}
F \quad : \quad (\Dduploid,\otimes,\1) \arrow[rr] &&  (\Eduploid,\otimes,\1)
\end{tikzcd}
\end{center}
between symmetric monoidal duploids is a duploid functor from $\Dduploid$ to $\Eduploid$ 
equipped with a family of thunkable and linear isomorphisms
\begin{center}
\begin{tikzcd}[column sep = 1.5em, row sep=-.2em]
{m_{X,Y}} \quad : \quad {FX\otimes FY} \arrow[rr] && {F(X\otimes Y)}
\\
\quad {m_{1}} \quad \hspace{.15em}  :  \hspace{.85em} \quad \quad {1} \quad \quad\hspace{.1em} \arrow[rr] && \quad\hspace{.2em} {F(1)} \quad
\end{tikzcd}
\end{center}
natural in each component~$X$ and~$Y$ independently,
and making the same coherence diagrams commute
as in the usual case of a strong monoidal functor between symmetric monoidal categories.
\end{definition}

\begin{definition}
A pair of strong monoidal functors $F:\Dduploid\to\Eduploid$ and $G:\Eduploid\to\Dduploid$
between symmetric monoidal duploids $(\Dduploid,\otimes,\1)$ and $(\Eduploid,\otimes,\1)$
is called a \define{monoidal equivalence}
when there exists two families of thunkable 
and linear isomorphisms $\nu_X : F(GX) \to X$ and $\nu'_X : G(FX) \to X$, both natural in $X$
and compatible with the respective $m_{X,Y}$ and $m_1$.
\end{definition}

This
leads us to the following definition of a dialogue duploid.
\begin{definition}\label[definition]{dialogueduploidsym}
A \define{dialogue duploid} is a duploid $\mathcal D$
equipped with a positive and negative symmetric monoidal duploid 
structure $(\mathcal D, \otimes, \1)$ and $(\mathcal D, \parr, \bot)$
related by a strong monoidal equivalence
   \[
      \begin{tikzcd}
        (\mathcal D, \otimes, \1) \arrow[r, bend left, "{\dupldual{(-)}}"] & \arrow[l, bend left, "{\dupldual{(-)}}"] (\mathcal D, \parr, \bot)^{\op}
      \end{tikzcd}
    \]
together with a family of adjunctions $-\otimes Y\vdash \dupldual{Y}\parr -$ 
between graph morphisms (called
currification)
$$\chi_{X, Y, Z} \quad : \quad \Dduploid (X \otimes Y, Z) \quad \xrightarrow{\iso} \quad \Dduploid (X, \dupldual{Y} \parr Z)$$
natural component-wise in~$X$, $Y$ and $Z$,
and subject up to monoidality, symmetry and associativity to the
coherence condition \ifbool{arxiv}{between $\chi$ and monoidality}{}
$\chi_{A,B\otimes C,D}=\chi_{A,B,C^{*}\parr D}\circ\chi_{A\otimes B,C,D}$.
\end{definition}

\noindent Note that an associative dialogue duploid is the same thing
as a $*$-autonomous category.
The theorem below establishes in what sense the notion of dialogue duploid
can be seen as a direct and computational counterpart to dialogue chiralities,
which provides an overarching mathematical framework for reasoning in direct style
about (linear and non-linear continuations),
while preserving the perfect symmetry between CBV and CBN evaluation paradigms.
\begin{theorem}\label{theorem/adjunctions-dialogue-duploids}
Every duploid~$\duploid{L}{R}$ 
associated to a dialogue chirality $L\dashv R$
comes equipped with a dialogue duploid structure.
Conversely, every dialogue duploid~$\Dduploid$
induces a dialogue chirality structure on the
adjunction~\eqref{equation/adjunction-PN}, whose associated dialogue
duploid is equivalent to~$\Dduploid$ via strong monoidal duploid
functors that also preserve the duality.
\end{theorem}

\sectioncaps{The linear classical $L$-calculus}\label{section/classical-L-calculus}

\begin{figure*}[!t]
\begin{framed}
\vspace*{-1.5ex}
\begin{small}
\centering
\hspace*{\fill}\begin{small}\parbox[c]{18.8em}{The grammar,
conversions and typing rules are those from \cref{figure/syntaxCBPV},
with the constructors $\mu(a\push\beta).c$ and $V\push S$ removed and
the following added:}\end{small}
\hspace*{\fill}\hspace*{0.5em}
\subfloat[Additional Grammar]{
\begin{varwidth}{\linewidth}
$
 \stretcharray
 \begin{array}{rcccccccccccccccc}V, W & ::= & \dots &\bmid& \mu[].\command c &\bmid& \mu(\alpha \smallparr \beta).\command c&\bmid& [S] &\bmid& \mu[a].\command c\\
S & ::= & \dots &\bmid& [] &\bmid& S \smallparr S'&\bmid& \tmu[\alpha].\command c &\bmid& [V]\\
 \end{array}
$
\end{varwidth}
}\hspace*{\fill}

\hspace*{\fill}\subfloat[Additional conversions (Reduction and expansion rules)]{
 $
 \stretcharray
 \begin{array}{lccc}
   (R\bot) & \perfectcut{\mu[].\command c}{[]}\moins & \triangleright_R & c\\
   (R\parr) & \perfectcut{\mu(\alpha\smallparr\beta).\command c}{S \smallparr S'}\moins & \triangleright_R & c[S/\alpha,S'/\beta]\\
   (R\negN) & \perfectcut{[S]}{\tmu[\alpha].\command c}\plus & \triangleright_R & c[S/\alpha]\\
   (R\negP) & \perfectcut{\mu[a].\command c}{[V]}\moins & \triangleright_R & c[V/a]
 \end{array}
 \quad\quad\quad\quad
 \begin{array}{lccc}
   (E\bot) & V & \triangleright_E & \mu[].\perfectcut{V}{[]}\moins\\
   (E\parr) & V & \triangleright_E & \mu(\alpha\smallparr\beta).\perfectcut{V}{\alpha\smallparr\beta}\moins\\
   (E\negN) & S & \triangleright_E & \tmu[\alpha].\perfectcut{[\alpha]}{S}\plus \\
   (E\negP) & V & \triangleright_E & \mu[a].\perfectcut{V}{[a]}\moins 
 \end{array}
 $
}\hspace*{\fill}

\hspace*{\fill}\subfloat[Additional typing rules]{
\begin{varwidth}{\linewidth}
\centering
$
 \stretcharraymore
 \begin{array}{cc}
 \begin{prooftree}
   \hypo{}
   \infer1[$(\bot\vdash)$]{|\ []:\bot \vdash}
 \end{prooftree}
 &
 \begin{prooftree}
   \hypo{\command c:(\Gamma\vdash\Delta)}
   \infer1[$(\vdash\bot)$]{\Gamma \vdash \mu[].\command c : \bot \ |\ \Delta}
 \end{prooftree}
 \\
 \begin{prooftree}
   \hypo{\Gamma\ |\ S : A\vdash \Delta}
   \hypo{\Gamma'\ |\ S' : B\vdash \Delta'}
   \infer2[$(\parr\vdashf)$]{\Gamma,\Gamma'\ |\ S \smallparr S':A\parr B\vdash \Delta,\Delta'}
 \end{prooftree}
 &
 \begin{prooftree}
   \hypo{\command c:(\Gamma\vdash \alpha : A, \beta : B, \Delta)}
   \infer1[$(\vdash\parr)$]{\Gamma \vdash \mu(\alpha \smallparr \beta).\command c : A \parr B\ |\ \Delta}
 \end{prooftree}
 \\
 \begin{prooftree}
   \hypo{\Gamma\ |\ S : N\vdash\Delta}
   \infer1[$(\vdashf\negN)$]{\Gamma\vdash[S]:\dupldual{N}\ |\ \Delta}
 \end{prooftree}
 &
 \begin{prooftree}
   \hypo{\command c:(\Gamma\vdash\alpha:N, \Delta)}
   \infer1[$(\negN\vdash)$]{\Gamma\ |\ \tilde\mu[\alpha].\command c:\dupldual{N}\vdash\Delta}
 \end{prooftree}
 \\
 \begin{prooftree}
   \hypo{\Gamma\vdash V : P\ |\ \Delta}
   \infer1[$(\negP\vdashf)$]{\Gamma\ |\ [V]:\dupldual{P}\vdash\Delta}
 \end{prooftree}
 &
 \begin{prooftree}
   \hypo{\command c:(a:P,\Gamma\vdash\Delta)}
   \infer1[$(\vdash\negP)$]{\Gamma\vdash\mu[a].\command c:\dupldual{P}\ |\ \Delta}
 \end{prooftree}
 \end{array}
$
\end{varwidth}
}\hspace*{\fill}
\end{small}
\end{framed}
\caption{Syntax of the linear classical $L$-calculus}
\label{figure/syntaxDialogue}
\end{figure*}
In this section, we expand the calculus of \cref{section/CBPV-L-calculus} with a disjunction
and an involutive negation, whose constructors and rules are presented in \cref{figure/syntaxDialogue}.
We call this expanded calculus the linear classical $L$-calculus.

\paragraphcaps{Syntax}

From the types of the linear call-by-push-value $L$-calculus, we
remove $\linarrow$ and add the disjunction $\parr$ as well as its the
corresponding unit $\bot$. Furthermore, we split negation
$\dupldual{(-)}$ into two type constructors, one for each polarity (as
in \citep{danos95new, Munch14Involutive}). We write the two negation
connectives the same, for simplicity and due to lack of ambiguity.
\begin{align*}
\text{Types } A, B, A_\veps &::=
\underbrace{X\plus\;\bmid\; 1\;\bmid\; A\otimes B\;\bmid\;\dupldual{N}}_{\text{positive } (P, Q, A_+)} \;\bmid\;
\underbrace{X\moins\;\bmid\; \bot\;\bmid\; A \parr B\;\bmid\;\dupldual{P}}_{\text{negative } (N, M, A_-)}
\end{align*}
Corresponding to the unit $\bot$ and to the disjunction $\parr$,
dualizing the unit $1$ and the conjunction $\otimes$, we have the
nullary and binary binders $\mu[].\command c$ and
$\mu(\alpha\smallparr \beta).\command c$ and the constructors $[]$ and
$S \smallparr S'$.

\paragraphcaps{Constructors for an involutive negation}

In order to model the rules of negation, we also have the unary
binders $\mu[a].\command c$ and $\tmu[\alpha].\command c$, as well as
the constructions $[V]$ and $[S]$ which turn terms into duals.
The computational intuition for $|[V]:\dupldual{P}\vdash$ is that of a
stack with a single positive argument and \emph{no return}, where
$\dupldual{P}$ is a negative type of continuations expecting $P$. On
the other hand, the intuition for $\vdash[S]:\dupldual{N}|$ is that of
a value denoting a stack that has been captured (by a control
operator) where $\dupldual{N}$ is a positive type of (inspectable)
captured stacks~\cite{Levy2004, Munch14Involutive}.

The rules for negation can create right-hand side contexts $\Delta$
with strictly more or less than one formula. The underlying logic of
typing judgments is not intuitionistic anymore and is a polarised
multiplicative linear logic.

\paragraphcaps{The closure in the linear $L$-calculus}
In the linear classical $L$-calculus, the closure is definable as:
\[
  A \linarrow B = \dupldual A \parr B
\]
Moreover, the rules associated with it presented
in \cref{figure/syntaxCBPV} can be deduced from the rules for the
disjunction and the negation. This is why we do not include
$\linarrow$ in the definition of the linear classical $L$-calculus.

\paragraphcaps{Recovering non-focused rules}
Recovering the non-focused rules $(\vdash\negN)$ and $(\negP\vdash)$
is a crucial and subtle part about the involutive negation.
\begin{definition}
For $e$ a negative context, we define $[e] := \mu\alpha\plus.\perfectcut{\mu\beta\moins.\perfectcut{[\beta]}{\alpha}\plus}{e}\moins$. Symmetrically, for t a positive term, we define $[t] := \tilde\mu a\moins.\perfectcut{t}{\tilde\mu b\plus.\perfectcut{a}{[b]}\moins}\plus$. The following rules can be derived:
  \[
    \begin{prooftree}
      \hypo{\Gamma\ |\ e : N\vdash\Delta}
      \infer[double]1[$(\vdash\negN)$]{\Gamma\vdash[e]:\dupldual N\ |\ \Delta}
    \end{prooftree}
    \quad    \quad    \quad
    \begin{prooftree}
      \hypo{\Gamma\vdash t : P\ |\ \Delta}
      \infer[double]1[$(\negP\vdash)$]{\Gamma\ |\ [t]:\dupldual P\vdash\Delta}
    \end{prooftree}
  \]
\end{definition}

In words, the reduction of $[e]$ and $[t]$ proceeds inside the terms
using what looks like let-expansions:\vspace{-1ex}
\begin{align*}
\perfectcut{[e]}{S}\plus & \triangleright_R\perfectcut{\mu\beta\moins.\perfectcut{[\beta]}{S}\plus}{e}\moins & \text{(\ensuremath{e} not a stack)}\\
\perfectcut{V}{[t]}\moins & \triangleright_R\perfectcut{t}{\tilde\mu b\plus.\perfectcut{V}{[b]}\moins}\plus & \text{(\ensuremath{t} not a value)}
\end{align*}
Notice that these expansions involve cuts of both polarities in the
right-hand side. It is therefore not possible to obtain this
computational behaviour in calculi that are globally call-by-value or
call-by-name, in which negation is a suspension
(\emph{e.g.}~\cite{CH00Duality,Wad03Dual,Laurent2008} among others).

A straightforward but nevertheless useful property states that the
involutive negation internalises the duality between expressions and
contexts:
\begin{lemma}\label[lemma]{lemma/negSwitchSide}
For $e$ a context and $\command c$ a command, one has
$
  \perfectcut{[e]}{\tmu[\alpha].\command c}\plus\simeq_{RE}\perfectcut{\mu\alpha\moins.\command c}{e}\moins
$.
Likewise, for $t$ an expression and $\command c$ a command, one has:
$
  \perfectcut{\mu [a].\command c}{[t]}\moins\simeq_{RE}\perfectcut{t}{\tmu a\plus.\command c}\plus
$.
\end{lemma}

\begin{arxiv}
\begin{proof}
  Let $e$ be a negative context and $\command c$ a command.
\begin{align*}
  & \perfectcut{\underline{\mu\beta\plus}.\perfectcut{\mu\gamma\moins.\perfectcut{[\gamma]}{\beta}\plus}{e}\moins}{\tmu[\alpha].\command c}\plus\\
  &\simeq_{RE} \perfectcut{\mu\gamma\moins.\perfectcut{[\gamma]}{\underline{\tmu[\alpha]}.\command c}\plus}{e}\moins & (R\mu\plus)\\ 
  &\simeq_{RE} \perfectcut{\mu\gamma\moins.\command c[\gamma/\alpha]}{e}\moins &(R\negN)\\ 
  &\simeq_{RE} \perfectcut{\mu\alpha\moins.\command c}{e}\moins
\end{align*}
The case of $e=S$ is straightforward.
The other case is similar.
\end{proof}
\end{arxiv}

\paragraphcaps{Soundness of the calculus}

The results for the linear call-by-push-value $L$-calculus extend to
the linear classical $L$-calculus. \ifbool{arxiv}{The reader will
find detailed proofs in the Appendix~\ref{cohsoundcomp}.}{}

\begin{theorem}[Subject reduction]
  If $\command c \to_{RE} \command c'$ and $\command c :
  (\Gamma \vdash \Delta)$, then $\command c' :
  (\Gamma \vdash \Delta)$.
\end{theorem}

\begin{theorem}[Soundness of the linear classical $L$-calculus]\label{thm/soundness}
\ifbool{arxiv}{
Given some assigment $\rho$ of atoms to objects respecting the
polarity, the interpretation of typed terms of the linear classical
$L$-calculus in any dialogue duploid, sending sequents
$A_1,\dots,A_n\vdash B_1,\dots,B_m$ to hom-sets
$\Dduploid(A^\rho_1\otimes\cdots\otimes
A^\rho_n,B^\rho_1\parr\cdots\parr B^\rho_m)$, is invariant modulo
typed reductions and expansions.
}{
The interpretation of typed terms of the linear classical $L$-calculus
in any dialogue duploid is invariant modulo typed reductions and expansions.
}
\end{theorem}

\begin{popl}
\paragraphcaps{Variant: the one-sided $L$-calculus}
It is possible to present a simplified linear classical $L$-calculus,
with all formulae on the right-hand side of sequents, as in the
original presentations of linear logic and polarised classical logic
$\LC$~\citep{Gir87,girardnew}. (This was the approach originally
retained in \citet{munchmonolateral}.) In the extended version of this
paper~\citep{MMMM2025}, we explain the relationship between the two-sided and the
one-sided linear classical $L$-calculus in light of the coherence
theorem relating dialogue chiralities and dialogue
categories~\citep{melliesdialogue}.
\end{popl}

\sectioncaps{The syntactic dialogue duploid}\label{section/syntactic-duploid}
We now construct a dialogue duploid whose objects are the types of the
\ifbool{arxiv}{(two-sided)}{} linear classical $L$-calculus and whose morphisms $\command c : A\to
B$ between two types $A$ and $B$ are the commands $\command c : (a :
A \vdash \beta : B)$ quotiented by the rewriting
relation~$\simeq_{RE}$.
The composite of two maps
$$\mathsf{c} : (a : A \vdash \beta : B_\varepsilon)
\quad\quad\quad
\command c' : (b : B \vdash \gamma : C)$$
with respective typing derivations $\pi_1$ and $\pi_2$, 
is defined as the command of the $L$-calculus:
$$\perfectcut{\mu \beta\eps.\command c}{\tmu b\eps.\command c'}\eps$$
with typing derivation:
  \[
    \begin{prooftree}
      \hypo{\pi_2}
      \infer1{\command c' : (b : B \vdash \gamma : C)}
      \infer1[$(\tmu\vdash)$]{|\ \tmu b\eps.\command c' : B \vdash \gamma : C}
      \hypo{\pi_1}
      \infer1{\command c : (a : A \vdash \beta : B)}
      \infer1[$(\vdash\mu)$]{a : A \vdash \mu \beta\eps.\command c : B\ |}
      \infer2[$(\mathsf{cut})$]{\perfectcut{\mu \beta\eps.\command c}{\tmu b\eps.\command c'}\eps : (a : A \vdash \gamma : C)}
    \end{prooftree}
  \]

\begin{theorem}\label{thm/classical-syntactic-duploid}
  The construction just described
  defines a dialogue duploid called the syntactic dialogue duploid.
\end{theorem}

In order to establish the theorem, we give the following characterizations of thunkable maps
and of central maps in the non-associative category of commands.
Linear maps are characterized symmetrically.
\begin{lemma}[Adapted from \citep{CFM2015}]\label{lemma/syntacticallythunkable}
  Let $a : A \vdash t : B\ |$ be an expression. The two following properties are equivalent :
\begin{enumerate}
\item For all commands $\command c : (b : B_\veps \vdash \gamma : C)$,
$\perfectcut{t}{\tmu b\eps.\command c}^{\veps} \simeq_{RE} \command c[t/b]$;
\item For all commands $\command c : (b : B_\veps \vdash \gamma : C_{\veps'})$ and contexts $|\ e : C_{\veps'}\vdash \delta : D$,
\[
  \perfectcut{t}{\tmu b^{\veps}.\perfectcut{\mu \gamma^{\smash{\veps'}}.c}{e}^{\smash{\veps'}}}^{\veps} \simeq_{RE} \perfectcut{\mu \gamma^{\smash{\veps'}}.\perfectcut{t}{\tmu b^{\veps}.c}^{\veps}}{e}^{\veps'}\;.
\]
\end{enumerate}
\noindent
We say that an expression~$t$ is \define{syntactically thunkable}
  when it satisfies one of the above equivalent properties.
\end{lemma}
 
\begin{lemma}\label[lemma]{lemma/thunkable}
A command $\command c : (a : A \vdash \beta : B)$ is thunkable
if and only if $\mu\beta^{\veps_B}.\command c$ is syntactically thunkable.
\end{lemma}

\noindent
This characterization based on the intuition that thunkable expression behave like values
plays a fundamental role in the proof that the syntactic polarity $\veps$ of a type $A_{\veps}$ in the $L$-calculus
coincides with its semantic polarity as an object of the non-assocative category,
as it is defined in \cref{definition/polarity}.

\begin{definition}\label[definition]{definition/syntacticallycentral}
An expression $t$ is \define{syntactically central} when the equality up to reduction and expansion
is satisfied
\[
\perfectcut{t}{\tmu q_1.\perfectcut{u}{\tmu q_2.c}^{\veps_2}}^{\veps_1}\simeq_{RE} \perfectcut{u}{\tmu q_2.\perfectcut{t}{\tmu q_1.c}^{\veps_1}}^{\veps_2}
\]
for all commands $\command c$, expressions $u$ and binders $q_1$ and $q_2$ (i.e.\ 
either $a$, $a \smallotimes b$, $()$ or $[\alpha]$) of polarity $\veps_1$
and $\veps_2$ respectively.
\end{definition}

\begin{lemma}\label[lemma]{lemma/central}
A command $\command c : (a : A \vdash \beta : B)$ is central
if and only if the expression $\mu\beta^{\veps_B}.\command c$ is syntactically central.
\end{lemma}

This characterization of central commands in the $L$-calculus is the
basis of the proof that (positive) thunkable commands are central, and
thus, that the tensor $\tensor$ of the syntax defines a positive
monoidal structure. \ifbool{arxiv}{The interested reader will find the proofs of the two lemmas in the Appendix.}{}

\sectioncaps{The \FH{} theorem}\label{section/fh-theorem}

In this section, we formulate and establish the \FH{} theorem in the
language of dialogue duploids.
We have seen in \cref{lemma/thunkableimpliescentral}
that every thunkable map is central in a symmetric monoidal duploid,
and that the converse property is not true in general.
We establish now that the two notions coincide in a dialogue duploid.
\begin{theorem}[\FH{}]
  In a dialogue duploid, a morphism is central for $\otimes$ if and
  only if it is thunkable.
\end{theorem}
A direct proof by equational reasoning is given
in \ifbool{arxiv}{Appendix~\ref{section/detailedFH}}{the extended
version of this paper~\citep{MMMM2025}}.
It relies on the crucial observation that the composite $g \ccomp f$
for every pair of maps $f:A\to B$ and $g:B\to C$
can be expressed in every dialogue duploid by internal duality:
  \[
    g \ccomp f =  \varphi_C^{-1}(\varphi_B(f)
    \bullet (A \rtensortimes \dupldual g)) \hspace{.5em} : \hspace{.5em} A \to C
  \]
where $\varphi_D:\Dduploid(A,D)\xrightarrow{\iso}\Dduploid(A\otimes \dupldual D,\bot)$
is obtained from $\chi^{-1}_{A,\dupldual D,\bot}$, the duality $\dupldoubledual{D}\iso D$,
and unitors.
The proof derives in fact from our earlier
observation~(\S\ref{section/introFH}) that the two distinct
commutation properties characterising thunkability~(\ref{eq:FH-thunk})
and centrality~(\ref{eq:FH-central}) in sequent calculus,
respectively \cref{lemma/syntacticallythunkable} and
\cref{definition/syntacticallycentral}, coincide via duality.

One benefit of the linear classical $L$-calculus is that the same statement
can be also established by syntactic means in sequent calculus, thanks
to its equational theory and the
soundness theorem.  This allows us to make the explanation from \S\ref{section/introFH}
rigorous with a (brief!) proof of the \FH{} theorem that directly
relates by internal duality the characterisations of thunkability and
centrality.

\begin{theorem}[Syntactic \FH{} theorem]\label{theorem/fh-syntactic}
An expression of the linear classical $L$-calculus is syntactically central
for $\otimes$ if and only if it is syntactically thunkable.
\end{theorem}

\begin{proof}
As we have already seen, an expression which is syntactically
thunkable is also syntactically central. Now assume that $t$ is
syntactically central. To prove that $t$ is syntactically thunkable,
let $\command c$ be a command and $e$ be a context as
per \cref{lemma/syntacticallythunkable}. We need to establish
\begin{align*}
  \perfectcut{t}{\tmu b.\perfectcut{\mu\gamma.\command c}{e}{}}{} &
  \simeq_{RE} \perfectcut{\mu \gamma.\perfectcut{t}{\tmu b.\command c}{}}{e}{}
\end{align*}
The only difficult case is when $t$ is positive and $e$ is negative.
Using internal duality (\cref{lemma/negSwitchSide}) twice, one indeed
has:
\begin{align*}
  & \perfectcut{t}{\tmu b\plus.\perfectcut{\mu\gamma\moins.\command c}{e}{}\moins}{}\plus & &\\
  & \simeq_{RE} \perfectcut{t}{\tmu b\plus.\perfectcut{[e]}{\tmu[\gamma].\command c}{}\plus}{}\plus & & \mbox{by\ \cref{lemma/negSwitchSide}}\\
  & \simeq_{RE} \perfectcut{[e]}{\tmu[\gamma].\perfectcut{t}{\tmu b\plus.\command c}{}\plus}{}\plus & & \mbox{by centrality of } t\\
  & \simeq_{RE} \perfectcut{\mu \gamma\moins.\perfectcut{t}{\tmu b\plus.\command c}{}\plus}{e}{}\moins & & \mbox{by\ \cref{lemma/negSwitchSide}}\qedhere
\end{align*}
\end{proof}

\noindent
Now recall that the general situation of a duploid~$\Dduploid$ associated
to an adjunction~$L\dashv R$, one has that\vspace*{0.5ex}
\begin{center}
\fbox{
\begin{tabular}{c}
\emph{the monad $R\circ L$ is idempotent
if and only if}
\\
\emph{every morphism of the duploid~$\Dduploid$ is thunkable.}
\end{tabular}}
\end{center}\vspace*{0.5ex}
Also, it is not difficult to see that in the situation described in~\S\ref{section/sm-duploids}
of a symmetric monoidal duploid~$\Dduploid$ associated to 
an adjunction~$L\dashv R$ where~$\Acategory$ is symmetric monoidal
and where the monad~$T=R\circ L$ is strong, one has that\vspace*{-0.5ex}
\begin{center}
\fbox{
\begin{tabular}{c}
\emph{the monad $T$ is commutative if and only if}
\\
\emph{every morphism of the duploid~$\Dduploid$ is central.}
\end{tabular}}
\end{center}\vspace*{0.5ex}
In the case of a dialogue duploid~$\Dduploid$ associated to a dialogue
category, this proves as a corollary of
\cref{theorem/fh-syntactic} the following statement, attributed to Hasegawa
in \citet{melliestabareau}.
\begin{corollary}
The continuation monad of a dialogue category is commutative if and only if it is idempotent.
\end{corollary}

It is natural to wonder if we could not weaken the assump\-tions of
structure on duploids. Removing negation from
\cref{figure/syntaxDialogue} leads to consider a linearly distributive
structure on duploids:
\begin{definition}
  A \define{linearly distributive duploid} is a duploid equipped with
  a pair of positive and negative symmetric monoidal structures
  related by a family of mappings
$A \otimes (B \parr C) \to (A \otimes B) \parr C$ natural
  component-wise and that respects the usual coherence diagrams for a
  linearly distributive category~\cite{Cockett_1997,Mellies2017micrological}.
  (Note in particular that a linearly distributive duploid that
is associative is the same thing as a linearly distributive category.)
\end{definition}

\noindent A variant of the syntactic argument given in \citet[p.262]{munchthese}
then suggests the following refinement of the \FH{} theorem (in the
dual): in any linearly distributive duploid which is closed
(in the sense of an isomorphism $\Dduploid (X \otimes Y, Y'
\parr Z) \iso \Dduploid (X, (Y\multimap Y') \parr Z)$ natural in $X,Y',Z$ component-wise), a morphism is
central for $\parr$ if and only if it is linear.

\begin{arxiv}

\sectioncaps{Variant: the one-sided classical $L$-calculus}\label{section/onesided}

\begin{figure*}[!t]
\begin{framed}
\vspace*{-2ex}
\begin{small}
\centering

\hspace*{\fill}\subfloat[Formulae]{
$
  \stretcharray
  \begin{array}{lccl}
    \mbox{Types} & A, B, A_\veps & ::= & P\bmid N\\
    \mbox{Positives:} & P, Q, A_+ & ::= & X\bmid 1\bmid A\otimes B\\
    \mbox{Negatives:} & N, M, A_- & ::= & \orth{X}\bmid \bot\bmid A\parr B
  \end{array}
$
}\hspace*{\fill}

\hspace*{\fill}\subfloat[Grammar]{
$
 \stretcharray
 \begin{array}{lrcccccccccccccccc}
   \mbox{(Co)Values:} & V, W & ::= & x,y,\dots &\bmid& \mu x\moins.\command c &\bmid& () &\bmid& V \smallotimes W &\bmid& \mu().\command c &\bmid& \mu(x \smallparr y).\command c\\
   \mbox{(Co)Expressions:} & t,u & ::= & V &\bmid& \mu x\plus.\command c\\
   \mbox{Commands:} & \command c & ::= & \perfectcut{t}{V}
 \end{array}
$
}\hspace*{\fill}

\hspace*{\fill}\subfloat[Conversions (Reduction and expansion rules)]{
 $
 \stretcharray
 \begin{array}{lccc}
   (R\mu\plus) & \perfectcut{\mu x\plus.\command c}{V} & \triangleright_R & \command c[V/x]\\
   (R\mu\moins) & \perfectcut{V}{\mu x\moins.\command c} & \triangleright_R & \command c[V/x]\\
   (R1/\bot) & \perfectcut{()}{\mu().\command c} & \triangleright_R & \command c\\
   (R{\otimes}/{\parr}) & \perfectcut{V\smallotimes W}{\mu(x\smallparr y).\command c} & \triangleright_R & \command c[V/x,W/y]\\
 \end{array}
 \quad\quad\quad\quad
 \begin{array}{lccc}
   (E\mu\plus) & t & \triangleright_E & \mu x\plus.\perfectcut{t}{x} \\
   (E\mu\moins) & V & \triangleright_E & \mu x\moins.\perfectcut{x}{V}\\
   (E1/\bot) & V & \triangleright_E & \mu().\perfectcut{()}{V} \\
   (E{\otimes}/{\parr}) & V & \triangleright_E & \mu(x\smallparr y).\perfectcut{x\smallotimes y}{V} \\
 \end{array}
 $
}\hspace*{\fill}

\hspace*{\fill}\subfloat[Judgements]{
$
 \command c : (\vdash \Gamma) \quad\quad\quad \vdash t : A\ |\ \Gamma
$
}\hspace*{\fill}

\hspace*{\fill}\subfloat[Typing rules (Identity and structural groups)]{
\begin{varwidth}{\linewidth}
\centering
 $
   \begin{prooftree}
     \infer0[$(\mathbf{ax})$]{\vdash x : A\ |\ x:\orth{A}}
   \end{prooftree}
   \quad\quad
   \begin{prooftree}
     \hypo{\command c:(\vdash x : A_\veps, \Gamma)}
     \infer1[$(\boldsymbol{\mu}\eps)$]{\vdash \mu x\eps.\command c:A_\veps\ |\ \Gamma}
   \end{prooftree}
   \quad\quad
   \begin{prooftree}
     \hypo{\vdash t : P\ |\ \Gamma}
     \hypo{\vdash V : \orth{P}\ |\ \Gamma'}
     \infer2[$(\mathbf{cut})$]{\perfectcut{t}{V} : (\vdash \Gamma, \Gamma')}
   \end{prooftree}
 $
\bigbreak
$
 \forall\sigma\in\Sigma(\Gamma',\Gamma)\;: \quad\quad
 \begin{prooftree}
   \hypo{\vdash t : A\ |\ \Gamma}
   \infer1[$(\vdash\sigma)$]{\vdash t[\sigma] : A\ |\ \Gamma'}
 \end{prooftree}
 \quad\quad
 \begin{prooftree}
   \hypo{\command c:(\vdash \Gamma)}
   \infer1[$(\sigma)$]{\command{c}{[\sigma]}:(\vdash \Gamma)}
 \end{prooftree}
$
\end{varwidth}
}\hspace*{\fill}

\hspace*{\fill}\subfloat[Typing rules (Logic group)]{
\begin{varwidth}{\linewidth}
\centering
$
 \begin{prooftree}
   \hypo{\vphantom{\command c : (\vdash\Gamma)}}
   \infer1[$(1)$]{\vdash() : 1\ |}
 \end{prooftree}
 \quad\quad
 \begin{prooftree}
   \hypo{\command c : (\vdash\Gamma)}
   \infer1[$(\bot)$]{\vdash\mu().\command c : 1\ |\ \Gamma}
 \end{prooftree}
$
\bigbreak
$
 \begin{prooftree}
   \hypo{\vdash V : A\ |\ \Gamma}
   \hypo{\vdash W : B\ |\ \Gamma'}
   \infer2[$(\otimes)$]{\vdash V \smallotimes W:A\otimes B\ |\ \Gamma,\Gamma'}
 \end{prooftree}
 \quad\quad
 \begin{prooftree}
   \hypo{\command c : (\vdash x : A, y : B, \Gamma)}
   \infer1[$(\parr)$]{\vdash\mu(x \smallparr y).\command c : A\parr B\ |\ \Gamma}
 \end{prooftree}
$
\end{varwidth}
}\hspace*{\fill}
\end{small}
\end{framed}
\caption{Syntax of the one-sided classical $L$-calculus}
\label{figure/syntaxDialogueOneSided}
\end{figure*}

It is possible to present a simplified linear classical $L$-calculus
with all formulae in negative normal form and placed on the right-hand
side of sequents, as in the original presentations of linear logic and
the polarised classical logic $\LC$~\citep{Gir87,girardnew}. This is
the presentation retained in the third author's one-sided classical
$L$-calculus \citep{munchmonolateral}.

Formulae being in negative normal form means that negation is no
longer a connective except in front of atoms. Negation is now an
operation on formulae, written $\orth{\,\cdot\,\vphantom{A}}$, defined
by De Morgan laws:
\begin{align*}
\orth{A\otimes B} & \defeq\orth A\parr\orth B & \orth 1 & \defeq\bot&\orth{\orth X} & \defeq X\\
\orth{A\parr B} & \defeq\orth A\otimes\orth B & \orth{\bot} & \defeq 1
\end{align*}
In particular, negation is strictly involutive rather than only up to
type isomorphism:
\[
\orth{\orth A} = A
\]
Sequents are of the form $\vdash\Gamma$ and left-introduction rules
then coincide with the right-introduction rules of the dual
connective.

One goal of this paper has been to explain $\LC$'s involutive negation
by giving a categorical and syntactic account where duality is present
as an explicit connective. In this section we go further and describe
a one-sided calculus as a strictification of the two-sided calculus,
by relating this simplification of the syntax to the coherence theorem
between dialogue chiralities and dialogue categories presented
in~\citet{melliesdialogue} (via
thms.~\ref{theorem/adjunctions-monoidal-duploids}
and~\ref{thm/soundness} which provide an interpretation of the direct
syntax into the indirect models). The one-sided calculus internalises
the \FH{} theorem, in the sense that the strictification makes the
sequent calculus derivations characterising
thunkability~(\ref{eq:FH-thunk}) and centrality~(\ref{eq:FH-central})
coincide formally.

\paragraphcaps{The one-sided linear classical $L$-calculus}

We present the one-sided linear classical $L$-calculus
in \cref{figure/syntaxDialogueOneSided}.
Informally, one can think of a one-sided sequent
\[
\vdash \Gamma
\]
as any of its two-sided unfoldings, chosen as deemed convenient:
\begin{equation}\label[equation]{equation/unfolding}
\Gamma_1 \vdash \Gamma_2 \quad\textrm{where}\quad \orth{\Gamma_1},\Gamma_2 \textrm{ is a reordering of } \Gamma
\end{equation}
as obtained by free applications of exchange and the following informal negation rules:
\begin{equation*}
 \begin{prooftree}
   \hypo{\Gamma\vdash A,\Delta}
   \infer1{\Gamma,\orth{A}\vdash\Delta}
 \end{prooftree}
 \quad\quad\quad\quad
 \begin{prooftree}
   \hypo{\Gamma, A\vdash\Delta}
   \infer1{\Gamma\vdash\orth{A},\Delta}
 \end{prooftree}
\end{equation*}
Similarly, the one-sided calculus can be thought of as the two-sided
calculus where the term formers of negation are omitted:
\begin{equation}\label[equation]{equation/rules-negation}
 \begin{prooftree}
   \hypo{\Gamma\vdash t:A\ |\ \Delta}
   \infer1{\Gamma\ |\ t:\orth{A}\vdash\Delta}
 \end{prooftree}
 \quad\quad\quad\quad
 \begin{prooftree}
   \hypo{\Gamma\ |\ t:A\vdash\Delta}
   \infer1{\Gamma\vdash t:\orth{A}\ |\ \Delta}
 \end{prooftree}
 \quad\quad\quad\quad
 \begin{prooftree}
   \hypo{c:(\Gamma\vdash x:A,\Delta)}
   \infer1{c:(\Gamma,x:\orth{A}\vdash\Delta)}
 \end{prooftree}
 \quad\quad\quad\quad
 \begin{prooftree}
   \hypo{c:(\Gamma, x:A\vdash\Delta)}
   \infer1{c:(\Gamma\vdash x:\orth{A},\Delta)}
 \end{prooftree}
\end{equation}
As a matter of fact, if we introduce the following notations:
\[
\perfectcut{V}{t}\moins \defeq \perfectcut{t}{V}\plus \defeq \perfectcut{t}{V}
\]
then we are free to use whichever notation is more convenient
regarding the computational interpretation of sequents.

For instance, as is well known, the call-by-name $\lambda\mu$-calculus
can be recovered in the negative fragment of polarised classical
logic. The Krivine abstract machine is recovered in the one-sided
classical $L$-calculus with the previous notation and the following
ones:
\begin{align*}
& \lambda x.V \defeq \mu(x\smallparr y).\perfectcut{V}{y}\moins
&& V\,W \defeq \mu x\moins.\perfectcut{V}{W\smallotimes x}\moins
\end{align*}
Indeed, one has with these notations:
\begin{align*}
& \perfectcut{V\,W}{S} \triangleright_R \perfectcut{V}{W\smallotimes S}
&& \perfectcut{\lambda x.V}{W\smallotimes S}\moins \triangleright_R \perfectcut{V[W/x]}{S}\moins
\end{align*}
which correspond to the rules of the Krivine abstract machine, where
$V,W$ denote negative expressions (which are indeed values) and $S$ a
stack: a positive value of the form $V_1\smallotimes\cdots\smallotimes
V_n\smallotimes x$ where $x$ is a (co)variable denoting the end of the
stack.

It is a nice exercise to reproduce using similar notations the sequent
calculus derivations characterising thunkability~(\ref{eq:FH-thunk})
and centrality~(\ref{eq:FH-central}), as this shows that they indeed
coincide in the one-sided calculus.

\paragraphcaps{Correspondence with the two-sided $L$-calculus}

Formally, terms from the two-sided $L$-calculus correspond to terms
from the one-sided $L$-calculus via a surjective
mapping~$\fold{\;\cdot\;}$ from the latter to the former, which we
call \emph{folding}, obtained by erasing term formers of negation:
{\allowdisplaybreaks
\begin{align*}
\fold{\perfectcut tS^{+}} & =\perfectcut{\fold{t}}{\fold{S}} &&& \fold{\perfectcut Ve^{-}} & =\perfectcut{\fold{e}}{\fold{V}}\\
\fold{\mu\alpha^{+}.c} & =\mu\fold{\alpha}^{+}.\fold c &&& \fold{\tmu x^{-}.c} & =\mu x^{+}.\fold c\\
\fold{\mu\alpha^{-}.c} & =\mu\fold{\alpha}^{-}.\fold c &&& \fold{\tmu x^{+}.c} & =\mu x^{-}.\fold c\\
\fold{()} & =() &&& \fold{[]} & =()\\
\fold{V\smallotimes W} & =\fold V\smallotimes\fold W &&& \fold{S\smallparr S'} & =\fold S\smallotimes\fold{S'}\\
\fold{\mu[].c} & =\mu().\fold c &&& \fold{\tmu().c} & =\mu().\fold c\\
\fold{\mu(\alpha\smallparr\beta).c} & =\mu(\fold{\alpha}\smallparr\fold{\beta}).\fold c &&& \fold{\tmu(x\smallotimes y).c} & =\mu(x\smallparr y).\fold c\\
\fold{[S]} & =\fold S &&& \fold{[V]} & =\fold V\\
\fold{\mu[x].c} & =\mu x^{-}.\fold c &&& \fold{\tmu[\alpha].c} & =\mu \fold{\alpha}^{-}.\fold c
\end{align*}}Folding moreover sends the reduction and expansion rules into
corresponding reduction and expansion rules, except for the rules
$R\dupldual\pm$ and $E\dupldual\pm$ which are sent into the equality
of terms. Noticing that those reductions and expansions related to
negation are linear in their metavariables and therefore do not affect
computation, folding preserves many computational properties of the
term. (For instance, it is easy to see that an untyped command
$\command c$ is strongly normalising if and only if $\fold{\command
c}$ is strongly normalising.)

This relates to the biequivalence between (2-categories of) dialogue
categories and chiralities~\cite{melliesdialogue} in the following
way. It so happens that continuation semantics are themselves
``one-sided'', in the sense that they do not distinguish between the
two-sided and the one-sided calculi. This is first reflected in the
fact that a dialogue category
\begin{equation*}
\begin{tikzcd}[column sep = 1em]
\Ccategory \arrow[rr,"{\neg}", bend left] & \bot & \Ccategory^{\mathrlap{\op}}\arrow[ll,"{\neg}",bend left]
\end{tikzcd}
\end{equation*}
seen as a dialogue chirality
\vspace{-1em}
\begin{equation*}
\begin{tikzcd}[column sep = 1em]
  (\Acategory,\tensorialand,\tensorialtrue) \arrow[rr,"\chirdual{(-)}", bend left] & \simeq &
  (\Bcategory,\tensorialor,\tensorialfalse)^{\mathrlap{\op}}\arrow[ll,"\chirdual{(-)}",bend left]
\end{tikzcd}
\end{equation*}
that is, with
\[
  (\Acategory,\tensorialand,\tensorialtrue)=(\Ccategory,\otimes,1)\quad\quad
  (\Bcategory,\tensorialor,\tensorialfalse)=(\Ccategory^\op,\otimes,1)\;,
\]
where dualities $\chirdual{(-)}$ are given by the identity functors on
$\Ccategory$ and $\Ccategory^\op$, and where negation $\neg$ is seen
as a pair of (covariant) functors $L:\Acategory\rightarrow\Bcategory$
and $R:\Bcategory\rightarrow\Acategory$, is indeed a dialogue
chirality that (trivially) statisfies strict identities
$\chirdoubledual{A}=A$, $\chirdual{(A\tensorialand
B)}=\chirdual{A}\tensorialor \chirdual{B}$, etc. (This chirality
$(\Ccategory,\Ccategory^\op)$ is written $\mathcal{F}\Ccategory$ in
\citet{melliesdialogue}.) This lets us refine our
interpretation of the two-sided calculus into dialogue chiralities
(via dialogue duploids) into an interpretation of the one-sided
calculus into dialogue categories.

This furthermore leads to the observation that the interpretation of
typed terms of the two-sided $L$-calculus into any dialogue category
$\Ccategory$, seen as dialogue chirality $(\Ccategory,\Ccategory^\op)$,
factors into folding $\fold{\;\cdot\;}$ followed by the
interpretation of the one-sided $L$-calculus into dialogue categories;
in other words the following diagram always commutes:
$$
\begin{tikzcd}[column sep = 1.5em, row sep = 1em]
\mbox{2-sided $L$}\arrow[dd,""{swap}]\arrow[rr,"{\fold{\;\cdot\;}}"]
&&
\mbox{1-sided $L$}\arrow[dd,""]
\\
\\
(\Ccategory,\Ccategory^\op) && \Ccategory\arrow[ll,"{\mathcal{F}}"]
\end{tikzcd}
$$
Traditional continuation semantics, by being intrinsically
``one-sided'' in this way---with every object $A$ being for instance
identified with its negation $\dupldual{A}$ in the opposite
category---obscure the algebraic description
of the involutive negation of $\LC$,
which might explain why this involutive negation has at times been
misunderstood.

At this point, we find convenient to
offer an alternative perspective based on seeing the correspondence
between dialogue categories and chiralities as a coherence
result~\cite{melliesdialogue}. On the semantic side, we observe that
dialogue chiralities (and dialogue duploids), better than dialogue
categories, provide an accurate and explicit account the involutive
negation of $\LC$ in a polarised setting, and thus,\vspace*{1ex}
\begin{center}
\fbox{
\begin{tabular}{c}
\emph{on the semantic side, dialogue chiralities and dialogue duploids}
\\
\emph{offer an explicit and more accurate view on continuations.}
\end{tabular}}\vspace*{1ex}
\end{center}
Things are reversed on the syntactic side: the one-sided syntax is
worth considering, as it brings a considerable syntactic
simplification---if we keep in mind that one-sided sequents stand for
arbitrary unfoldings \eqref{equation/unfolding} and free applications
of the negation rules \eqref{equation/rules-negation}. Thus,\vspace*{1ex}
\begin{center}
\fbox{
\begin{tabular}{c}
\emph{on the syntactic side, the one-sided classical $L$-calculus}
\\
\emph{offers an accurate and more implicit view on continuations.}
\end{tabular}}\vspace*{1ex}
\end{center}

\end{arxiv}

\sectioncaps{Classical notions of computations: turning around \joyallemma}\label{section/classical-history}
Andr{\'e} Joyal made the important observation (recalled below, see \cref{thm/joyal})
that it is not possible to develop a proof-theoretic account of classical logic using
the language of usual (associative) cartesian categories.
A simple argument shows that every return object~$\bot$
in a symmetric monoidal category~$\Ccategory$ induces a family of canonical maps
\begin{equation}\label[equation]{equation/A-implies-not-not-A}
{\eta_A} \;\; : \;\; A\;\longrightarrow{}\; \lnot\lnot A
\end{equation}
indexed by the objects~$A$ of the category~$\Ccategory$, 
which reflects the logical principle that every formula~$A$ implies its double negation~$\lnot\lnot A$.
This family of maps is the unit of the self-adjunction of negation with itself,
mentioned in~\eqref{equation/adjunction-neg-neg}.
A return object~$\bot$ is called \emph{dualizing}
when the canonical map~\eqref{equation/A-implies-not-not-A}
is an isomorphism for every object~$A$.
A natural direction to resolve the quest for a proof-theoretic interpretation of classical logic
would be to look for a cartesian category~$(\Ccategory,\times,1)$ equipped 
with a dualizing object~$\bot$.
Unfortunately, Joyal observed that the search for such a simple solution cannot succeed:
\begin{theorem}[\joyallemma]\label{thm/joyal}
Any cartesian category $(\Ccategory,\times,1)$ with a dualizing object~$\bot$ is a preorder,
and thus defines a boolean algebra (up to equivalence).
\end{theorem}
\ifbool{arxiv}{
\begin{proof}
See Appendix~\ref{sec:joyallemma}.
\end{proof}
}{}
\noindent
For a long time, this observation has been widely accepted 
as evidence that classical logic cannot be interpreted 
in a denotational and proof-relevant way.
The situation changed in the early 1990s
when \citet{Griffin90aformulae-as-types} and
\citet{Murthy91EvaluationSemantics} 
observed a fundamental and unexpected relationship
between proof systems for classical logic, 
and programs written with the control operator $\mathcal{C}$, 
a variant of Scheme's \emph{call-cc}.
Since then,
a large number of investigations have been made
to define a clean denotational and proof-theoretic
interpretation of classical logic.
Interestingly, each of the two main directions taken
can be seen as providing a specific way
to relax one of the hypothesis of Joyal's obstruction theorem:
\vspace{.3em}

\noindent
1)\, \emph{Classical linear logic~\cite{Gir87}:}
the idea is to relax the cartesianity condition
and to work with $\ast$-autonomous categories,
defined as symmetric monoidal categories~$(\Ccategory,\tensor,1)$
equipped with a dualizing object~$\bot$,
possibly supplemented with an exponential modality~$A\mapsto{!A}$
to deal with non-linearity,

\vspace{.3em}

\noindent
2)\, \emph{Continuation models:} the idea is to relax the dualizing
condition, and work with categories
where~\eqref{equation/A-implies-not-not-A} has a section or a retraction, obtained from continuation-passing style (CPS) constructions over
cartesian categories~$(\Ccategory,\times,1)$
equipped with a return object~$\bot$.

\vspace{.3em}

\noindent
In these two directions, influential and most notable works have been the
Lafont-Reus-Streicher translation~\cite{LRS93} as well as the later
works by \citet{Hof02ComplLambdaMu} and by \citet{Sel01Control}.
Another important and early work has been the introduction of two dual
sequent calculi $\LKT$ and $\LKQ$ for classical logic, and their
translation in linear logic
by \citet{danos93structure,DJS95LKQLKT,danos95new}, which turned out
to rephrase respectively the CBV and CBN CPS
semantics~\cite{Ogata2000}. Interestingly, all these models ``break
the symmetry'' of classical logic by giving precedence at some stage
to the CBV or CBN side.
The symmetry between the two sides remains however, as a categorical
duality observed
by \citet{StreicherReus98ContinuationAbstractMachines} and made
manifest by \citet{Sel01Control} and \citet{CH00Duality} (predated by,
and in the spirit of, Filinski's \emph{symmetric
$\lambda$-calculus}~\citep{filinski89declarative}).

Curien and Herbelin's original $L$-calculus explored in particular a
syntactic symmetry between the CBN and CBV calculi which reflects the
categorical duality. It was discovered through the reunion of two
research lines---the one we just mentioned around the connection
between constructive classical logic and
CPS~\cite{DJS95LKQLKT,danos95new,Ogata2000}, and the one that
investigated well-behaved $\lambda$-calculi for classical logic
\cite{Parigot92} and sequent calculus \cite{Herbelin1994}.
Curien and Herbelin's $L$-calculus was also inspired by Barbanera and
Bernardi's \emph{symmetric
$\lambda$-calculus}~\cite{BarbaneraBerardi1996} whose reduction is
non-deterministic.
In a similar line of research, an order-enriched categorical interpretation 
of classical logic~\cite{fuhrmann2006order} based on a non-deterministic 
calculus~\cite{urban2000classical} was described in~\cite{bellin2006categorical}.
The connections between this line of research based on a
non-deterministic interpretation of cut elimination, and the present
paper based on the polarised and deterministic linear classical
$L$-calculus, remain to be clarified.

\vspace{.3em}

\noindent
3)\, \emph{Preserving the symmetries of classical logic at the expense of associativity:}
at about the same time as \citet{Griffin90aformulae-as-types} and
\citet{Murthy91EvaluationSemantics}, in the early 1990s, an elegant and third
direction, inspired by linear logic, was explored by \citet{girardnew}
with the classical logic~$\LC$.
The goal was to preserve the symmetries of logic---in particular, an
involutive negation and various De Morgan identities present as type
isomorphisms---by giving a formal status to the notion of polarity of a formula.
Girard's work on $\LC$ inspired many later works
(\citet{Murthy92LC,Quatrini96polarisationdes,Lau02PhD,Zeil2008Unity,Liang2009a,melliestabareau}
among others) including in fact some of the works we already mentioned
\citep{LRS93,DJS95LKQLKT,danos95new}.

The solution, which involves giving up the associativity of
composition precisely in the way which we have described, had not seen
much exploration from the angle of categorical proof theory.
In fact, the question of categorical proof for classical logic theory
was essentially mentioned as open in \citet{Hyland2002}.
This is the direction we took in the present paper.
By applying the duploid construction to dialogue categories with
linear classical logic in mind, we showed that the approach from
Girard's $\LC$ makes sense from a semantic point of view with dialogue
duploids, related to the syntactic point of view of the $L$-calculus.

\sectioncaps{Conclusion and future work}\label{section/conclusion}
We have introduced the syntax and semantics of linear $L$-calculus,
and developed theories of symmetric monoidal duploids with closure and
with involutive negation.
We see the framework
as a solid foundation for the study of non-associative and effectful
logical systems and term calculi for linearity, effects, and classical
logic, integrating the lessons of linear logic, continuation models
and functorial game semantics. The perspective of an encompassing
theory of non-associative direct models for effectful programs and
proofs is promising, but much remains to be done in this respect.

One interesting application area for such a classical calculus is the
study of subtle design issues in continuations for programming
languages.
For instance, \citet{Cong_2019} characterise a restriction to the usage of
continuations suitable for compilation, which is not as strong as
linearity: crucially, its still permits to copy and discard
continuations. This is beyond the scope of dialogue duploids and could
involve the notion of linearly distributive duploid just introduced.

\begin{acks}
This work has received funding from the European Research Council
under the European Union’s Horizon 2020 research and innovation
programme (Synergy Project Malinca, ERC Grant Agreement No 670624).
\end{acks}

\printbibliography

\ifbool{arxiv}{}{\end{document}}

\appendix

\allowdisplaybreaks[4]

\sectioncaps{A proof of \joyallemma}\label{sec:joyallemma}

\global\long\def\pair#1#2{\langle#1,#2\rangle}
\begin{proof}[Proof of \joyallemma] Let $(\Ccategory,\times,1)$ a cartesian
category with a dualizing object~$\bot$. One has natural bijections
$\Ccategory(A,\bot)\iso \Ccategory(A\times
1,\bot) \iso \Ccategory(A,\bot^1)$ hence $\bot\iso\bot^{1}$. Observe
then that the object $\bot^{\bot}\iso\bot^{\bot^{1}}\iso 1$ is
the terminal object. Consequently, the set
$\Ccategory(\bot\times\bot,\bot)$, in bijection with
$\Ccategory(\bot,\bot^{\bot})$, is a singleton; in particular one has
$\pi_{1}=\pi_{2}\in\Ccategory(\bot\times\bot,\bot)$. Now consider the
pairs $\pair fg\in\Ccategory(A,\bot\times\bot)$ for
$f,g\in\Ccategory(A,\bot)$. By the identity of projections, one has
$f=g$ for any such pair of morphisms, in other words any
$\Ccategory(A,\bot)$ has at most one element. Thus, any hom-set
$\Ccategory(B,C)$ has at most one element as well, as witnessed by the
bijections:
$\Ccategory(B,C)\iso\Ccategory(B,\bot^{\bot^{C}})\iso\Ccategory(B\times\bot^{C},\bot)$.
\end{proof}

\sectioncaps{Chasing and rewriting triangulated commutative diagrams}\label{section/chasing}
As we mentioned, very basic principles of usual associative categories 
are not necessarily true anymore in non-associative categories.
In particular, 
the fact that the triangulated diagram on the left commutes
in the sense that $h_Y\ccomp f = f' \ccomp h_X$ and $h_Z\ccomp g = g' \ccomp h_Y$
\begin{center}
\begin{tabular}{ccccc}
\begin{tikzcd}[row sep = 1.5em, column sep = 2em]
X\arrow[rr,"f"] \arrow[dd,"h_X"{swap}]
\arrow[rrdd,"(1)"{description}]
&&
Y \arrow[dd,"h_Y"]\arrow[rr,"g"]
\arrow[rrdd,"(2)"{description}]
&& 
Z \arrow[dd,"h_Z"]
\\
\\
X'\arrow[rr,"f'"{swap}] 
&&
Y'\arrow[rr,"g'"{swap}] && Z'
\end{tikzcd}
& \quad\quad &
\begin{tikzcd}[row sep = 1.5em, column sep = 4em]
X\arrow[rr,"{g\circ f}"] \arrow[dd,"h_X"{swap}]
\arrow[rrdd,"(3)"{description}]
&&
Z \arrow[dd,"h_Z"]
\\
\\
X'\arrow[rr,"{g'\circ f'}"{swap}] && Z'
\end{tikzcd}
\end{tabular}
\end{center}
\emph{does not} imply 
that the diagram~{(3)} commutes in the sense that 
$h_Z\ccomp (g\ccomp f) = (g'\ccomp f') \ccomp h_X$.
The reason is that one cannot apply a series of flips
to transform the triangular decomposition~$(a)$ 
of the hexagon on the left to the decomposition~$(b)$
of the same hexagon on the right:
\begin{center}
\begin{small}
\begin{tabular}{ccccc}
$(a) \, =$ &
\begin{tikzcd}[column sep = 1.5em, row sep = .7em]
&&
Y \arrow[rrdd,"g"]\arrow[dddddddd,"h_Y"]\arrow[rrdddddd,,"(2)"{description}]
&&
\\
\\
X\arrow[rruu,"f"]\arrow[dddd,"h_X"{swap}] \arrow[rrdddddd,"(1)"{description}] && && Z \arrow[dddd,"h_Z"]
\\
\\
\\
\\
X'\arrow[rrdd,"f'"{swap}] && && Z'
\\
\\
&& Y' \arrow[rruu,"g'"{swap}]
\end{tikzcd}
& \hspace{3em} &
$(b) \, =$ & 
\begin{tikzcd}[column sep = 1.5em, row sep = .7em]
&&
Y \arrow[rrdd,"g"]
&&
\\
\\
X\arrow[rruu,"f"]\arrow[dddd,"h_X"{swap}] \arrow[rrrr,"g\ccomp f"] \arrow[rrrrdddd,"(3)"{description}] 
&& && Z \arrow[dddd,"h_Z"]
\\
\\
\\
\\
X'\arrow[rrdd,"f'"{swap}] \arrow[rrrr,"g'\ccomp f'"{swap}] && && Z'
\\
\\
&& Y' \arrow[rruu,"g'"{swap}]
\end{tikzcd}
\end{tabular}
\end{small}
\end{center}
However, the sequence of computations below establishes that diagram~{(3)} commutes
when the three paths of length three of the diagram associate:
\begin{align}
h_Z\circ (g\circ f) & = (h_Z\circ g) \circ f & \hspace{2em} & \mbox{the path $(f,g,h_z)$ associates}
\\
& = (g'\circ h_Y) \circ f & & \mbox{equation (2)}
\\
& = g'\circ (h_Y \circ f) & & \mbox{the path $(f,h_Y,g')$ associates}
\\
& = g'\circ (f'\circ h_X) & & \mbox{equation (1)}
\\
& = (g'\circ f')\circ h_X & & \mbox{the path $(h_X,f',g')$ associates}
\end{align}
The equation can be also established diagrammatically
by flipping the map~$h_{Y}$ into the map~$(3)$
using the fact that the path $(f,h_Y,g')$ associates, in order to obtain
the commutative triangulation, from which it is easy
to obtain the commutative triangulation~$(b)$ by flipping $(1)$ and $(2)$
into $g\ccomp f$ and $g'\ccomp f'$ using the fact that 
the two paths $(f,g,h_Z)$ and $(h_X,f',g')$ associate.
\begin{center}
\begin{small}
\begin{tabular}{cc}
$(c) \, =$
& \begin{tikzcd}[column sep = 1.5em, row sep = .7em]
&&
Y \arrow[rrdd,"g"]\arrow[rrdddddd,,"(2)"{description}]
&&
\\
\\
X\arrow[rruu,"f"]\arrow[dddd,"h_X"{swap}] \arrow[rrrrdddd,"(3)"{description}]
\arrow[rrdddddd,"(1)"{description}] && && Z \arrow[dddd,"h_Z"]
\\
\\
\\
\\
X'\arrow[rrdd,"f'"{swap}] && && Z'
\\
\\
&& Y' \arrow[rruu,"g'"{swap}]
\end{tikzcd}
\end{tabular}
\end{small}
\end{center}

\sectioncaps{Non functoriality of the shift operator: an illustration}\label{section/non-functoriality-illustration}
We consider the duploid associated to the finite distribution monad $T$
already discussed in \cref{paragraph/proba-example-i} and explain
why the positive shift~$X\mapsto\nDownarrow X$ is not functorial in that specific example.
The positive shift~$\nDownarrow X$ associated to an object~$X$ of the duploid
is defined by case analysis on the polarity of~$X$:

\begin{itemize}
  \item If $X = (0,A)$, then the shifted object $\nDownarrow (0,A) := (0,A)$ is equal to the original object. 
  \item If $X = (1,B)$, then the shifted object $\nDownarrow (1,B) := (0,TB)$ is the set~$TB$
  of finite probability distributions of the set~$B$, seen as a positive object.
\end{itemize}
Since the definition of $\nDownarrow X$ depends on the polarity as we have just seen,
it is also the case for the definition of the map $\omega_X:X\to\nDownarrow X$.
\begin{itemize}
\item $\omega_{(0,A)}:(0,A)\to (0,A)$ is the identity, 
\item $\omega_{(1,B)}:(1,B)\to (0,TB)$ is the map $d \mapsto \dirac d$ 
which associates to every distribution $d\in TB$ the Dirac distribution of distributions $\dirac d \in TTB$.
\end{itemize}
Now, suppose given a map $f:X\to Y$ in the duploid.
In the same way as for $\omega_X:X\to \nDownarrow X$,
the construction of the linear map $f^\dagger:\nDownarrow X\to Y$ depends on the polarity of $X$.
\begin{itemize}
  \item If $X = (0,A)$ then the linear map $f^{\dagger}:(0,A)\to Y$ is defined as $f^{\dagger}=f$,
\item If $X = (1,B)$ then the map $f:(1,B)\to Y$ is the same thing as a stochastic map $f:TB\to Y$ 
  where we identity $Y$ with its underlying set.
The linear map $f^{\dagger}:(0,TB)\to Y$ may be thus defined as the stochastic map underlying~$f$.
\end{itemize}
\noindent
Suppose given a map $f:X\to Y$ in the duploid. 
The definition of the map $\nDownarrow f$ depends on the polarity of $X$ and $Y$.
We give the precise description for $X$ and $Y$ both positive and $X$ and $Y$ both negative, 
as the other cases can be easily deduced.
\begin{itemize}
  \item When $X = (0,A)$ and $Y = (0,A')$, the map $\nDownarrow f:(0,A)\to(0,A')$ is equal to $f$.
  \item When $X = (1,B)$ and $Y = (1,B')$, the map $f$ is a stochastic map $f:TB\to B'$
 of the general form 
\[
f = d \mapsto \psum{i}{p_i(d)}{f_i(d)}.
\]
 The map $\nDownarrow f:(0,TB)\to(0,TB')$ is the stochastic map $\nDownarrow f :TB\to TB'$
 which transports every distribution to the Dirac distribution of its image by~$f$:
    \[
      \nDownarrow f = d \mapsto \dirac[\big]{\psum{i}{p_i(d)}{f_i(d)}}
    \]
\end{itemize}
So, one should think of the positive shift as a \emph{wrap} operation obtained 
by turning every output distribution~$d'\in TB$ into the Dirac distribution $\dirac{d'}\in TTB$.
In particular, if we define the two maps~$f$ and~$g$ as the following maps
\[
  \begin{array}{c} f : (0,A) \to (0,B) \\ f := a \mapsto \psum{i}{p_i(a)}{f_i(a)}\end{array} \quad\quad \mbox{and} \quad\quad \begin{array}{c} g : (0,B) \to (1,B) \\ g : b \mapsto \dirac{b}\end{array},
\]
then, we have
\[
  \nDownarrow (g \ccomp f) := \dirac[\big]{\psum{i}{p_i(a)}{f_i(a)}} \quad\quad \mbox{and} \quad\quad \nDownarrow g \pcomp \nDownarrow f := \psum[\big]{i}{p_i(x)}{\dirac{f_i(x)}}
\]
This implies that $\nDownarrow (g \ccomp f) = \nDownarrow g \pcomp \nDownarrow f$
precisely when~$f$ transports every input~$x\in A$ to a Dirac distribution.
Note that we have established in the introduction (\S\ref{subsection/intro-thunkable-linear})
that this property characterizes the thunkable maps
of the duploid.

\sectioncaps{Graph morphisms and adjunctions between them}
\begin{definition}
Let $\Eduploid$ a non-associative category and $F,G:\Dduploid\graphto\Eduploid$
two graph morphisms into $\Eduploid$. A \define{natural transformation}
$\tau$ from $F$ to $G$ is given by, for every object $A\in\Dduploid$
a morphism $\tau_{A}:FA\rightarrow GA$, such that for all $f\in\Dduploid(A,B)$
one has $\tau_{B}\ccomp Ff=Gf\ccomp\tau A$.
\end{definition}
In the following, $\Eduploid$ will often be a category such as $\Set$.
\begin{definition}
For $\Dduploid$ a non-associative category, its \define{graph hom-morphism}
${\Dduploid}:\Dduploid^{\op}\boxtimes\Dduploid\graphto\Set$
is the graph morphism defined component-wise with:
\[
{\Dduploid}(f,A)(g)=g\ccomp f\qquad{\Dduploid}(A,f)(g)=f\ccomp g
\]
\end{definition}
\begin{definition}
Let $F:\Dduploid\graphto\Eduploid$ and $G:\Eduploid\graphto\Dduploid$
two graph morphisms between non-associative categories.
An \define{adjunction between graph morphisms} (notation
$\graphadjoint F \Dduploid \Eduploid G$) is given by a natural isomorphism of graph
morphisms:
\[
\varphi:{\Eduploid}(F-,=)\xrightarrow{\iso}{\Dduploid}(-,G{=}):\Dduploid^{\op}\boxtimes\Eduploid\graphto\Set
\]
(i.e. natural component-wise). $F$ is said to be left adjoint to
$G$ and $G$ right adjoint to $F$.
\end{definition}
The following proposition provides an alternative characterisation
of an adjunction between graph morphisms:
\begin{proposition}
A adjunction between graph morphisms $\graphadjoint F \Dduploid \Eduploid G$
is the same thing as two pairs of natural transformations $\eta:\Id_{\Dduploid}\graphto GF$
and $\varepsilon:FG\graphto\Id_{\Eduploid}$ satisfying the following
conditions:
\begin{enumerate}
\item \emph{inverses at a distance:} for all $f\in\Dduploid(A,GB)$ and
$g\in\Eduploid(FA,B)$,
\begin{align*}
G\varepsilon_{B}\ccomp(\eta_{GB}\ccomp f) & =f\\
(g\ccomp\varepsilon_{FA})\ccomp F\eta_{A} & =g\:,
\end{align*}
\item \emph{$G$-thunkability of $\eta$:} for all morphisms $f\in\Eduploid(FA,B),g\in\Eduploid(B,C)$
one has:
\[
G(g\ccomp f)\ccomp\eta_{A}=Gg\ccomp(Gf\ccomp\eta_{A})\:,
\]
\item \emph{$F$-linearity of $\varepsilon$:} for all morphisms $f\in\Dduploid(A,B),g\in\Dduploid(B,GC)$
one has:
\[
\varepsilon_{C}\ccomp F(g\ccomp f)=(\varepsilon_{C}\ccomp Fg)\ccomp Ff\:.
\]
\end{enumerate}
\end{proposition}
\begin{proposition}
\label{prop/graph-adjunctions-props}Let an adjunction between graph morphisms $\varphi:\componentwise{\Eduploid}(F-,=)\xrightarrow{\iso}\componentwise{\Dduploid}(-,G{=})$.
\begin{enumerate}
\item \label{enu:adj-pres-thunk-lin}$F$ preserves thunkability, and $G$
preserves linearity.
\item For $f,g$ morphisms in $\Dduploid$, one has $F(f\ccomp g)=Ff\ccomp Fg$
if and only if $\eta_{X}\ccomp f\ccomp g$ associates (for instance,
whenever $g$ is thunkable and whenever the codomain of $f$ is positive).
\item \label{enu:G-functorial}For $f,g$ morphisms in $\Eduploid$, one
has $G(f\ccomp g)=Gf\ccomp Gg$ if and only if $f\ccomp g\ccomp\varepsilon_{X}$
associates (for instance, whenever $f$ is linear and whenever the
domain of $g$ is negative).
\end{enumerate}
\end{proposition}
\begin{proposition}
Let an adjunction between graph morphisms $\varphi:\componentwise{\Eduploid}(F-,=)\xrightarrow{\iso}\componentwise{\Dduploid}(-,G{=})$.
\begin{enumerate}
\item The following conditions are equivalent: $F$ is functorial, $\varphi$
preserves linearity, $\eta$ is linear.
\item The following conditions are equivalent: $G$ is functorial, $\varphi^{-1}$
preserves thunkability, $\varepsilon$ is thunkable.
\item If $F$ is full and surjective on objects, then $\varphi^{-1}$ preserves
linearity (in particular $\varepsilon$ is linear).
\item If $G$ is full and surjective on objects, then $\varphi$ preserves
thunkability (in particular $\eta$ is thunkable).
\end{enumerate}
\end{proposition}
\begin{proposition}
\label{prop:right-adjoint-construction}Let $\Dduploid,\Eduploid$
two non-associative categories together with:
\begin{itemize}
\item for each $A\in\Dduploid$ an object $G_{0}A\in\Eduploid$,
\item a graph morphism $F:\Eduploid\graphto\Dduploid$ satisfying the following
condition:
\begin{equation}
\forall f\in\Eduploid(A,G_{0}B),Ff\text{ is thunkable,}\label[equation]{eq:right-adjoint-condition}
\end{equation}
\item for each $A\in\Eduploid$ and $B\in\Dduploid$ a family of bijections
\[
\varphi:\Dduploid(FA,B)\xrightarrow{\iso}\Eduploid(A,G_{0}B)
\]
natural in $A$.
\end{itemize}
The graph morphism $G:\Dduploid\graphto\Eduploid$ defined with:
\begin{align*}
Gg & \defeq\varphi(g\ccomp\inv{\varphi}(\id{G_{0}B}))
\end{align*}
is right adjoint to $F$.
\end{proposition}
Note that the condition (\ref{eq:right-adjoint-condition}) holds:
\begin{itemize}
\item whenever $F$ preserves thunkability and $G_{0}A$ is negative for
every $A\in\Dduploid$,
\item whenever $F$ is negative.
\end{itemize}
\begin{proof}[Proof of Proposition~\ref{prop:right-adjoint-construction}.]
The definition $Gg=\varphi(g\ccomp\inv{\varphi}(\id{G_{0}B}))$ defines
$G$ as a graph morphism $\Dduploid\graphto\Eduploid$ such that for
each $A\in\Eduploid$ and $B\in\Dduploid$ there is a family of bijections
\[
\varphi:\Dduploid(FA,B)\xrightarrow{\iso}\Eduploid(A,GB)
\]
natural in $A$. Naturality in $B$ then follows by hypothesis:
\begin{align*}
Gg\ccomp\varphi(f) & =\varphi(g\ccomp\inv{\varphi}(\id{G_{0}B}))\ccomp\varphi(f) &  & \text{by definition}\\
 & =\varphi((g\ccomp\inv{\varphi}(\id{G_{0}B}))\ccomp F\varphi(f)) &  & \text{by naturality in }A\text{ of }\varphi\\
 & =\varphi(g\ccomp(\inv{\varphi}(\id{G_{0}B})\ccomp F\varphi(f))) &  & \text{since }F\varphi(f)\text{ is thunkable}\\
 & =\varphi(g\ccomp(\inv{\varphi}(\id{G_{0}B}\ccomp\varphi(f)))) &  & \text{by naturality in }A\text{ of }\varphi\\
 & =\varphi(g\ccomp f) &  & \qedhere
\end{align*}
\end{proof}
\begin{proposition}
~
\begin{itemize}
\item A positive shift on a non-associative category $\Dduploid$ is the
same thing as a graph adjunction
$\graphadjoint \nDownarrow \Dduploid \Dduploid {\Id_{\Dduploid}}$ such
that $\nDownarrow A$ is positive every object $A$.
\item A negative shift on a non-associative category $\Dduploid$ is the
same thing as a graph adjunction $ \graphadjoint {\Id_{\Dduploid}} \Dduploid \Dduploid \nUparrow$
such that $\nUparrow A$ is negative every object $A$.
\end{itemize}
\end{proposition}
A duploid $\Dduploid$ is therefore the same thing as a non-associative
category $\Dduploid$ where all objects are either positive or negative
(or both), together with graph adjunctions:
\[
\componentwise{\Dduploid}(\nDownarrow\mathord{-},=)\iso\componentwise{\Dduploid}(-,=)\iso\componentwise{\Dduploid}(-,\nUparrow\mathord{=}):\Dduploid^{\op}\boxtimes\Dduploid\graphto\Set
\]
such that $\nDownarrow A$ is positive and $\nUparrow A$ is negative
for every object A.

Given a non-associative category $\Dduploid$, it is an instructive
exercise to check that the data of a functor $\nDownarrow:\Dduploid_{t}\rightarrow\Dduploid_{l}$
together with a natural isomorphism
\[
\Dduploid_{l}(\nDownarrow-,\mathord{=})\iso\Dduploid(-,\mathord{=}):\Dduploid_{t}^{\op}\times\Dduploid_{l}\rightarrow\Set
\]
does not suffice to define a positive shift on $\Dduploid$, despite
being implied by it.

\sectioncaps{Symmetric monoidal closed duploids}\label{sec:SMCD-appendix}

\begin{definition}
A (positive) symmetric monoidal duploid is \define{closed} when the
graph morphism $\mathord{-}\otimes A:\Dduploid\rightarrow\Dduploid$
has a right adjoint (written $A\linarrow\mathord{-}$) for every object
$A$, such that $A\linarrow B$ is negative for every objects $A$
and $B$.
\end{definition}
As we have previously seen, this definition does not imply the functoriality
for $\linarrow$, nor is such a requirement expected. However, the
adjunction requirement is quite strong as it implies the following
functoriality properties:
\begin{proposition}
Let $\Dduploid$ a closed symmetric monoidal duploid. $\mathord{\linarrow}$
extends into a graph morphism $\Dduploid^{\op}\boxtimes\Dduploid\graphto\Ncategory$
which is functorial on the following sub-categories: $\Pcategory^{\op}\boxtimes\Ncategory$
(i.e. $P\linarrow\mathord{-}:\Ncategory\rightarrow\Ncategory$ and
$-\linarrow N:\Pcategory^{\op}\rightarrow\Ncategory$) and $\Dduploid_{t}^{\op}\times\Dduploid_{l}$
preserving linearity (i.e.: $\mathord{\linarrow}:\Dduploid_{t}^{\op}\times\Dduploid_{l}\rightarrow\Ncategory_{l}$).
\end{proposition}
\begin{proof}
Let $\Dduploid$ a closed symmetric monoidal duploid and let us note
\[
\varphi_{A}:\componentwise{\Dduploid}(\mathord{-}\otimes A,\mathord{=})\xrightarrow{\iso}\componentwise{\Dduploid}(\mathord{-},A\linarrow\mathord{=}):\Dduploid\boxtimes\Dduploid\graphto\Set
\]
For all $A$, $A\linarrow-$ preserves linearity by \ref{prop/graph-adjunctions-props}
(\ref{enu:adj-pres-thunk-lin}) and is functorial on both $\Dduploid_{l}$
and $\Ncategory$ (separately) by \ref{prop/graph-adjunctions-props}
(\ref{enu:G-functorial}).

The action $f\linarrow C$ on the left for a morphism $f\in\Dduploid(A,B)$
is defined with
\[
\varphi_{A,B\linarrow C,C}(\ev_{B,C}\ccomp((B\linarrow C)\otimes f))\in\Dduploid(B\linarrow C,A\linarrow C)
\]
where $\ev_{B,C}=\varphi_{B,B\linarrow C,C}^{-1}(\id{B\linarrow C})$.
This defines a graph morphism $\mathord{-}\linarrow C:\Dduploid^{\op}\graphto\Dduploid$
for every object $C$ which is adjoint to itself on the right:
\[
\componentwise{\Dduploid}(-,\mathord{=}\linarrow C)\iso\componentwise{\Dduploid}^{\op}(-\linarrow C,\mathord{=}):\Dduploid^{\op}\boxtimes\Dduploid^{\op}\graphto\Set
\]
Thus by \ref{prop/graph-adjunctions-props} again it preserves linearity
in $\Dduploid^{\op}$ (thunkability in $\Dduploid$) and is functorial
on both $\Dduploid_{t}^{\op}$ and $\Pcategory^{\op}$ (separately).

Lastly, for $f\in\Dduploid(A,B)$ and $g\in\Dduploid(C,D)$ one has
\[
(f\linarrow D)\ccomp(B\linarrow g)=(A\linarrow g)\ccomp(f\linarrow C)
\]
whenever $f$ is thunkable or $g$ is linear. Therefore $\linarrow$
gives rise to a bifunctor $(-\linarrow\mathord{=}):\Dduploid_{t}^{\op}\times\Dduploid_{l}\rightarrow\Dduploid_{l}$.\end{proof}
Now recall that the call-by-value arrow is usually defined in call-by-push-value
and in polarised logics with $\nDownarrow(P\linarrow\nUparrow Q)$
where $P$ and $Q$ are positive. We recover in this way the a closed
structure on $\Pcategory$ in a usual sense.
\begin{proposition}
Every closed symmetric monoidal structure on a duploid $\Dduploid$
gives rise to a closed structure on the symmetric monoidal Freyd structure
(in the sense of \citet{power2002premonoidal}): a (categorical) right
adjoint to the functor $\iota\mathord{-}\otimes P:\Pcategory_{t}\rightarrow\Pcategory$.
\end{proposition}
\begin{proof}
Let $\Dduploid$ a closed symmetric monoidal duploid and define $P\rightarrowtriangle^{+}\mathord{-}\defeq\nDownarrow(P\linarrow I\mathord{-}):\Pcategory\graphto\Pcategory_{t}$.
We write $I$ for inferrable inclusion functors. \begin{comment}
As we have seen, $P\linarrow\mathord{-}:\Dduploid_{l}\rightarrow\Ncategory_{l}$
is functorial and therefore this defines a functor $P\rightarrowtriangle^{+}\mathord{-}:\Pcategory\rightarrow\Pcategory_{t}$.
\end{comment}
{} One \begin{comment}
also
\end{comment}
{} has the following natural isomorphisms of graph morphisms:
\begin{align*}
\Pcategory(\iota\mathord{-}\otimes P,\mathord{=}) & =\componentwise{\Dduploid}(I\mathord{-}\otimes P,I\mathord{=}) &  & \text{by definition}\\
 & \iso\componentwise{\Dduploid}(I\mathord{-},P\linarrow\nUparrow\mathord{=}) &  & \text{by adjunction}\\
 & =\Dduploid_{t}(I-,P\linarrow\nUparrow\mathord{=}) &  & \text{by definition}\\
 & \iso\Dduploid_{t}(I-,\nDownarrow(P\linarrow\nUparrow\mathord{=})) &  & I\dashv\nDownarrow:\Dduploid_{t}\rightarrow\Pcategory_{t}\text{ when restricted to thunkables}\\
 & =\Pcategory_{t}(-,\nDownarrow(P\linarrow\nUparrow\mathord{=})) &  & \text{by definition}
\end{align*}
An adjunction between graph morphisms between categories is an
adjunction in the usual sense and therefore we precisely have an
adjunction $\Pcategory(\iota\mathord{-}\otimes
P,\mathord{=})\iso\Pcategory_{t}(-,P\linarrow^{+}\mathord{=})$.
\end{proof}
\begin{definition}
A linear effect adjunction is given by a symmetric monoidal category
$\Acategory$ and a $\Psh{\Acategory}$-category $\underline{\Bcategory}$
with powers of representable presheaves, together with an $\Psh{\Acategory}$-adjunction
$\underline{L}\dashv\underline{R}:\underline{\Acategory}\rightarrow\underline{\Bcategory}$.
\end{definition}
We also recall the following useful characterisation of linear effect adjunction
given in \citet{Mellies2012parametric}.
\begin{proposition}
A linear effect adjunction is the same thing as a symmetric monoidal
category $\Acategory$ together with an adjunction $L\dashv R:\Bcategory\rightarrow\Acategory$,
a pseudo-action $\mathord{\linarrow}:\Acategory^{\op}\times\Bcategory\rightarrow\Bcategory$
of $\Acategory^{\op}$ on $\Bcategory$, and a family of adjunctions
\begin{equation}
L(-\otimes A)\dashv R(A\linarrow-):\Bcategory\rightarrow\Acategory\:.\label[equation]{eq:param-adj}
\end{equation}
\end{proposition}
We recall the definition of a pseudo-action of a monoidal category
$\Acategory$ on a category $\Bcategory$ from \citet{Mellies2012parametric}.
\begin{definition}
A pseudo-action of a monoidal category $(\Acategory,\otimes,I)$ on
a category $\Bcategory$ is a functor
\[
\ast:\Acategory\times\Bcategory\rightarrow\Bcategory
\]
 together with two natural isomorphisms
\begin{align*}
\delta_{A,A',B}^{2} & :(A\otimes A')\ast B\rightarrow A'\ast(A\ast B) & \delta_{A}^{0} & :I\ast A\rightarrow A
\end{align*}
subject to three coherence laws: of $\delta^{2}$ with the associator
for $\otimes$, and of $\delta^{0}$ with each unitor of $\otimes$.
\end{definition}

\begin{theorem}
For every linear effect adjunction, its associated symmetric
monoidal duploid $\duploid LR$ is closed.
\end{theorem}

\begin{proof}
Given a linear effect adjunction and $\Dduploid=\duploid LR$ its
associated symmetric monoidal duploid, the adjunctions
(\ref{eq:param-adj}) correspond to a family of bijections between
hom-sets
\begin{equation}
\Dduploid(A\otimes_{\Dduploid}B,C)\iso\Dduploid(A,B\linarrow_{\Dduploid}C)\label[equation]{eq:closure-restr}
\end{equation}
where we define
\[
B\linarrow_{\Dduploid}C\defeq B^{+}\linarrow C^{-}
\]
and where we recall that $A\otimes_{\Dduploid}B$ is $A^{+}\otimes B^{+}$.
This isomorphism is natural in $A\in\Dduploid$.
Since $-\otimes_{\Dduploid}B$ preserves thunkability and $B\linarrow_{\Dduploid}C$
is negative, by Proposition \ref{prop:right-adjoint-construction}
the family of objects $B\linarrow_{\Dduploid}C$ extends into a graph
morphism $(B\linarrow_{\Dduploid}-):\Dduploid\graphto\Dduploid$ and
the family of bijections (\ref{eq:closure-restr}) into an adjunction
$\graphadjoint {-\otimes_{\Dduploid} B} \Dduploid \Dduploid
{B\linarrow_\Dduploid -}$.
\end{proof}
In analogy with the $L$-calculus, Proposition \ref{prop:right-adjoint-construction}
extends the stack constructor $V\push S$ from the CBPV target into
a context constructor $V\push e$ by $\varsigma$-expansion (focusing):
$V\push e\defeq\tmu x.\cut{\mu\alpha.\cut x{V\push \alpha}}e$.

Conversely, given $\Dduploid$ a symmetric monoidal duploid, we have
seen that the family of adjunctions $\graphadjoint
{-\otimes_{\Dduploid} B} \Dduploid \Dduploid {B\linarrow_\Dduploid -}$
gives rise to a bifunctor
$\mathord{\linarrow}:\Dduploid_{t}^{\op}\times\Dduploid_{l}\rightarrow\Dduploid_{l}$.
In fact, one easily sees from this adjunction that
$\mathord{\linarrow}$ has the structure of a pseudo-action of
$\Dduploid_{t}^{\op}$ on $\Dduploid_{l}$. Moreover, one has
$\nUparrow(-\otimes
A)\dashv\nDownarrow(A\linarrow-):\Dduploid_{l}\rightarrow\Dduploid_{t}$
since $\nUparrow$ (resp. $\nDownarrow$) is equivalent to
$\Id_{\Dduploid_{l}}$ on $\Dduploid_{l}$ (resp. $\Dduploid_{t}$).
Thus:
\begin{theorem}
Every symmetric monoidal closed duploid gives rise to an $\polsystem{IMLL}$
model on the adjunction (\ref{equation/adjunction-PN}).
\end{theorem}
We expect that the monoidal structure can be obtained similarly as a
left adjoint; however this would require either a suitable notion of
``multi''-duploids or of closed duploid. Then the structure of shifts
on duploids can be derived with $\nDownarrow A=A\otimes1$ and
$\nUparrow A=1\linarrow A$ rather than assumed.

\sectioncaps{Linearly distributive duploids}\label{section/linearly-distributive-duploids}
We want to describe the structure inherited by a duploid associated to
an adjunction of the form~\eqref{equation/adjunction-LR} where both
categories $\Acategory$ and $\Bcategory$ come equipped with symmetric
monoidal structures noted $(\Acategory, \tensorialand,
\tensorialtrue)$ and $(\Bcategory, \tensorialor, \tensorialfalse)$,
generalising linearly-distributive categories \cite{Cockett_1997}, in
the sense that there are four distributivity laws (or commutators)
\[
\begin{array}{ccc}
    ldistr^\tensorialand_{A_1,A_2,B} & \hspace{-.7em} :  \hspace{-.7em} & A_1 \tensorialand R(L(A_2) \tensorialor B) \to R(L(A_1 \tensorialand A_2) \tensorialor B)
        \vspace{.4em}
        \\
ldistr^\tensorialor_{A,B_1,B_2} &  \hspace{-.7em} :  \hspace{-.7em} & L(R(B_1 \tensorialor B_2) \tensorialand A) \to B_1 \tensorialor L(R(B_2) \tensorialand A)
        \vspace{.4em}
        \\
rdistr^\tensorialand_{A_1,A_2,B} &  \hspace{-.7em} :  \hspace{-.7em} & R(B \tensorialor L(A_1)) \tensorialand A_2 \to R(B \tensorialor L(A_1 \tensorialand A_2))
\vspace{.4em}
\\
rdistr^\tensorialor_{A,B_1,B_2} &  \hspace{-.7em} :  \hspace{-.7em} & L(A \tensorialand R(B_1 \tensorialor B_2)) \to L(A \tensorialand R(B_1)) \tensorialor B_2\\
  \end{array}
\]
introduced in \citet{Mellies2017micrological} and assumed to make a number of coherence diagrams commute.
We note that the strengths for $\tensorialand$ and $\tensorialor$ can be deduced from the commutators.

When translating the commutators into the duploid framework, the four rules collapse into only two, as they were merely cases depending on the polarity of $A'$/$B$.
$ldistr^\tensorialand$ and $rdistr^\tensorialor$ become $\delta^l$ and $ldistr^\tensorialor$ and $rdistr^\tensorialand$ become $\delta^r$.

\begin{definition}
  A \define{linearly distributive duploid} $\mathcal D$ is a duploid equipped with a pair of 
  positive and negative symmetric monoidal structures related by two families of mappings:
  \[
    \begin{array}{ccc} 
      \delta^l_{A,B,C} & : & A \otimes (B \parr C) \to (A \otimes B) \parr C\\
      \delta^r_{A,B,C} & : & (A \parr B) \otimes C \to A \parr (B \otimes C)
    \end{array}
  \]
  natural for each component and that respects the usual coherence diagrams for a linearly distributive category.
\end{definition}

\begin{definition}\label[definition]{definition/thunkable-wrt-type}
  Let $\mathcal D$ be a linearly distributive duploid.
We say that a morphism $h \in \mathcal D(A \otimes B, C)$ is \define{linear wrt. $A$} (and we note it $h \in \mathcal D(\underline A \otimes B, C)$) when, for all $g \in \mathcal D(A',A)$ and $f \in \mathcal D(A'',A')$, we have
  \[
    h \pcomp ((g \ccomp f) \ltensortimes B) = h \pcomp (g \ltensortimes B) \pcomp (f \ltensortimes B).
  \]
  Dually, we say that a morphism $f$ from $A$ to $B \parr C$ is \define{thunkable wrt. $B$} (noted $f \in \mathcal D(A, \underline B \parr C)$) when, for all $g \in \mathcal D(B, B')$ and $h \in \mathcal D(B',B'')$, we have
  \[
    ((h \ccomp g) \lparrtimes C) \ncomp f = (h \lparrtimes C) \ncomp (g \lparrtimes C) \ncomp f.
  \]
\end{definition}

\begin{proposition}
  Let $h$ be a morphism from $A \otimes B$ to $C$. If $A$ is positive, then $h$ is linear wrt. $A$. Symmetrically, if $B$ is negative, then $f \in \Dduploid(A, B \parr C)$ is thunkable wrt. $B$.
\end{proposition}

\sectioncaps{Dialogue duploids and dialogue functors}
We start by describing explicitly the commuting diagrams expressing
the naturality and coherence conditions in the definition of dialogue duploid
given in \cref{dialogueduploidsym}.
\begin{equation}\label[equation]{chinaturality}
\begin{array}{ccc}
\begin{tikzcd}[column sep = -0.3em, row sep = 2em]
A'\arrow[rrrr, "{h_A}"] \arrow[rrd, "\chi_{A',B,C}(f \pcomp (h_A \ltensortimes B))"'] 
&&&&
A \arrow[lld, "\chi_{A,B,C}(f)"]
\\
& & \dupldual{B} \parr C
\end{tikzcd}
&
\hspace{-1em}
\begin{tikzcd}[column sep = 0.15em, row sep = 1.2em]
&&
A 
\arrow[rrdd, "\chi_{A,B',C}(f \pcomp (A \rtensortimes h_B))"]
\arrow[lldd,"\chi_{A,B,C}(f)"{swap}]
&&
\\ 
\\
\dupldual{B} \parr C 
\arrow[rrrr,"\dupldual{h_B} \lparrtimes C"]
&&&&
\dupldual{B'} \parr C
\end{tikzcd}
&
\hspace{-.5em}
\begin{tikzcd}[column sep = 0.15em, row sep = 1.2em]
 &&
A \arrow[lldd, "\chi_{A,B,C}(f)"{swap}]
 \arrow[rrdd, "\chi_{A,B,C'}(h_C \ccomp f)"] 
\\
\\
\dupldual{B} \parr C \arrow[rrrr, "\dupldual{B}\rparrtimes h_C"]
&&&&
\dupldual{B} \parr C'
\end{tikzcd}
\end{array}
\end{equation}
\begin{equation}\label[equation]{chicoherence}
    \begin{tikzcd}
      \mathcal D(A \otimes (B \otimes C), D) \arrow[rrrr, "\chi_{A,B \otimes C,D}"] \arrow[dd, "\text{associativity}"'] & & & & \mathcal D(A, \dupldual{(B \otimes C)} \parr D) \arrow[dd, "\begin{array}{c}\text{monoidality}\\\text{symmetry}\\\text{associativity}\end{array}"]\\ \\
      \mathcal D((A \otimes B)\otimes C, D) \arrow[rr, "\chi_{A\otimes B, C, D}"] & & \mathcal D(A \otimes B, \dupldual{C} \parr D) \arrow[rr, "\chi_{A,B,\dupldual{C} \parr D}"] & & \mathcal D(A, \dupldual{B} \parr (\dupldual{C} \parr D)) 
    \end{tikzcd}
  \end{equation}
We define the notion of dialogue duploid functors.
\begin{definition}
  A \define{dialogue duploid functor} $F : \mathcal D \to \mathcal D'$ is a duploid functor, lax monoidal for $\otimes$ and colax monoidal for $\parr$ (i.e.\ $F^\op$ is lax monoidal) and equipped with a family of natural central invertible morphisms $\widetilde F_A : F(\dupldual A) \iso (FA)^\star$ such that the following coherence diagram commutes:
  \[
    \begin{tikzcd}
      \mathcal D(A \otimes B, C) \arrow[rr, "\chi_{A,B,C}"] \arrow[d, "F"'] & & \mathcal D(B, \dupldual A \parr C)\arrow[d, "F"]\\
      \mathcal D'(F(A \otimes B), FC) \arrow[dd, "\begin{array}{c}\text{monoidality}\\\text{($\tensor$) of }F\end{array}"'] & & \mathcal D'(FB, F(\dupldual A \parr C)) \arrow[d, "\begin{array}{c}\text{monoidality}\\\text{($\parr$) of }F\end{array}"]\\ 
                                                                            & & \mathcal D'(FB, F(\dupldual A) \parr' FC) \arrow[d, "\widetilde F_A"]\\\
      \mathcal D'(FA \otimes' FB, FC) \arrow[rr, "\chi'_{FA, FB, FC}"] & & \mathcal D'(FB, (FA)^\star \parr' FC)
    \end{tikzcd}
  \]
\end{definition}
\begin{definition}
  $Dia\dupl$ is the category whose objects are the dialogue duploids and whose morphisms are the dialogue duploid functors.
\end{definition}

\sectioncaps{Interpretation of the syntax}

This section and the one that follows adapts to the classical case the
technique used in \citep{CFM2015} which is further detailed in
\citet{Munch-Maccagnoni2017curry}.
A context on the left $(a_1 : A_1, a_2 : A_2,\dots,a_n : A_n)$ is interpreted as $A_1 \otimes A_2 \otimes \dots \otimes A_n$ and a context on the right $(\beta_1 : B_1, \beta_2 : B_2, \dots, \beta_n : B_n)$ is interpreted as $B_1 \parr B_2 \parr \dots \parr B_n$.

Let $\Gamma$ and $\Gamma'$ be two contexts on the left and $\sigma$ an element of $\Sigma(\Gamma,\Gamma')$. We note $\llbracket \sigma \rrbracket$ the associated canonical isomorphism of $\Dduploid(\Gamma,\Gamma')$ constructed by composing symmmetries of $\otimes$. We stress on the fact that, as a composition of thunkable morphisms, $\llbracket \sigma \rrbracket$ is thunkable.
Dually, let $\Delta$ and $\Delta'$ be two contexts on the right and $\tsigma$ an element of $\Sigma(\Delta',\Delta)$. The associated canonical isomorphism of $\Dduploid(\Delta', \Delta)$ obtained by composing symmetries of $\parr$ is noted $\llbracket\tsigma \rrbracket$ and is linear. 

\subsectioncaps{Interpretation of judgements}

\begin{itemize}
  \setlength{\itemsep}{4pt}
  \item $\llbracket \Gamma \vdash t : A\ |\ \Delta\rrbracket \in \mathcal D(\Gamma, A \parr \Delta)$
  \item $\llbracket \Gamma \vdash V : A\ |\ \Delta\rrbracket \in \mathcal D(\Gamma, \underline A \parr \Delta)$
  \item $\llbracket \Gamma\ |\ e : A \vdash \Delta\rrbracket \in \mathcal D(\Gamma \otimes A, \Delta)$
  \item $\llbracket \Gamma\ |\ S : A \vdash \Delta\rrbracket \in \mathcal D(\Gamma \otimes \underline A, \Delta)$
  \item $\llbracket \command c : (\Gamma \vdash \Delta)\rrbracket \in \mathcal D(\Gamma, \Delta)$
\end{itemize}
See \cref{definition/thunkable-wrt-type} for the meaning of the notation $\underline A$.
\subsectioncaps{Interpretation of typing rules}

\paragraphcaps{Identity rules}
\begin{itemize}
  \setlength{\itemsep}{4pt}
  \item $\llbracket a : A \vdash a : A\ |\rrbracket = \mathsf{id}_{A} \in \mathcal D_t(A,A)$
  \item $\llbracket |\ \alpha : A \vdash \alpha : A\rrbracket = \mathsf{id}_{A} \in \mathcal D_l(A,A)$
  \item $\llbracket \Gamma\ |\ \tmu a\eps.\command c : A_\veps\vdash \Delta\rrbracket = \llbracket \command c : (\Gamma, x:A_\veps\vdash\Delta) \rrbracket \in \mathcal D(\Gamma \otimes A_\veps, \Delta)$
  \item $\llbracket \Gamma\vdash \mu \alpha\eps.\command c : A_\veps\ |\ \Delta\rrbracket = \llbracket \command c : (\Gamma \vdash\alpha:A_\veps,\Delta) \rrbracket \in \mathcal D(\Gamma, A_\veps \parr \Delta)$
  \item $\begin{aligned}[t]
& \llbracket \perfectcut{t}{e}\eps : (\Gamma, \Gamma' \vdash \Delta, \Delta')\rrbracket\\
&= (\llbracket \Gamma\ |\ e : A \vdash \Delta\rrbracket \lparrtimes \Delta') \ncomp (\delta^l_{\Gamma,A,\Delta'} \pcomp (\Gamma \rtensortimes \llbracket \Gamma' \vdash t : A\ |\ \Delta'\rrbracket)) \in \mathcal D(\Gamma \otimes \Gamma', \Delta \parr \Delta')
\end{aligned}$\\
where $\delta^l_{\Gamma,A,\Delta'} : \Gamma \otimes (A \parr \Delta') \to (\Gamma \otimes A) \parr \Delta'$ is the distributor.
\end{itemize}

\paragraphcaps{Structural rules}

$\forall\sigma\in\Sigma(\Gamma',\Gamma),\ \forall\tsigma\in\Sigma(\Delta,\Delta')$
\begin{itemize}
  \setlength{\itemsep}{4pt}
  \item $\llbracket \Gamma'\vdash t[\sigma,\tsigma]: A\ |\ \Delta'\rrbracket = (A \rparrtimes \llbracket \tsigma \rrbracket) \ncomp (\llbracket \Gamma \vdash t : A\ |\ \Delta\rrbracket \ccomp \llbracket \sigma\rrbracket) \in \Dduploid(\Gamma', A \parr \Delta')$
  \item $\llbracket \Gamma'\ |\ e[\sigma,\tsigma] : A\vdash\Delta'\rrbracket = \llbracket \tsigma \rrbracket \ccomp (\llbracket \Gamma\ |\ e : A\vdash\Delta \rrbracket \pcomp (\llbracket \sigma\rrbracket \ltensortimes A)) \in \Dduploid(\Gamma' \tensor A, \Delta')$
  \item $\llbracket c[\sigma,\tsigma]:(\Gamma'\vdash \Delta')\rrbracket = \llbracket \tsigma \rrbracket \ccomp (\llbracket \command c:(\Gamma \vdash \Delta) \rrbracket \ccomp \llbracket \sigma\rrbracket) \in \Dduploid(\Gamma', \Delta')$
\end{itemize}

\paragraphcaps{Conjunction rules}
\begin{itemize}
  \setlength{\itemsep}{4pt}
  \item $\llbracket \vdash () : \1\ |\rrbracket = \mathsf{id}_\1 \in \mathcal D_t(\1,\1)$
  \item $\llbracket \Gamma\ |\ \tmu ().\command c\vdash \Delta \rrbracket = \llbracket \command c : (\Gamma\vdash\Delta)\rrbracket \ccomp \rho_\Gamma \in \mathcal D(\Gamma \otimes \underline{\1},\Delta)$\\
    where $\rho_\Gamma : \Gamma \otimes 1 \to \Gamma$ is the right unitor of $\otimes$.
  \item $\begin{aligned}[t]
&\llbracket \Gamma,\Gamma'\vdash V \smallotimes W : A \otimes B\ |\ \Delta, \Delta'\rrbracket \\
&= ((((\sigma_{B,A} \lparrtimes \Delta) \ncomp \delta^l_{B,A,\Delta}) \pcomp \sigma_{A\parr \Delta, B})\lparrtimes \Delta')\ncomp\delta^l_{A\parr \Delta, B, \Delta'}) \pcomp (\llbracket \Gamma\vdash V : A\ |\ \Delta\rrbracket \otimes \llbracket \Gamma'\vdash W : B\ |\ \Delta'\rrbracket) \\
&\in \mathcal D(\Gamma \otimes \Gamma',\underline{(A \otimes B)} \parr \Delta \parr \Delta')
\end{aligned}$
  \item $\llbracket \Gamma\ |\ \tmu(a \smallotimes b).\command c : A\otimes B\vdash\Delta \rrbracket = \llbracket \Gamma, a : A, b : B\vdash\Delta\rrbracket \in \mathcal D(\Gamma \otimes \underline{A \otimes B},\Delta)$
\end{itemize}

\paragraphcaps{Disjunction rules}
\begin{itemize}
  \setlength{\itemsep}{4pt}
  \item $\llbracket |\ [] : \bot \vdash \rrbracket = \mathsf{id}_\bot \in \mathcal D_l(\bot,\bot)$
  \item $\llbracket \Gamma \vdash \mu [].\command c\ |\ \Gamma \rrbracket = \lambda'_\Delta \ccomp \llbracket \command c : (\Gamma\vdash\Delta)\rrbracket \in \mathcal D(\Gamma,\underline\bot \parr \Delta)$\\
    where $\lambda'_\Delta : \Delta \to \bot\parr\Delta$ is the left unitor of $\parr$.
  \item $\begin{aligned}[t]
&\llbracket \Gamma,\Gamma' \ |\ S \smallparr S' : A \parr B \vdash\Delta, \Delta'\rrbracket\\
&\mkern-40mu= \llbracket \Gamma\ |\ S : A \vdash \Delta\rrbracket \parr\llbracket \Gamma'\ |\ S' : B\vdash \Delta'\rrbracket \circ (\delta^l_{\Gamma,A,(\Gamma'\otimes B)} \bullet (\Gamma \otimes (\sigma'_{(\Gamma'\otimes B), A} \circ \delta^l_{\Gamma',B,A})) \bullet (\Gamma \otimes \Gamma' \otimes \sigma'_{A,B}))\\
&\in \mathcal D(\Gamma \otimes \Gamma' \otimes \underline{(A \parr B)}, \Delta \parr \Delta')
\end{aligned}$
  \item $\llbracket \Gamma \vdash \mu(\alpha \smallparr \beta).\command c : A\parr B\ |\ \Delta \rrbracket = \llbracket \command c:(\Gamma\vdash \alpha : A, \beta : B, \Delta) \rrbracket \in \mathcal D(\Gamma, \underline{A \parr B} \parr \Delta)$
\end{itemize}

\paragraphcaps{Negation rules}
\begin{itemize}
  \setlength{\itemsep}{4pt}
  \item $\llbracket \Gamma \vdash [S] : \dupldual N\ |\ \Delta\rrbracket = \chi_{\Gamma,N,\Delta}(\llbracket \Gamma\ |\ S : N \vdash \Delta \rrbracket) \in \mathcal D(\Gamma, \underline{\dupldual N} \parr \Delta)$
  \item $\llbracket \Gamma\ |\ [V] : \dupldual P \vdash \Delta\rrbracket = \chi^{-1}_{\Gamma,\dupldual P,\Delta}((\nu_P \lparrtimes \Delta) \ncomp \llbracket\Gamma \vdash V : P\ |\ \Delta\rrbracket) \in \mathcal D(\Gamma \otimes \underline{\dupldual P},\Delta)$
  \item $\llbracket \Gamma\ |\ \tmu[\alpha].\command c : \dupldual N \vdash \Delta\rrbracket = \chi^{-1}_{\Gamma,\dupldual N,\Delta}((\nu_N \lparrtimes \Delta) \ncomp \llbracket \command c : (\Gamma\vdash \alpha : N, \Delta)\rrbracket) \in \mathcal D(\Gamma \otimes \underline{\dupldual N}, \Delta) $
  \item $\llbracket \Gamma\vdash \mu[a].\command c : \dupldual P\ |\ \Delta\rrbracket = \chi_{\Gamma,P,\Delta}(\llbracket \command c : (\Gamma, a : P \vdash \Delta)\rrbracket) \in \mathcal D(\Gamma, \underline{\dupldual P} \parr \Delta)$
\end{itemize}
where $\nu_A : A \to \dupldoubledual A$.
\sectioncaps{Soundness of the interpretation}\label{cohsoundcomp}

We follow again \citet{Munch-Maccagnoni2017curry} which we adapt to
classical logic with an involutive negation. We leave implicit the
assignment of objects to atoms, and the interpretation of types as
objects.
We start by proving coherence properties of the interpretation.
We say that two derivations are \define{equivalent} if their
interpretation are equal in all dialogue duploids.

\begin{lemma}\label[lemma]{exaonestruct}
  For any typing derivations, there is an equivalent derivation starting by one structural rule.
\end{lemma}
\begin{proof}
  We treat the case of a typing derivation of $\Gamma \vdash t : A\ |\ \Delta$;
the other cases are similar. We look at the smallest equivalent
  typing derivation of $\Gamma \vdash t :A\ |\ \Delta$ in terms of
  number of rules used. If it starts by two structural rules
  $\tau,\tilde\tau$ and $\sigma, \tsigma$, then the derivation where
  the two first rules are replaced by the structural rule
  $\tau\circ\sigma, \tsigma\circ\tilde\tau$ is equivalent and uses
  strictly less rules, which is impossible by hypothesis.
So we have a derivation starting with at most one structural rule.
  If there is none, we can always add the structural rule of the
  identity, which is interpreted as the identity.
\end{proof}

For a term $\mathfrak g$, we will note $\fvwo \mathfrak g$ the set of free variables of $\mathfrak g$ and $\fcvwo \mathfrak g$ the set of free co-variables of $\mathfrak g$. For $\Gamma$ a context and $X$ a subset of the domain of $\Gamma$, we will note the restriction of $\Gamma$ to $X$ as $\Gamma_{\upharpoonright X}$.

\begin{lemma}
  For any derivation $\command c: (\Gamma\vdash\Delta)$, one has $\fvwo \command c = \dom \Gamma$ and $\fcvwo \command c = \dom \Delta$ and similarly for $t$ and $e$ replacing $\command c$.
\end{lemma}
\begin{proof}
  By induction on the derivation.
\end{proof}

We prove a \emph{coherent generation lemma} which says that, from the
form of the term, we can deduce the first rules of a derivation, or,
at least, find an equivalent derivation starting by those rules.

\begin{lemma}\label[lemma]{cohgen}

$(\vdash \mathbf{ax})$ : Any derivation of $\Gamma\vdash x : A\ |\ \Delta$ satisfies $\Gamma = (x : A)$ and $\Delta = \emptyset$ and is equivalent to the derivation:
\[
  \begin{prooftree}
    \infer0[$(\vdash\mathbf{ax})$]{x:A \vdash x : A\ |\ }
  \end{prooftree}
\]

\medskip

$(\mathbf{cut}\eps)$ : For any derivation of $\perfectcut{t}{e}\eps : (\Gamma\vdash\Delta)$, there exists $A\eps$ and an equivalent derivation ending with:
\[
  \begin{prooftree}
    \hypo{\Gamma_{\fv e}\ |\ e : A\eps \vdash\Delta_{\fcv e}}
    \hypo{\Gamma_{\fv t}\vdash t : A\eps \ |\ \Delta_{\fcv t}}
    \infer2[$(\mathbf{cut}\eps)$]{\perfectcut{t}{e}\eps : (\Gamma_{\fv e},\Gamma_{\fv t}\vdash\Delta_{\fcv e},\Delta_{\fcv t})}
    \infer1[$(\sigma,\tsigma)$]{\perfectcut{t}{e}\eps : (\Gamma\vdash\Delta)}
  \end{prooftree}
\]
where $\sigma \in \Sigma(\Gamma,(\Gamma_{\fv t},\Gamma_{\fv e}))$ is the unique permutation without renaming from $\Gamma$ to $(\Gamma_{\fv t},\Gamma_{\fv e})$ and $\tsigma \in \Sigma((\Delta_{\fcv t},\Delta_{\fcv e}),\Delta)$ is the unique permutation without renaming from $(\Delta_{\fcv t},\Delta_{\fcv e})$ to $\Delta$.

\medskip

$(\vdash\negN)$ : For any derivation of $\Gamma\vdash [S] : A\ |\ \Delta$, one has $A$ of the form $\dupldual N$ and an equivalent derivation ending with:
\[
  \begin{prooftree}
    \hypo{\Gamma\ |\ S : N\vdash \Delta}
    \infer1[$(\vdash\negN)$]{\Gamma\vdash [S] : \dupldual N\ |\ \Delta}
  \end{prooftree}
\]

The other cases are similar.
\end{lemma}

\begin{proof}
  $(\vdash \mathsf{ax})$ By using the previous lemma, we have that $\dom\Gamma = \{A\}$ and $\Delta$ is empty. We know from \cref{exaonestruct} that we can assume that it starts with one structural rule but it is a renaming which is interpreted as the identity. Finally, the only non-structural rule that can be applied to $x : A\vdash x : A\ |$ is $(\vdash \mathsf{ax})$.

  $(\vdash \mathsf{cut}\eps)$ From the previous lemma, we know that $\dom\Gamma = \fvwo \perfectcut{t}{e}\eps = \fvwo t \uplus \fvwo e$, so $\sigma$ is well defined. We can say the same about $\Delta$ and $\tsigma$. By using \cref{exaonestruct} and the fact that there is only one non-structural rule that can be applied to $\perfectcut{t}{e}\eps$, we have a type $A\eps$ and a derivation of $\perfectcut{t}{e}\eps : (\Gamma \vdash \Delta)$ of the form:
  \[
    \begin{prooftree}
      \hypo{\Gamma_1\ |\ e[\tau,\tilde\tau] : A\eps\vdash \Delta_1}
      \hypo{\Gamma_2 \vdash t[\tau,\tilde\tau] : A\eps\ |\ \Delta_2}
      \infer2[$(\mathsf{cut}\eps)$]{\perfectcut{t[\tau,\tilde\tau]}{e[\tau,\tilde\tau]}\eps : (\Gamma_1, \Gamma_2 \vdash \Delta_1, \Delta_2)}
      \infer1[$(\tau,\tilde\tau)$]{\perfectcut{t}{e}\eps : (\Gamma \vdash \Delta)}
    \end{prooftree}
  \]
  We can add the structural rules $\sigma,\tsigma$ and $\sigma^{-1},\tsigma^{-1}$ and, by centrality of symmetries and by coherence between symmetries and distributors, we can commute the cut rule and the structural rules to obtain the following equivalent derivation:
  \[
    \begin{prooftree}
      \hypo{\Gamma_1\ |\ e[\tau,\tilde\tau] : A\eps\vdash \Delta_1}
      \infer1[$(\sigma^{-1}\circ\tau,\tilde\tau\circ\sigma^{-1})$]{\Gamma_{\fv e}\ |\ e : A\eps \vdash\Delta_{\fcv e}}
      \hypo{\Gamma_2 \vdash t[\tau,\tilde\tau] : A\eps\ |\ \Delta_2}
      \infer1[$(\sigma^{-1}\circ\tau,\tilde\tau\circ\sigma^{-1})$]{\Gamma_{\fv t}\vdash t : A\eps \ |\ \Delta_{\fcv t}}
      \infer2[$(\mathsf{cut}\eps)$]{\perfectcut{t}{e}\eps : (\Gamma_{\fv e},\Gamma_{\fv t}\vdash\Delta_{\fcv e},\Delta_{\fcv t})} 
      \infer1[$(\sigma,\tilde\sigma)$]{\perfectcut{t}{e}\eps : (\Gamma \vdash \Delta)}
    \end{prooftree}
  \]

  $(\vdash\negN)$ : By using \cref{exaonestruct} and the fact that
  only the rule $(\vdash\negN)$ can be applied, we have a negative
  type $N$ and a derivation of the form:
  \[
    \begin{prooftree}
      \hypo{\Gamma'\ |\ S[\tau,\tilde\tau] : N\vdash \Delta'}
      \infer1[$(\vdash\negN)$]{\Gamma'\vdash [S[\tau,\tilde\tau]] : \dupldual N\ |\ \Delta'}
      \infer1[$(\tau,\tilde\tau)$]{\Gamma\vdash [S] : \dupldual N\ |\ \Delta}
    \end{prooftree}
  \]
  We can commute the negation and the structural rule by naturality
  component-wise of $\chi$ and we obtain the equivalent derivation we
  seek:
  \[
    \begin{prooftree}
      \hypo{\Gamma'\ |\ S[\tau,\tilde\tau] : N\vdash \Delta'}
      \infer1[$(\tau,\tilde\tau)$]{\Gamma\ |\ S : N\vdash \Delta}
      \infer1[$(\vdash\negN)$]{\Gamma\vdash [S] : \dupldual N\ |\ \Delta}
    \end{prooftree}  \]
  The other cases are similar and rely on the two previous lemmas and
  the coherence between the operations we are using.
\end{proof}

\noindent Thanks to the previous lemma, we can now reason on
derivations up to equivalence by doing an induction on the structure
of the term.

\begin{lemma}
  We consider a derivation of $\Gamma \vdash V : A\ |\ \Delta$ and its interpretation $\llbracket V\rrbracket \in \Dduploid(\Gamma,\underline A \parr \Delta)$.
  \begin{itemize}
    \item For any derivation of $\command c : (\Gamma',a:A\vdash\Delta')$, there exists a derivation of $\command c[V/a] : (\Gamma',\Gamma\vdash\Delta',\Delta)$ such that:
      \[
        \llbracket c[V/a]\rrbracket = (\llbracket c\rrbracket \lparrtimes \Delta) \ncomp (\delta^l_{\Gamma',A,\Delta} \pcomp (\Gamma' \rtensortimes \llbracket V\rrbracket))
      \]
    \item For any derivation of $\Gamma', a:A\vdash t : B\ |\ \Delta'$, there exists a derivation of $\Gamma',\Gamma\vdash t[V/a] : B\ |\ \Delta',\Delta$ such that:
      \[
        \llbracket t[V/a]\rrbracket = (\llbracket t\rrbracket \lparrtimes \Delta) \ncomp (\delta^l_{\Gamma',A,\Delta} \pcomp (\Gamma' \rtensortimes \llbracket t\rrbracket))
      \]
    \item For any derivation of $\Gamma',a:A\ |\ e : B\vdash \Delta'$, there exists a derivation of $\Gamma',\Gamma\ |\ e[V/a] : B\vdash \Delta',\Delta$ such that:
      \[
        \llbracket e[V/x]\rrbracket = ((\llbracket e\rrbracket \pcomp (\Gamma' \rtensortimes \sigma^{-1}_{A,B})) \lparrtimes \Delta) \ncomp (\delta^l_{\Gamma'\otimes B,A,\Delta} \pcomp ((\Gamma' \otimes B) \rtensortimes \llbracket V\rrbracket) \pcomp (\Gamma' \rtensortimes \sigma_{\Gamma,B}))
      \]
  \end{itemize}
\end{lemma}
\begin{proof}
  We reason by induction on $\command c$, $t$, $e$ by using \cref{cohgen}.
In the case where the last rule used is $(\mathbf{cut}\eps)$ and $\command c$ is of the form $\perfectcut{t}{e}\eps$ with derivations $\Gamma'_{\fv t} \vdash t : B\ |\ \Delta'_{\fcv t}$ and $\Gamma'_{\fv e}\ |\ e : B \vdash \Delta'_{\fcv e}$:
If $a \in \fvwo t$, by induction, we know that we have a derivation of $\Gamma'_{\fv t}, \Gamma \vdash t[V/a] : B\ |\ \Delta'_{\fcv t}, \Delta$ and that :
  \[
    \llbracket t[V/a]\rrbracket = (\llbracket t\rrbracket \lparrtimes \Delta) \ncomp (\delta^l_{\Gamma'_{\fv t},A,\Delta} \pcomp (\Gamma'_{\fv t} \rtensortimes \llbracket V\rrbracket))
  \]
  So,

\vspace*{\medskipamount}
\begin{adjustbox}{width=\columnwidth}
$\begin{aligned}
& \llbracket \perfectcut{t}{e}\eps[V/a]\rrbracket \\
    &= \llbracket\tsigma\rrbracket \ncomp (((\llbracket e\rrbracket \lparrtimes (\Delta'_{\fcv t}\parr \Delta)) \ncomp (\delta^l_{\Gamma'_{\fv e},B,\Delta'_{\fcv t}\parr \Delta} \pcomp (\Gamma'_{\fv e} \rtensortimes \llbracket t[V/a]\rrbracket))) \pcomp \llbracket\sigma\rrbracket)\\ 
    &= \llbracket\tsigma\rrbracket \ncomp (((\llbracket e\rrbracket \lparrtimes (\Delta'_{\fcv t}\parr \Delta)) \ncomp (\delta^l_{\Gamma'_{\fv e},B,\Delta'_{\fcv t}\parr \Delta} \pcomp (\Gamma'_{\fv e} \rtensortimes ((\llbracket t\rrbracket \lparrtimes \Delta) \ncomp (\delta^l_{\Gamma'_{\fv t},A,\Delta} \pcomp (\Gamma'_{\fv t} \rtensortimes \llbracket V\rrbracket)))))) \pcomp \llbracket\sigma\rrbracket)\\
    &= \llbracket\tsigma\rrbracket \ncomp ((((\llbracket e\rrbracket \lparrtimes \Delta'_{\fcv t}) \ncomp (\delta^l_{\Gamma'_{\fv e},B,\Delta'_{\fcv t}} \pcomp (\Gamma'_{\fv e} \rtensortimes \llbracket t\rrbracket))) \lparrtimes \Delta) \ncomp (\delta^l_{(\Gamma'_{\fv e},\Gamma'_{\fv t}),A,\Delta} \pcomp ((\Gamma'_{\fv e},\Gamma'_{\fv t}) \rtensortimes \llbracket V\rrbracket) \pcomp \llbracket\sigma\rrbracket))\\
    & \hspace{4em}\mbox{by thunkability of }V\mbox{ and }\delta^l\\
    &= \llbracket\tsigma\rrbracket \ncomp ((((\llbracket e\rrbracket \lparrtimes \Delta'_{\fcv t}) \ncomp (\delta^l_{\Gamma'_{\fv e},B,\Delta'_{\fcv t}} \pcomp (\Gamma'_{\fv e} \rtensortimes \llbracket t\rrbracket))) \lparrtimes \Delta) \ncomp ((\llbracket\sigma'\rrbracket \lparrtimes \Delta) \ncomp (\delta^l_{\Gamma',A,\Delta} \pcomp (\Gamma' \rtensortimes \llbracket V\rrbracket))))\\
    &\hspace{4em}\mbox{by centrality and compatibility with the distributor of symmmetries}\\
    &= (\llbracket\tsigma'\rrbracket \ncomp ((\llbracket e\rrbracket \lparrtimes \Delta'_{\fcv t}) \ncomp (\delta^l_{\Gamma'_{\fv e},B,\Delta'_{\fcv t}} \pcomp (\Gamma'_{\fv e} \rtensortimes \llbracket t\rrbracket))) \lparrtimes \Delta) \ncomp ((\llbracket\sigma'\rrbracket \lparrtimes \Delta) \ncomp (\delta^l_{\Gamma',A,\Delta} \pcomp (\Gamma' \rtensortimes \llbracket V\rrbracket)))\\
    & \hspace{4em}\mbox{by linearity of }\llbracket \tsigma\rrbracket \\
    &= (\llbracket \perfectcut{t}{e}\eps \rrbracket \lparrtimes \Delta) \ncomp (\delta^l_{\Gamma',A,\Delta} \pcomp (\Gamma' \rtensortimes \llbracket V\rrbracket))
\end{aligned}$
\end{adjustbox}\vspace*{\medskipamount}
  where:
\begin{align*}
&\sigma \in \Sigma((\Gamma',\Gamma), (\Gamma'_{\fv e},\Gamma'_{\fv t\setminus\{x\}},\Gamma))\\
&\sigma' \in \Sigma((\Gamma',x:A), (\Gamma'_{\fv e},\Gamma'_{\fv t\setminus\{x\}},x:A))\\
&\tsigma\in \Sigma((\Delta'_{\fcv e},\Delta'_{\fcv t},\Delta),(\Delta', \Delta))\\
&\tsigma'\in \Sigma((\Delta'_{\fcv e},\Delta'_{\fcv t}),\Delta')
\end{align*}
  The other cases are also straightforward, by using induction and the
  compatibility of the operations we are using.
\end{proof}

The following lemma is exactly the symmetric of the previous and is proved accordingly.
\begin{lemma}\label[lemma]{soundstack}
  Let a derivation of $\Gamma\ |\ S : A\vdash \Delta$. We consider $\llbracket S\rrbracket \in \Dduploid(\Gamma \otimes A, \Delta)$ its interpretation.
  \begin{itemize}
    \item For any derivation of $\command c : (\Gamma'\vdash\alpha:A,\Delta')$, there exists a derivation of $\command c[S/\alpha] : (\Gamma,\Gamma'\vdash\Delta,\Delta')$ such that:
      \[
        \llbracket c[S/\alpha]\rrbracket = (\llbracket S\rrbracket \lparrtimes \Delta') \ncomp (\delta^l_{\Gamma,A,\Delta'} \pcomp (\Gamma \rtensortimes \llbracket c\rrbracket))
      \]
    \item For any derivation of $\Gamma'\vdash t : B\ |\ \alpha:A,\Delta'$, there exists a derivation of $\Gamma,\Gamma'\vdash t[S/\alpha] : B\ |\ \Delta,\Delta'$ such that:
      \[
        \llbracket t[S/\alpha]\rrbracket = ((\sigma'_{B, \Delta} \lparrtimes \Delta') \ncomp (\llbracket S\rrbracket \lparrtimes (B \parr \Delta')) \ncomp (\delta^l_{\Gamma,A,B \parr\Delta'} \pcomp (\Gamma \rtensortimes ((\sigma'_{A,B} \lparrtimes \Delta') \llbracket t\rrbracket))
      \]
    \item For any derivation of $\Gamma'\ |\ e : B\vdash \alpha:A,\Delta'$, there exists a derivation of $\Gamma,\Gamma'\ |\ e[S/\alpha] : B\vdash \Delta,\Delta'$ such that:
      \[
        \llbracket e[S/\alpha]\rrbracket = (\llbracket S\rrbracket \lparrtimes \Delta') \ncomp (\delta^l_{\Gamma,A,\Delta'} \pcomp (\Gamma \rtensortimes \llbracket e\rrbracket))
      \]
  \end{itemize}
\end{lemma}

\noindent We now prove the \emph{sound subject reduction} lemma.

\begin{lemma}
  $\triangleright_{RE}$ preserves typing, and, when restricted to typed terms, $\triangleright_{RE}$ preserves the interpretation.
\end{lemma}

\begin{proof}
  We reason by case analysis. We will treat in details the case of $(R\negN)$ and $(E\negN)$.

  $(R\negN)$ For any $\command c = \perfectcut{[S]}{\tmu[\alpha].\command c'}\plus \triangleright_R \command c'[S/\alpha]$ and derivation of $\command c :(\Gamma \vdash \Delta)$, by applying \cref{cohgen}, we have a negative type $N$ and an equivalent derivation of the form:
  \[
    \begin{prooftree}
      \hypo{\command c' : (\Gamma_{\fv \command c'}\vdash\alpha : N,\Delta_{\fcv \command c'\setminus\{\alpha\}})} 
      \infer1[$(\negN\vdash)$]{\Gamma_{\fv \command c'}\ |\ \tmu[\alpha].\command c' : \dupldual N \vdash \Delta_{\fcv \command c'\setminus\{\alpha\}}}
      \hypo{\Gamma_{\fv S}\ |\ S : N\vdash \Delta_{\fcv S}}
      \infer1[$(\vdash\negN)$]{\Gamma_{\fv S} \vdash [S] :\dupldual N\ |\ \Delta_{\fcv S}}
      \infer2[$(\mathsf{cut}\plus)$]{\perfectcut{[S]}{\tmu[\alpha].\command c'}\plus:(\Gamma_{\fv \command c'},\Gamma_{\fv S}\vdash\Delta_{\fcv \command c'\setminus\{\alpha\}},\Delta_{\fcv S})}
      \infer1[$(\sigma,\tsigma)$]{\perfectcut{[S]}{\tmu[\alpha].\command c'}\plus:(\Gamma\vdash\Delta)}
    \end{prooftree}
  \] 
  where $\sigma \in \Sigma(\Gamma,(\Gamma_{\fv S},\Gamma_{\fv \command c}))$ is the unique permutation without renaming from $\Gamma$ to $(\Gamma_{\fv S},\Gamma_{\fv \command c})$ and $\tsigma \in \Sigma((\Delta_{\fcv S},\Delta_{\fcv \command c\setminus\{\alpha\}}),\Delta)$ is the unique permutation without renaming from $(\Delta_{\fcv S},\Delta_{\fcv \command c\setminus\{\alpha\}})$ to $\Delta$.
  
  So, by the previous lemma, we have a derivation of $\command c'[S/\alpha] : (\Gamma\vdash\Delta)$. Moreover, one has:
  \[
    \llbracket\perfectcut{[S]}{\tmu[\alpha].\command c'}\plus:(\Gamma\vdash\Delta)\rrbracket = \llbracket \command c'|S/\alpha]:(\Gamma\vdash\Delta)\rrbracket
  \]
  The proof goes along the lines of the proof of \cref{lemma/commutationSemantic}.

  $(E\negN)$ For any $\tmu[\alpha].\perfectcut{[\alpha]}{S}\plus \triangleright_R S$ and derivation of $\Gamma\ |\ \tmu[\alpha].\perfectcut{[\alpha]}{S'}\plus : \dupldual N \vdash \Delta$, by applying \cref{cohgen}, we have an equivalent derivation of the form:
  \[
    \begin{prooftree}
      \hypo{\Gamma\ |\ S : \dupldual N\vdash \Delta}
      \infer0[$(\mathsf{ax})$]{|\ \alpha : N\vdash \alpha : N}
      \infer1[$(\vdash\negN)$]{\vdash [\alpha] : \dupldual N\ |\ \alpha : N}
      \infer2[$(\mathsf{cut}\plus)$]{\perfectcut{[\alpha]}{S}\plus : (\Gamma\vdash\alpha : N,\Delta)}
      \infer1[$(\negN\vdash)$]{\Gamma\ |\ \tmu[\alpha].\perfectcut{[\alpha]}{S}\plus : \dupldual N \vdash \Delta}
    \end{prooftree}
  \]
  So, one has:
  \begin{align*}
    & \llbracket\Gamma\ |\ \tmu[\alpha].\perfectcut{[\alpha]}{S}\plus : \dupldual N\vdash \Delta\rrbracket \\
     &= \chi^{-1}_{\Gamma,\dupldual N,\Delta}((\nu_N \lparrtimes \Delta) \ncomp \llbracket \perfectcut{[\alpha]}{S}\plus \rrbracket)\\
                                                                                                      &= \chi^{-1}_{\Gamma,\dupldual N,\Delta}((\nu_N \lparrtimes \Delta) \ncomp \sigma'_{\Delta,N} \ncomp (\llbracket S\rrbracket \parr N) \ncomp (\delta^l_{\Gamma,\dupldual N,N} \pcomp (\Gamma \rtensortimes \llbracket [\alpha]\rrbracket)))\\
                                                                                                      &= \chi^{-1}_{\Gamma,\dupldual N,\Delta}((\dupldoubledual N \rparrtimes \llbracket S\rrbracket) \ncomp (\nu_N \lparrtimes (\Gamma \otimes \dupldual N)) \ncomp \sigma'_{\Gamma \otimes \dupldual N,N} \ncomp (\delta^l_{\Gamma,\dupldual N,N} \pcomp (\Gamma \rtensortimes \llbracket [\alpha]\rrbracket)))\\
                                                                                                      &= \llbracket S\rrbracket \pcomp \chi^{-1}_{\Gamma,\dupldual N,\Gamma \otimes \dupldual N}((\nu_N \lparrtimes (\Gamma \otimes \dupldual N)) \ncomp \sigma'_{\Gamma \otimes \dupldual N,N} \ncomp (\delta^l_{\Gamma,\dupldual N,N} \pcomp (\Gamma \rtensortimes \llbracket [\alpha]\rrbracket))) & \mbox{ by naturality of }\chi^{-1}\\
                                                                                                      &= \llbracket\Gamma\ |\ S : \dupldual N\vdash \Delta\rrbracket
  \end{align*}
  The other cases are treated similarly, by using the coherent generation lemma and the sound value/stack substitution.
\end{proof}

\begin{theorem}
  $\rightarrow_{RE}$ preserves typing.
\end{theorem}
\begin{proof}
  We reason by induction on $\rightarrow_{RE}$. On the base case, we use the previous lemma. On other cases, we use \cref{cohgen} and the induction hypothesis.
\end{proof}

\sectioncaps{Syntactically thunkable and central expressions} 

In this section, we will prove the characterizations in the classical $L$-calculus of thunkable morphisms (\cref{lemma/thunkable}) and central morphisms (\cref{lemma/central}).

\subsectioncaps{Proof of the characterization of thunkable morphisms (\cref{lemma/thunkable})}\label{thunkableproofs}
  Let $t$ be an expression. We will prove that the following properties are equivalent:
  \begin{enumerate}
    \item For all $c$, $\perfectcut{t}{\tmu x^\veps.c}^\veps \simeq_{RE} c[t/x]$;
    \item For all $c$ and $e$, $\perfectcut{t}{\tmu x^{\veps_1}.\perfectcut{\mu \alpha^{\veps_2}.c}{e}^{\veps_2}}^{\veps_1} \simeq_{RE} \perfectcut{\mu \alpha^{\veps_2}.\perfectcut{t}{\tmu x^{\veps_1}.c}^{\veps_1}}{e}^{\veps_2}$;
    \item For all $c$, $e$, $q$ of polarity $\veps_1$ and $\tilde q$ of polarity $\veps_2$, $\perfectcut{t}{\tmu q.\perfectcut{\mu \tilde q.c}{e}^{\veps_2}}^{\veps_1} \simeq_{RE} \perfectcut{\mu \tilde q.\perfectcut{t}{\tmu q.c}^{\veps_1}}{e}^{\veps_2}$.
  \end{enumerate}
  We added the property $3$, as it is an intermediate step that simplifies the proof.
\begin{proof}
  ($1 \Rightarrow 2$) Let $c$ be a command and $e$ a context.
  \begin{align*}
    \perfectcut{t}{\tmu x^{\veps_1}.\perfectcut{\mu \alpha^{\veps_2}.\underline c}{e}^{\veps_2}}^{\veps_1} &\simeq_{RE} \perfectcut{t}{\tmu x^{\veps_1}.\perfectcut{\mu \alpha^{\veps_2}.\perfectcut{x}{\tmu x^{\veps_1}.c}^{\veps_1}}{e}^{\veps_2}}^{\veps_1} && (R\tmu^{\veps_1}) \\
                                                                                                           &\simeq_{RE} \perfectcut{\mu \alpha^{\veps_2}.\perfectcut{t}{\tmu x^{\veps_1}.c}^{\veps_1}}{e}^{\veps_2} && \mbox{by using 1.}
  \end{align*}

  ($2 \Rightarrow 3$) Let $c$ be a command, $e$ a context, $q$ of polarity $\veps_1$ and $\tilde q$ of polarity $\veps_2$.
  \begin{align*} 
    &\perfectcut{t}{\tmu q.\perfectcut{\underline{\mu \tilde q.c}}{e}^{\veps_2}}^{\veps_1}\\
    &\simeq_{RE} \perfectcut{t}{\tmu q.\underline{\perfectcut{\mu \alpha^{\veps_2}.\perfectcut{\mu \tilde q.c}{\alpha}^{\veps_2}}{e}^{\veps_2}}}^{\veps_1} && (E\mu^{\veps_2})\\ 
                                                                                          &\simeq_{RE} \perfectcut{t}{\underline{\tmu q.\perfectcut{q}{\tmu x^{\veps_1}.\perfectcut{\mu\alpha^{\veps_2}.\perfectcut{x}{\tmu q.\perfectcut{\mu \tilde q.c}{\alpha}^{\veps_2}}^{\veps_1}}{e}^{\veps_2}}^{\veps_1}}}^{\veps_1} && (Rq)\mbox{ and } (R\tmu^{\veps_1})\\
                                                                                          &\simeq_{RE} \perfectcut{t}{\tmu x^{\veps_1}.\perfectcut{\mu\alpha^{\veps_2}.\perfectcut{x}{\tmu q.\perfectcut{\mu \tilde q.c}{\alpha}^{\veps_2}}^{\veps_1}}{e}^{\veps_2}}^{\veps_1} && (Eq)\\
                                                                                          &\simeq_{RE} \perfectcut{\mu \alpha^{\veps_2}.\perfectcut{t}{\underline{\tmu x^{\veps_1}.\perfectcut{x}{\tmu q.\perfectcut{\mu \tilde q.c}{\alpha}^{\veps_2}}^{\veps_1}}}^{\veps_1}}{e}^{\veps_2} &&\mbox{by using 2.}\\
                                                                                          &\simeq_{RE} \perfectcut{\underline{\mu \alpha^{\veps_2}.\perfectcut{t}{\tmu q.\perfectcut{\mu \tilde q.c}{\alpha}^{\veps_2}}^{\veps_1}}}{e}^{\veps_2} && (E\tmu^{\veps_1})\\
                                                                                          &\simeq_{RE} \perfectcut{\mu \tilde q.\perfectcut{\underline{\mu \alpha^{\veps_2}}.\perfectcut{t}{\tmu q.\perfectcut{\underline{\mu \tilde q}.c}{\alpha}^{\veps_2}}^{\veps_1}}{\tilde q}^{\veps_2}}{e}^{\veps_2} && (E\tilde q)\\
                                                                                          &\simeq_{RE} \perfectcut{\mu \tilde q.\perfectcut{t}{\tmu q.c}^{\veps_1}}{e}^{\veps_2} && (R\mu^{\veps_2})\mbox{ and }(R\tilde q)
  \end{align*}

  ($3 \Rightarrow 1$) We first need to prove two intermediary results. We introduce the following notations :
  \[
  \{V\} := []\parr V \quad \mu\{\alpha\}.c := \mu(\beta\parr\alpha).\perfectcut{\mu[].c}{\beta}^-
  \]
  If $t$ verifies the property 3, then, for all context $e$, one has :
  \[
    \perfectcut{t}{\tmu x^\veps.\perfectcut{\mu\{\alpha\}.\perfectcut{x}{\alpha}^\veps}{e}^-}^\veps \simeq_{RE} \perfectcut{\mu\{\alpha\}.\perfectcut{t}{\alpha}^\veps}{e}^-
  \]
  \begin{proof}
    One has
    \begin{align*}
      &\perfectcut{t}{\tmu x^\veps.\perfectcut{\mu(\beta\parr\alpha).\perfectcut{\mu[].\perfectcut{x}{\alpha}^\veps}{\beta}^-}{e}^-}^\veps\\
      &\simeq_{RE} \perfectcut{\mu(\beta\parr\alpha).\perfectcut{t}{\tmu x^\veps.\perfectcut{\mu[].\perfectcut{x}{\alpha}^\veps}{\beta}^-}^\veps}{e}^- && \mbox{by using 3.}\\
                                                                                                                                          &\simeq_{RE} \perfectcut{\mu(\beta\parr\alpha).\perfectcut{\mu[].\perfectcut{t}{\underline{\tmu x^\veps.\perfectcut{x}{\alpha}^\veps}}^\veps}{\beta}^-}{e}^- && \mbox{by using 3.}\\
                                                                                                                                          &\simeq_{RE} \perfectcut{\mu(\beta\parr\alpha).\perfectcut{\mu[].\perfectcut{t}{\alpha}^\veps}{\beta}^-}{e}^- && (E\tmu^\veps)\qedhere
    \end{align*}
  \end{proof}

  For $u$ a term, we have the following equation :
  \[
    \mu\alpha^{\veps}.\perfectcut{\mu\{\beta\}.\perfectcut{u}{\beta}^{\veps}}{\{\alpha\}}^- \simeq_{RE} u
  \]
  \begin{proof}
    One has
    \begin{align*}
      &\mu\alpha^{\veps}.\perfectcut{\underline{\mu(\gamma \parr \beta)}.\perfectcut{\mu[].\perfectcut{u}{\beta}^{\veps}}{\gamma}^-}{[]\parr\alpha}^-\\
      &\simeq_{RE} \mu\alpha^\veps.\perfectcut{\underline{\mu[]}.\perfectcut{u}{\alpha}^\veps}{[]}^-&& (R\parr)\\
                                                                                                                                                     &\simeq_{RE} \underline{\mu\alpha^\veps.\perfectcut{u}{\alpha}^\veps} && (R\bot)\\
                                                                                                                                                   &\simeq_{RE} u &&(E\mu^\veps)\qedhere
    \end{align*}
  \end{proof}
  Now, we can prove the result.
  \begin{align*}
    \perfectcut{t}{\tmu x^\veps. c}^\veps &\simeq_{RE} \perfectcut{t}{\tmu x^\veps.\underline{c[\mu\alpha^{\veps}.\perfectcut{\mu\{\beta\}.\perfectcut{x}{\beta}^{\veps}}{\{\alpha\}}^-/x]}}^\veps\\
                                          &\simeq_{RE} \perfectcut{t}{\tmu x^\veps.\perfectcut{\mu\{\beta\}.\perfectcut{x}{\beta}^\veps}{\tmu y^-.c[\mu\alpha^\veps.\perfectcut{y}{\{\alpha\}}^-/x]}}^\veps && (R\tmu^-)\\
                                          &\simeq_{RE} \perfectcut{\mu\{\beta\}.\perfectcut{t}{\beta}^\veps}{\underline{\tmu y^-}.c[\mu\alpha^\veps.\perfectcut{y}{\{\alpha\}^-}/x]} && \mbox{proved above}\\
                                          &\simeq_{RE} c[\mu\alpha^\veps.\perfectcut{\mu\{\beta\}.\perfectcut{t}{\beta}^\veps}{\{\alpha\}}^-/x] && (R\tmu^-)\\
                                          &\simeq_{RE} c[t/x]&&\qedhere
  \end{align*}
\end{proof}

\subsectioncaps{Proof of the characterization of central morphisms (\cref{lemma/central})}\label{centralproofs}
\begin{proof}
  One has
  \begin{align*}
     &\perfectcut{t}{\tmu q.\perfectcut{u}{\underline{\tmu q'.c}}^{\veps_2}}^{\veps_1}\\
     &\simeq_{RE} \perfectcut{t}{\tmu q.\underline{\perfectcut{u}{\tmu x^{\veps_2}.\perfectcut{x}{\tmu q'.c}^{\veps_2}}^{\veps_2}}}^{\veps_1} && (E\tmu^{\veps_2})\\
                                                                                     &\simeq_{RE} \perfectcut{t}{\underline{\tmu q.\perfectcut{q}{\tmu y^{\veps_1}. \perfectcut{u}{\tmu x^{\veps_2}. \perfectcut{y}{\tmu q. \perfectcut{x}{\tmu q'.c}^{\veps_2}}^{\veps_1}}^{\veps_2}}^{\veps_1}}}^{\veps_1} && (Rq) \mbox{ and } (R\tmu^{\veps_1})\\
                                                                                     &\simeq_{RE} \perfectcut{t}{\tmu y^{\veps_1}.\perfectcut{u}{\tmu x^{\veps_2}. \perfectcut{y}{\tmu q. \perfectcut{x}{\tmu q'.c}^{\veps_2}}^{\veps_1}}^{\veps_2}}^{\veps_1} && (Eq)\\
                                                                                     &\simeq_{RE} \perfectcut{u}{\tmu x^{\veps_2}.\perfectcut{t}{\underline{\tmu y^{\veps_1}. \perfectcut{y}{\tmu q. \perfectcut{x}{\tmu q'.c}^{\veps_2}}^{\veps_1}}}^{\veps_1}}^{\veps_2} && \mbox{by centrality of }t\\
                                                                                     &\simeq_{RE} \perfectcut{u}{\underline{\tmu x^{\veps_2}.\perfectcut{t}{\tmu q. \perfectcut{x}{\tmu q'.c}^{\veps_2}}^{\veps_1}}}^{\veps_2} &&(E\tmu^\veps_1)\\
                                                                                     &\simeq_{RE} \perfectcut{u}{\tmu q'.\perfectcut{q'}{\underline{\tmu x^{\veps_2}}. \perfectcut{t}{\underline{\tmu q}. \perfectcut{x}{\tmu q'.c}^{\veps_2}}^{\veps_1}}^{\veps_2}}^{\veps_2} && (Eq')\\
                                                                                     &\simeq_{RE} \perfectcut{u}{\tmu q'.\perfectcut{t}{\tmu q.c}^{\veps_1}}^{\veps_2} && (R\tmu^{\veps_2})\mbox{ and }(Rq')\qedhere
  \end{align*}
\end{proof}

\sectioncaps{A direct equational proof of the \FH{} theorem}\label{section/detailedFH}
\begin{lemma}\label[lemma]{lemma/commutationSemantic}
  Let $f \in \mathcal D(B,C)$ and $g \in \mathcal D(A,B)$ be two morphisms of $\mathcal D$. Let us note
\[
  \varphi_D = \chi^{-1}_{A,\dupldual D,\bot} : \Dduploid(A,\dupldoubledual D\parr\bot)\xrightarrow{\iso}\Dduploid(A\otimes \dupldual D, \bot)
\]
 We have:
  \[
    f \ccomp g = \nu_C \ccomp (\lambda'_{\dupldoubledual C} \circ \varphi_C^{-1}(\varphi_B((\nu^{-1}_B \lparrtimes \bot) \circ (\lambda'^{-1}_B \ccomp g)) \bullet (A \rtensortimes \dupldual f)))
  \]
\end{lemma}
\begin{proof} One has:
  \begin{align*}
    f \ccomp g &= f \ccomp ((\lambda'_B \circ \lambda'^{-1}_B) \ccomp g)\\ 
    &= f \ccomp \lambda'_B \circ (\lambda'^{-1}_B \ccomp g) && \mbox{by linearity of }\lambda'\\ 
    &= (\nu_C \ccomp (\dupldoubledual f \ccomp \nu^{-1}_B)) \ccomp \lambda'_B \circ (\lambda'^{-1}_B \ccomp g) && \mbox{by naturality of }\nu\\ 
    &= \nu_C \ccomp ((\dupldoubledual f \ccomp \nu^{-1}_B) \ccomp \lambda'_B \circ (\lambda'^{-1}_B \ccomp g)) && \mbox{by linearity of }\nu\\ 
    &= \nu_C \ccomp (\lambda'_{\dupldoubledual C} \circ ((\dupldoubledual f \ccomp \nu^{-1}_B) \lparrtimes \bot) \circ (\lambda'^{-1}_B \ccomp g))\\ 
    &= \nu_C \ccomp (\lambda'_{\dupldoubledual C} \circ (\dupldoubledual f\lparrtimes \bot) \circ (\nu^{-1}_B\lparrtimes \bot) \circ (\lambda'^{-1}_B \ccomp g))\\ 
    &= \nu_C \ccomp (\lambda'_{\dupldoubledual C} \circ (\dupldoubledual f \lparrtimes \bot) \circ \varphi_B^{-1}(\varphi_B((\nu^{-1}_B \lparrtimes \bot) \circ (\lambda'^{-1}_B \ccomp g))))\\ 
    &= \nu_C \ccomp (\lambda'_{\dupldoubledual C} \circ \varphi_C^{-1}(\varphi_B((\nu^{-1}_B \lparrtimes \bot) \circ (\lambda'^{-1}_B \ccomp g)) \bullet (A \rtensortimes \dupldual f))) && \mbox{by \cref{chinaturality}}\qedhere
  \end{align*}
\end{proof}
\begin{theorem}
  A morphism of $\mathcal D$ is thunkable if and only if it is central for $\otimes$.
\end{theorem}
\begin{proof}
  We know by definition that thunkable morphisms are central for
  $\otimes$, so we only have to prove that central morphisms are
  thunkable.
Let $A,B,C,D \in|\mathcal D|$ and $f \in \mathcal D(C,D)$,
  $g \in \mathcal D(B,C)$ and $h \in \mathcal D(A,B)$ such that $h$ is
  central for $\otimes$. Let us note as before
\[
  \varphi_E = \chi^{-1}_{B,\dupldual E,\bot} : \Dduploid(B,\dupldoubledual E\parr\bot)\xrightarrow{\iso}\Dduploid(B\otimes \dupldual E, \bot)
\]
  One has:
  \begin{align*}
    & (f \ccomp g) \ccomp h \\
    &= (\nu_D \ccomp (\lambda'_{\dupldoubledual D} \circ \varphi_D^{-1}(\varphi_C((\nu^{-1}_C \lparrtimes \bot) \circ (\lambda'^{-1}_C \ccomp g)) \bullet (B \rtensortimes\dupldual f)))) \ccomp h && \mbox{by the previous lemma}\\
    &= \nu_D \ccomp ((\lambda'_{\dupldoubledual D} \circ \varphi_D^{-1}(\varphi_C((\nu^{-1}_C \lparrtimes \bot) \circ (\lambda'^{-1}_C \ccomp g)) \bullet (B \rtensortimes \dupldual f))) \ccomp h) && \mbox{by linearity of }\lambda'\\ 
    &= \nu_D \ccomp (\lambda'_{\dupldoubledual D} \circ (\varphi_D^{-1}(\varphi_C((\nu^{-1}_C \lparrtimes \bot) \circ (\lambda'^{-1}_C \ccomp g)) \bullet (B \rtensortimes\dupldual f)) \ccomp h))  && \mbox{by linearity of }\lambda'\\ 
    &= \nu_D \ccomp (\lambda'_{\dupldoubledual D} \circ (\varphi_D^{-1}(\varphi_C((\nu^{-1}_C \lparrtimes \bot) \circ (\lambda'^{-1}_C \ccomp g)) \bullet (B \rtensortimes \dupldual f) \bullet (h \ltensortimes \dupldual D))))  && \mbox{by \cref{chinaturality}}\\ 
    &= \nu_D \ccomp (\lambda'_{\dupldoubledual D} \circ (\varphi_D^{-1}(\varphi_C((\nu^{-1}_C \lparrtimes \bot) \circ (\lambda'^{-1}_C \ccomp g)) \bullet (h \ltensortimes \dupldual D) \bullet (B \rtensortimes \dupldual f)))) && \mbox{by centrality of }h\\ 
    &= \nu_D \ccomp (\lambda'_{\dupldoubledual D} \circ (\varphi_D^{-1}(\varphi_C(((\nu^{-1}_C \lparrtimes \bot) \circ (\lambda'^{-1}_C \ccomp g)) \ccomp h) \bullet (B \rtensortimes \dupldual f)))) && \mbox{by \cref{chinaturality}}\\ 
    &= \nu_D \ccomp (\lambda'_{\dupldoubledual D} \circ (\varphi_D^{-1}(\varphi_C((\nu^{-1}_C \lparrtimes \bot) \circ ((\lambda'^{-1}_C \ccomp g) \ccomp h)) \bullet (B \rtensortimes \dupldual f)))) && \hspace{-1.75mm} \begin{array}{l} \mbox{by lin. of }\nu^{-1} \\ \mbox{preserved by } \lparrtimes \end{array}\\ 
    &= \nu_D \ccomp (\lambda'_{\dupldoubledual D} \circ (\varphi_D^{-1}(\varphi_C((\nu^{-1}_C \lparrtimes \bot) \circ (\lambda'^{-1}_C \ccomp (g \ccomp h))) \bullet (B \rtensortimes \dupldual f)))) && \mbox{by linearity of }\lambda'^{-1}\\ 
    &= f \ccomp (g \ccomp h)  && \mbox{by the previous lemma}
  \end{align*}
  So $h$ is thunkable. This concludes the proof.
\end{proof}

\newpage
\addtocontents{toc}{\protect\setcounter{tocdepth}{-1}}\tableofcontents{}
\addtocontents{toc}{\protect\setcounter{tocdepth}{3}}
\vspace*{-15ex}
\vfill
\listoffigures{}
\vspace*{-5ex}

\end{document}